\PassOptionsToPackage{table,xcdraw}{xcolor}
\documentclass[11pt,a4paper]{article}
\pdfoutput=1
\usepackage{jheppub}
\usepackage{gensymb}
\DeclareSymbolFont{matha}{OML}{txmi}{m}{it}
\DeclareMathSymbol{\varv}{\mathord}{matha}{118}
\usepackage{amsmath}
\usepackage{amssymb}
\usepackage{graphicx}
\usepackage{subfigure}
\usepackage{pstricks}
\usepackage{bm}
\usepackage{pbox}
\usepackage{placeins}
\usepackage{booktabs}
\usepackage[T1]{fontenc}
\usepackage{footnote}
\usepackage{pdfpages}
\usepackage{hhline}
\usepackage{multirow}
\usepackage{multicol}
\usepackage{enumitem}
\usepackage[toc,page]{appendix}
\allowdisplaybreaks
\usepackage{comment}
\usepackage{float}
\usepackage{slashed}
\usepackage{xcolor}
\usepackage{textcomp}
\usepackage{multirow}
\usepackage[numbers]{natbib}
\usepackage{notoccite}
\usepackage{soul}
\usepackage{extarrows}

\usepackage{physics}
\usepackage{comment}
\usepackage{dsfont}
\usepackage{blindtext}
\usepackage{csquotes}
\usepackage{soul}
\usepackage[none]{hyphenat}
\DeclareUnicodeCharacter{0304}{ }
\usepackage{rotating}
\usepackage{tabularx}

\usepackage[many]{tcolorbox}

\renewcommand{\thetable}{\Roman{table}}

\definecolor{MyDarkBlue}{rgb}{0.1, 0.1, 0.8} 
\definecolor{MyLightBlue}{rgb}{0.22,0.51,0.9}
\definecolor{MyGreen}{rgb}{0.0, 0.5, 0.0}
\definecolor{BrickRed}{rgb}{0.8, 0.25, 0.33}
\definecolor{magenta}{rgb}{1.0,0.0,1.0}
\definecolor{brightgreen}{rgb}{0.4,1,0}
\definecolor{orange}{rgb}{1.0, 0.65, 0.0}

\newcommand{\BG}{\textcolor{teal}}

\newcommand{\MDB}{\textcolor{MyDarkBlue}}

\usepackage{tikz}

\newif\ifstartedinmathmode
\newcommand\encircled[1]{%
  \relax\ifmmode\startedinmathmodetrue\else\startedinmathmodefalse\fi%
  \tikz[baseline,anchor=base]{%
  \node[color=red,draw=MyDarkBlue,circle,thick,outer sep=0pt,inner sep=.2ex]
    {\ifstartedinmathmode$#1$\else#1\fi};}%
}
\newif\ifstartedinmathmode
\newcommand\twincircled[1]{%
  \relax\ifmmode\startedinmathmodetrue\else\startedinmathmodefalse\fi%
  \tikz[baseline,anchor=base]{%
    \draw [MyDarkBlue, thick] circle [radius=.32](0,0) ;
  \node[color=red,draw=brightgreen,circle,thick,outer sep=0pt,inner sep=0]
    {\ifstartedinmathmode$#1$\else#1\fi};}%
}

\RequirePackage{hyperref}
\hypersetup{colorlinks=true, citecolor=MyGreen,linkcolor=BrickRed, urlcolor=MyLightBlue,pagecolor=false}

\title{\bf Neutrino masses and magnetic moments of electron and muon in the Zee Model }
\author[a]{Rahool Kumar Barman,}
\author[b]{Ritu Dcruz,}
\author[c]{Anil Thapa}

\affiliation[a,b]{Department of Physics, Oklahoma State University, Stillwater, OK 74078, USA}
\affiliation[c]{Department of Physics, University of Virginia, Charlottesville, Virginia 22904-4714, USA}
\emailAdd{rahool.barman@okstate.edu}\emailAdd{rdcruz@okstate.edu}\emailAdd{wtd8kz@virginia.edu}

\abstract{We explore parameter space in the Zee Model to resolve the long-standing tension of the electron and muon anomalous magnetic moment (AMM). The model comprises a second Higgs doublet and a charged singlet at electroweak scale and generates Majorana neutrino masses at one-loop level; the neutral partner of the $SU(2)_L$ doublet contributes to the AMM of electron and muon via one loop and two-loop corrections. We propose two minimal flavor structures that can explain these anomalies while fitting the neutrino oscillation data. We find that the neutral Higgs resides in the mass range of roughly 10-300 GeV or 1-30 GeV, depending on the flavor structures. The model is consistent with constraints from colliders, electroweak precision data, and lepton flavor violation. To be comprehensive, we examine the constraints from the electric dipole moment (EDM) and find a region of parameter space that gives a sizable contribution to muon EDM while simultaneously giving corrections to muon AMM. In addition to the light scalar, the two charged scalars with masses as low as 100 GeV can induce nonstandard neutrino interactions $\varepsilon_{ee}$ as large as $8\%$, potentially hinting at new physics. We also investigate the projected capability of future lepton colliders to probe the currently allowed parameter space consistent with both electron and muon AMMs via direct searches in the $\ell^{+}\ell^{-}\to \ell^{+}\ell^{-}(H \to \ell^{+}\ell^{-})$ channel.   
}

\begin{document}
\maketitle

\clearpage
\begin{sloppypar}

\section{Introduction\label{sec:intro}}

Understanding the origin of neutrino masses to explain the firmly established observed neutrino oscillation data \cite{ParticleDataGroup:2020ssz} stands out among the many reasons to consider Physics beyond the Standard Model (BSM). The tiny neutrino mass can be realized by the dimension-five Weinberg operator \cite{Weinberg:1979sa} that the breaks lepton number by two units and gives Majorana masses to neutrinos after electroweak symmetry breaking. This operator can be realized at tree level by adding SM-singlet fermions via a type-I seesaw mechanism \cite{Minkowski:1977sc, Mohapatra:1979ia, Yanagida:1980xy, Gell-Mann:1979vob, Glashow:1979nm}, adding an $SU(2)_L$-triplet scalar \cite{Schechter:1980gr, Cheng:1980qt, Mohapatra:1980yp, Lazarides:1980nt} via type-II, or $SU(2)_L$ triplet fermion \cite{Foot:1988aq} via type-III. An alternative and interesting scenario where small neutrino masses arise naturally are through quantum corrections \cite{Zee:1980ai, Cheng:1980qt, Zee:1985id, Babu:1988wk} (For a review, see Ref. \cite{Cai:2017jrq, Babu:2019mfe}). The new degrees of freedom that generate neutrino mass in these radiative models cannot be too heavy, and therefore, they can be accessible for the experimental test at colliders. Moreover, these new particles typically give rise to enhanced lepton flavor violating (LFV) signals such as $\mu \to e \gamma$ and $\tau \to 3 \mu$. Here we revisit the Zee model \cite{Zee:1980ai}, the simplest extension of SM that contains an extra doublet scalar and a singly-charged scalar that can generate Majorana neutrino mass at a one-loop level. The new Higgs doublet present in the model can also play an important role in explaining persistent experimental anomalies, viz., the anomalous magnetic moment (AMM) of muon ($\Delta a_\mu$) and electron ($\Delta a_e$). 

The long-standing discrepancy between the experiment and theory in the anomalous magnetic moment of lepton $a_\ell$ hints at physics beyond the SM, where $a_\ell = (g_\ell-2)/2$ in SM is calculated from perturbative expansion in the fine-structure constant $\alpha_{em}$. For instance, the one-loop QED effect~\cite{Schwinger:1948iu, Kusch:1948mvb} gives a deviation of $0.1\%$ from the Dirac prediction in the Land\'e $g$-factor at tree level $g=2$. The contribution to $a_{\ell}^{SM}$ arises from loops containing Quantum Electrodynamics (QED) corrections, hadronic (QCD) processes, and electroweak (EW) pieces. The QED calculations~\cite{Sommerfield:1957zz,Petermann:1957hs,Kinoshita:1981vs,Kinoshita:1990wp,Laporta:1996mq,Degrassi:1998es,Czarnecki:1998nd,Kinoshita:2004wi,Kinoshita:2005sm, Passera:2006gc,Kataev:2006yh,Aoyama:2007mn,Karshenboim:2008zz,Aoyama:2012wk,Schnetz:2017bko,Aoyama:2017uqe,Volkov:2017xaq,Volkov:2018jhy} have been carried out up to and including $\mathcal{O}(\alpha_{em}^5$) while electroweak corrections~\cite{Czarnecki:1995wq,Czarnecki:1995sz,Czarnecki:1996if,Czarnecki:2002nt,Heinemeyer:2004yq,Gribouk:2005ee,Gnendiger:2013pva} have been evaluated at full two-loop order with negligible uncertainty, arising mainly from nonperturbative effects in two-loop diagrams involving the light quarks. Note that the dominant sources of theoretical uncertainty in AMM arise from the hadronic contributions~\cite{Jegerlehner:1985gq,Lynn:1985sq,Swartz:1995hc,Martin:1994we,Eidelman:1998vc,Krause:1996rf,Davier:1998si,Jegerlehner:2003qp,deTroconiz:2004yzs,Davier:2007ua,Campanario:2019mjh}, in particular,
the $\mathcal{O}(\alpha_{em}^2$) Hadronic Vacuum Polarization~(HVP) term~\cite{Davier:2017zfy,Keshavarzi:2018mgv,Colangelo:2018mtw,Hoferichter:2019mqg,Davier:2019can,Keshavarzi:2019abf,Kurz:2014wya} and the $\mathcal{O}(\alpha_{em}^3$) hadronic light-by-light~(HLbL)~\cite{Bijnens:1995xf,Hayakawa:1997rq,Knecht:2001qf,Knecht:2001qg,Ramsey-Musolf:2002gmi,Melnikov:2003xd,Bijnens:2007pz,Prades:2009tw,Kataev:2012kn,Kurz:2014wya,Masjuan:2017tvw,Colangelo:2017fiz,Hoferichter:2018kwz,Gerardin:2019vio,Bijnens:2019ghy,Colangelo:2019uex,Colangelo:2014qya,Pauk:2014rta,Danilkin:2016hnh,Jegerlehner:2017gek,Knecht:2018sci,Eichmann:2019bqf,Roig:2019reh,Colangelo:2014qya,Blum:2019ugy} term. 

The current measurement of AMM of muon $a_\mu$ at Fermilab National Accelerator Laboratory~(FNAL) reports $a_\mu(\textnormal{FNAL})=116592040(54)\times  10^{-11}$ ~\cite{Muong-2:2021ojo}, which agrees with the previous Brookhaven National Laboratory~(BNL) E821 measurement~\cite{Muong-2:2006rrc,Muong-2:2001kxu}, while theoretical prediction finds it to be $a_\mu^{\textnormal{SM}}=116591810(43)\times 10^{-11}$ \cite{Davier:2019can, Aoyama:2020ynm,Davier:2010nc,Gerardin:2020gpp}. The difference, $\Delta a_\mu = a_\mu (\text{experiment})- a_\mu (\text{theory}) \simeq (251\pm59)\times 10^{-11}$, is a $4.2\sigma$ discrepancy. In addition to muon AMM, similar discrepancy of $\sim2.4\sigma$ between the experimental \cite{Hanneke:2008tm,Parker:2018vye} and theoretical prediction \cite{Aoyama:2020ynm,Aoyama:2012wj,Laporta:2017okg,Aoyama:2017uqe,Volkov:2019phy} of electron AMM has been observed, $\Delta a_e = (-8.8\pm3.6)\times 10^{-13}$.  These deviations may hint at new physics lying around or below the TeV scale. It is worth mentioning that a more recent measurement of the fine structure constant using Rubidium atoms~\cite{Morel:2020dww} instead of Cesium atoms has led to a $1.6\sigma$ discrepancy of the AMM, but in the opposite direction; $\Delta a_e=(4.8\pm3.0)\times 10^{-13}$. This is in complete disagreement with the previous result for unknown reasons. Since there is an ambiguity between the two measurements, we stick with the previous measurement in this work. Although it is not difficult to explain one of the $\Delta a_\ell$ in BSM models, it is challenging to explain both simultaneously because of opposite signs of the two AMMs. Various mechanisms have been proposed to explain these deviations; by invoking vector-like fermions~\cite{Crivellin:2018qmi,Hiller:2019mou,Chun:2020uzw,Chen:2020tfr,Hati:2020fzp,Escribano:2021css,Hernandez:2021tii,Borah:2021khc,Bharadwaj:2021tgp}, introducing new scalars~\cite{Davoudiasl:2018fbb,Liu:2018xkx,Han:2018znu,Bauer:2019gfk,Cornella:2019uxs,Dutta:2020scq,Endo:2020mev,Haba:2020gkr,Hernandez:2021xet,Mondal:2021vou,Adhikari:2021yvx,Bauer:2021mvw,Bharadwaj:2021tgp,De:2021crr,Hue:2021xzl}, leptoquarks~\cite{Keung:2021rps,Bigaran:2020jil,Dorsner:2020aaz,Bigaran:2021kmn}, extending the gauge symmetry~\cite{Abdullah:2019ofw,CarcamoHernandez:2020pxw,CarcamoHernandez:2019ydc,Bodas:2021fsy,Chowdhury:2021tnm,Hernandez:2021iss}, considering non-local QED effects~\cite{He:2019uvu} and in the context of supersymmetry~\cite{Badziak:2019gaf,Endo:2019bcj,Dong:2019iaf,Yang:2020bmh,Cao:2021lmj,Frank:2021nkq,Li:2021koa}. Moreover, there are various attempts in the literature that use Higgs doublet models~(THDM) to explain both the anomalies ~\cite{DelleRose:2020oaa,Botella:2020xzf,Jana:2020pxx,Fajfer:2021cxa}. We pursue similar in spirit as done in THDM, but unlike THDM, we address these anomalies in the context of radiative neutrino masses in the Zee model that gives direct connection to the neutrino massses and oscillations.

The flavor-changing nature of the second Higgs doublet that gives rise to large contributions to $a_\ell$ also plays a crucial role in neutrino masses and oscillations, which reveals insights into the flavor structure within the model. The opposite signs of the anomalies can be explained by either choosing Yukawa couplings with opposite signs or adjusting the mass splitting between the scalars so that $\Delta a_\mu$ is positive from the dominant one-loop diagram. In contrast, the two-loop Barr-Zee diagram provides negative correction to $a_e$. We propose two minimal Yukawa textures that achieve these while also providing excellent fits to the neutrino oscillation parameters. We find that the simultaneous explanation of the observed disparities in the lepton AMMs requires the scalar mass to be in the range of $10-300$ GeV for {\tt TX-I} and $1-30$ GeV for {\tt TX-II} while satisfying various LFV measurements such as $\ell_1\to\ell_2\gamma$, including other relevant constraints from low energy physics, collider searches, and fit to the neutrino oscillation data. It is also worth noting that the Yukawa couplings, if complex, can not only generate $a_\ell$ but could also have a sizable effect on the electric dipole moment (EDM) $d_\ell$. We find that an extremely small complex phase of $\mathcal{O}(10^{-5})$ is required to satisfy the current limits of electron EDM \cite{ACME:2018yjb} while simultaneously satisfying $a_e$. However, this is not the case for muon EDM \cite{Muong-2:2008ebm}, as there exist regions of parameter space that can give a sizable contribution to muon EDM, which can potentially be measured in future experiments \cite{Abe:2019thb,Sato:2021aor,Adelmann:2021udj}. The charged scalars in the model can induce nonstandard neutrino interactions (NSI) $\varepsilon_{ij}~ $\cite{Wolfenstein:1977ue,Wolfenstein:1979ni,Proceedings:2019qno} (for a recent review on NSI in the context of the Zee Model, see Ref.~\cite{Babu:2019mfe}), which, if observed, could be direct indicators for new physics. We find that diagonal NSI can be as large as $8\%$ for $\varepsilon_{ee}$.  

We have also evaluated constraints on the Yukawa couplings for both the textures from direct searches in the $e^{+}e^{-} \to \ell^{+}\ell^{-}(H \to \ell^{+}\ell^{-})$~($\ell = e, \mu$) channel at LEP. Moreover, we also explore the future sensitivity of the channel, as mentioned earlier, to the Yukawa couplings at the ILC operating at $\sqrt{s}=1~\text{TeV}$ with an integrated luminosity $\mathcal{L}=500~\text{fb}^{-1}$~\cite{Behnke:2013xla,Baer:2013cma,Adolphsen:2013jya,Adolphsen:2013kya,Behnke:2013lya}. We also study the projected sensitivity of the $\mu^{+}\mu^{-}\to \mu^{+}\mu^{-}\left(H_{2} \to \mu^{+}\mu^{-}\right)$ channel at a muon collider (MuC) with configuration:$\sqrt{s}=3~\text{TeV}$, $\mathcal{L}=1~\text{ab}^{-1}$~\cite{Delahaye:2019omf,Shiltsev:2019rfl}. These searches could preclude simultaneous explanation of both AMMs in the mass range of roughly $10-300$ GeV depending on the Yukawa coupings for a given texture within the model.  

The rest of the paper is organized as follows. In Sec.~\ref{sec:model} we present the basic description of the Zee Model, including the Yukawa lagrangian and radiative neutrino mass generation. In Sec.~\ref{sec:AMM} we discuss the anomalous magnetic moments and the possible ways of resolving the two anomalies simultaneously in the model. This section also points out the model predictions for muon EDM (cf. Sec.~\ref{sec:EDM}). Sec.~\ref{sec:NSI} briefly summarizes the NSI from charged scalars in the model, followed by a detailed study of the various textures of Yukawa coupling matrices which could incorporate both anomalies in Sec.~\ref{sec:flavor}. The low energy constraints such as $\ell_i \to \ell_j \gamma$, trilepton decays, T-parameter constraints, muonium oscillations, and direct experimental constraints are discussed in Sec.~\ref{sec:constraints}. In Sec.~\ref{sec:collider} we perform a detailed cut-based collider analysis at the detector level to estimate the projected capability of future lepton colliders to probe the Yukawa structure of the additional scalar Higgs boson considered in this work. The results of our analysis on the anomalous magnetic moment of lepton and neutrino oscillation fit are given in Sec.~\ref{sec:Fit}, followed by the conclusion in Sec.~\ref{sec:conclusion}.

\section{Model Description\label{sec:model}}
The Zee model~\cite{Zee:1980ai,Wolfenstein:1980sy}, built on the Two Higgs Doublet Model \cite{Lee:1973iz, Branco:2011iw}, is perhaps the simplest extension of the SM that can generate non-zero radiative neutrino mass at a one-loop level. The model is based on SM gauge symmetry $SU(3)_C \times SU(2)_L \times U(1)_Y$ and consists of a $SU(2)_L$ doublet scalar $H_2$ in addition to an SM Higgs doublet $H_1$ and a charged singlet scalar $\eta^\pm$. In Higgs basis~~\cite{Davidson:2005cw}, where only the neutral component of $H_1$ takes a vacuum expectation value~(VEV), $\langle H_1^0\rangle=v\simeq 246$ GeV, the doublets can be represented as
\begin{equation}
H_{1}=\left(\begin{array}{c}
G^{+} \\
\frac{1}{\sqrt{2}}\left(v+H_{1}^{0}+i G^{0}\right)
\end{array}\right), \quad H_{2}=\left(\begin{array}{c}
H_{2}^{+} \\
\frac{1}{\sqrt{2}}\left(H_{2}^{0}+i A\right)
\end{array}\right),
\end{equation}
where $(G^+,G^0)$ are the Goldstone modes, $(H^0_1,H^0_2)$ and $A$ are the neutral $\mathcal{CP}$-even and $\mathcal{CP}$-odd scalars, and $H_2^+$ is a charged scalar field. In the $\mathcal{CP}$ conserving limit, where the quartic couplings are real, the field $A$ decouples from $\{H_1^0, H_2^0 \}$. Then one can rotate the $\mathcal{CP}$-even states into a physical basis $\{h, H \}$, where the mixing angle parametrized as
\begin{equation}
\begin{aligned}
\begin{pmatrix}
h\\H
\end{pmatrix} = \begin{pmatrix}
\cos\widetilde{\alpha} & \sin \widetilde{\alpha} \\
-\sin\widetilde{\alpha}&  \cos\widetilde{\alpha} 
\end{pmatrix}\begin{pmatrix}
H_1^0\\H_2^0
\end{pmatrix},
\end{aligned}
\end{equation}
is given by 
\begin{equation}
    \sin2\widetilde{\alpha}=\frac{2\lambda_6v^2}{m_H^2-m_h^2} \, .
\end{equation}
Here $\lambda_6$ is the quartic coupling of the term $H_1^\dagger H_1H_1^\dagger H_2$. We base our analysis under the alignment/decoupling limit~\cite{Gunion:2002zf,Carena:2013ooa,BhupalDev:2014bir,Das:2015mwa}, when $\widetilde{\alpha}\to 0$, agreeing with the LHC Higgs data~\cite{Bernon:2015qea,Chowdhury:2017aav}, and identify $h$ as the observed $125$ GeV SM-like Higgs. Similarly, the charged scalars $\{H_2^+, \eta^+\}$ mix and give rise to the physical charged scalar mass eigenstates $\{ h^+, H^+ \}$
\begin{equation}
\begin{aligned}
\begin{pmatrix}
h^+\\H^+
\end{pmatrix} = \begin{pmatrix}
\cos\varphi & \sin\varphi \\
-\sin\varphi &  \cos\varphi 
\end{pmatrix}\begin{pmatrix}
\eta^+\\H_2^+
\end{pmatrix},
\end{aligned}
\label{eq:chargMix}
\end{equation}
with the mixing angle $\varphi$ defined as
\begin{equation}
    \sin2\varphi=\dfrac{-\sqrt{2}v\mu}{m^2_{H^+}-m^2_{h^+}},
    \label{eq:mixcharged}
\end{equation}
where, $\mu$ is the coefficient of the cubic coupling $H_{1}^{\alpha} H_{2}^{\beta} \epsilon_{\alpha \beta} \eta^{-}$ in the scalar potential, with $\epsilon_{\alpha \beta}$ being the $SU(2)_L$ antisymmetric tensor. This cubic term, along with Eq.~\eqref{eq:Yuk}, would break the lepton number by two units. 

The leptonic Yukawa interaction in the Higgs basis can be expressed  as
\begin{equation}
-\mathcal{L}_{Y} \supset f_{i j} L_{i}^{\alpha} L_{j}^{\beta} \epsilon_{\alpha \beta} \eta^{+}+\widetilde{Y}_{i j} \widetilde{H}_{1}^{\alpha} L_{i}^{\beta} \ell_{j}^{c} \epsilon_{\alpha\beta}+Y_{i j} \widetilde{H}_{2}^{\alpha} L_{i}^{\beta} \ell_{j}^{c} \epsilon_{\alpha\beta}+\text { h.c. } \, 
\label{eq:Yuk}
\end{equation}
where, $\{i,j\}$ are flavor indices, and $\widetilde{H}_a\equiv i \tau_{2} H_{a}^{\star}$, $\tau_2$ being the second Pauli matrix. $l^c$ and $L$ represent the left-handed antileptons and lepton doublets. Note, $f_{ij} = - f_{ji}$ is an antisymmetric matrix and can be made real by a phase redefinition $\hat{P} f \hat{P}$, where $\hat{P}$ is a diagonal phase matrix, whereas $\{Y,\widetilde{Y}\}$ are general complex asymmetric matrices. We assume that $H_2$ is leptophilic to avoid dangerous flavor violating processes in the quark sector, such as $\pi^+\to e^+\nu$ which would otherwise occur at unacceptably large decay rates for $Y_{ie}\sim\mathcal{O}(1)$. Here, after electroweak symmetry breaking, $M_\ell=\widetilde{Y}\frac{v}{\sqrt{2}}$ is the charged lepton mass matrix, and chosen to be diagonal without loss of generality.

\begin{figure}
    \centering
    \includegraphics[scale=0.4]{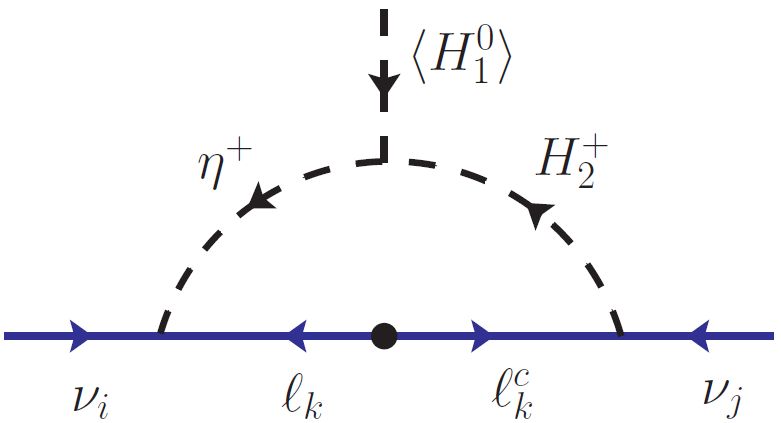}
    \caption{Radiative neutrino mass diagram in the Zee model. The dot $(\bullet)$ on the SM fermion line represents the mass insertion due to the SM Higgs VEV.}\label{fig:neutrino_mass}
\end{figure}

Neutrino masses in this model are zero at the tree level. However, due to explicit lepton number violation, a non-zero Majorana neutrino mass $M_\nu$ is induced as quantum corrections at the one-loop level, as shown in Fig.~\ref{fig:neutrino_mass}. The dot in the internal fermion line represents the mass insertion due to the SM Higgs VEV. The Yukawa couplings of Eq.~\eqref{eq:Yuk}, together with the $H_1 H_2 \eta$ trilinear term in the scalar potential, guarantees lepton number violation. These interactions result in an effective ($\Delta L = 2$) $d=7$ operator $\mathcal{O}_2 = L^i L^j L^k e^c H^l \epsilon_{ij} \epsilon_{kl}$ \cite{Babu:2019mfe, Babu:2001ex, deGouvea:2007qla, Cepedello:2017lyo, Gargalionis:2020xvt}. Note that a companion diagram can be obtained by reversing the arrows of the internal particles. Thus, the neutrino mass matrix is given by 
\begin{equation}
M_{\nu}=\kappa\left(f M_{\ell} Y+Y^{T} M_{\ell} f^{T}\right),
\label{eq:numass}
\end{equation}
where $\kappa$ is the one-loop factor \begin{equation}
    \kappa=\frac{1}{16 \pi^{2}} \sin 2 \varphi \log{ \left(\frac{m_{h^{+}}^{2}}{m_{H^{+}}^{2}}\right)} \, ,
    \label{eq:loopfactor}
\end{equation}
with $\varphi$ given in Eq.~\eqref{eq:mixcharged}. It is clear from Eq.~\eqref{eq:numass} that the product of couplings $Y$ and $f$ is constrained from neutrino oscillation data. For instance, a choice of $Y\sim \mathcal{O}(1)$ compels minuscule Yukawa couplings $f\ll 1$ to generate a tiny neutrino mass of $\mathcal{O}(0.1)$ eV consistent with the current measurements. Thus, such a choice of parameters can correctly reproduce the neutrino oscillation data (see Sec.~\ref{sec:Fit}), give the required corrections to the anomalous magnetic moments of electron and muon, and maximize the neutrino NSI in the model.

With the other possibility, namely, $Y \ll 1$, the stringent charge LFV (cLFV) constraints on f Yukawa coupling restrict the maximum NSI to $\leq 10^{-8}$ \cite{Herrero-Garcia:2017xdu}, well below any experimental sensitivity in the foreseeable future. Furthermore, the presence of a singly charged singlet leads to lepton universality violation which, for instance, would alter the decay rate of the muon. The Fermi constant extracted from the modified muon decay would be different from the SM. This new Fermi constant has constraints from the CKM unitarity measurements giving strong limits on the Yukawa couplings, $f$ \cite{Cai:2017jrq}. The charged current interactions leading to leptonic decays will also be modified so that such interactions are no longer universal, leading to further constraints \cite{Herrero-Garcia:2014hfa, Ghosal:2001ep}. Moreover, choosing $Y \ll 1$, one cannot get the required correction to $g-2$ of electron and muon, and $\eta^+$ always contributes the wrong sign to the $g-2$ of muon (cf. Sec.~\ref{sec:AMM} for details). The Yukawa couplings $f$ and $Y$ on the mass basis of charged leptons can be written as

\begin{equation}
 Y=\left(\begin{array}{lll}
Y_{e e} & Y_{e \mu} & Y_{e \tau} \\
Y_{\mu e} & Y_{\mu \mu} & Y_{\mu \tau} \\
Y_{\tau e} & Y_{\tau \mu} & Y_{\tau \tau}
\end{array}\right), \quad
f=\left(\begin{array}{ccc}
0 & f_{e \mu} & f_{e \tau} \\
-f_{e \mu} & 0 & f_{\mu \tau} \\
-f_{e \tau} & -f_{\mu \tau} & 0
\end{array}\right),
\end{equation}
where $Y$ is multiplied by $(\bar{\nu}_e,\bar{\nu}_\mu,\bar{\nu}_\tau)$ (or $(e_L,\mu_L,\tau_L)$) from the left and $(e_R,\mu_R,\tau_R)^T$ from the right in the charged scalar, $H_2^+$ (or neutral scalar, $\frac{1}{\sqrt{2}}(H_{2}^{0}+i A))$ interaction. It is worth mentioning that the Yukawa coupling $Y$ cannot be taken diagonal; otherwise, all the diagonal entries of the neutrino mass matrix would vanish, yielding neutrino mixing angles that are not compatible with the neutrino oscillation data \cite{Wolfenstein:1980sy, Koide:2001xy, He:2003ih, Babu:2013pma}.

\section{Anomalous Magnetic Moments and Related Process\label{sec:AMM}}
\begin{figure}
    \centering
    \includegraphics[scale=0.29]{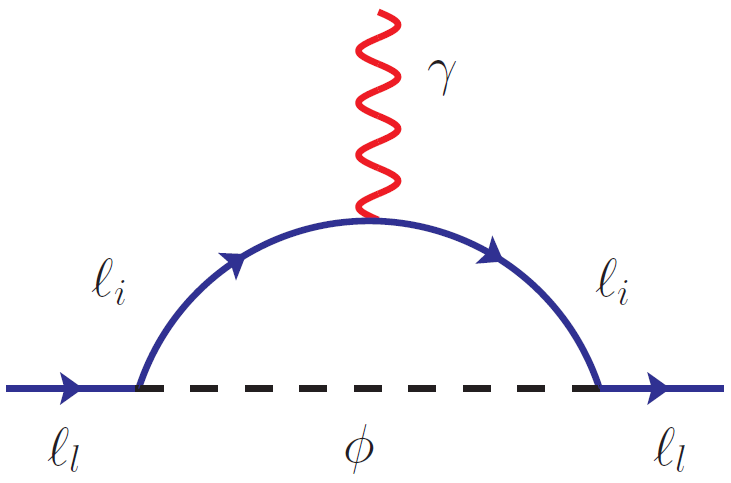}
    \includegraphics[scale=0.299]{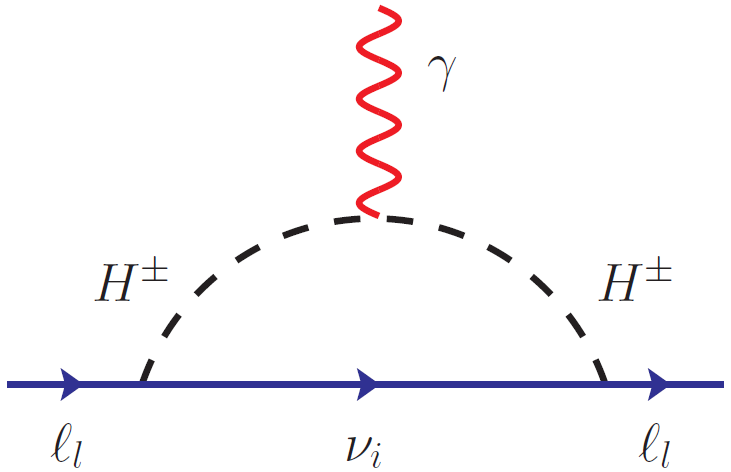}
    \includegraphics[scale=0.31]{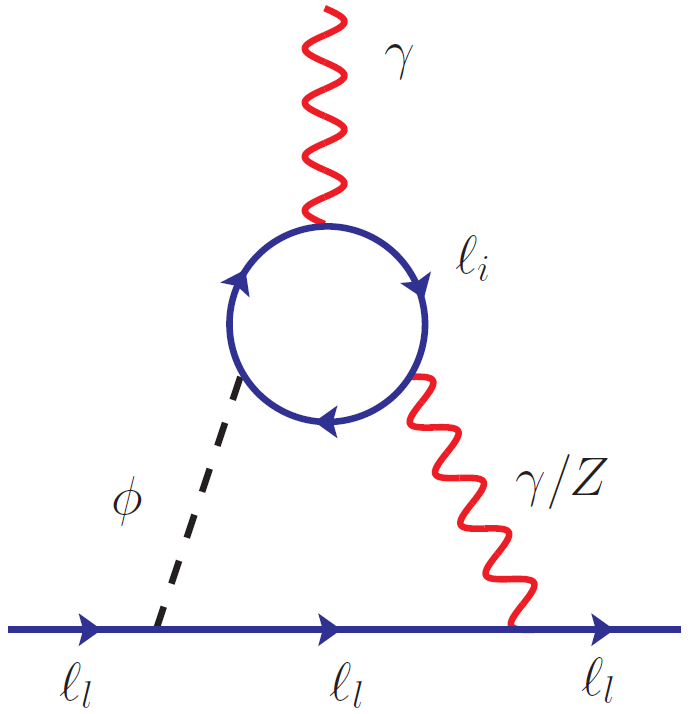} \\
   (a)~~~~~~~~~~~~~~~~~~~~~~~~~~~~~~~~~(b)~~~~~~~~~~~~~~~~~~~~~~~~~~~~~~~~~~(c)
    \caption{Dominant corrections to $a_\ell$ in Zee model. (a) and (b) are one-loop contributions from the scalar sector, and (c) is a typical two-loop Barr-Zee type correction to the AMMs. }
    \label{fig:AMM}
\end{figure}
Virtual corrections due to the new scalar fields present in the model can modify the electromagnetic interactions of the charged leptons. The contribution from $\widetilde{Y}$ of Eq.~\eqref{eq:Yuk} is part of the SM contribution, $a_\ell^{\text{SM}}$, in the decoupling limit $\tilde{\alpha} = 0$. The contribution from $f$ Yukawa coupling to AMMs is negligible due to the strong limit from cLFV and constrains from tiny neutrino mass, as aforementioned. Thus, the Yukawa $Y$ interactions of leptons with the physical scalars in the alignment limit is given by 
\begin{equation}
\begin{aligned}
     -\mathcal{L}_Y&\supset Y_{ij}\bar{\ell}_{L_i}\ell_{R_j}\phi + Y_{ij}\bar{\nu}_{L_i}\ell_{R_j} H^++\text{h.c.}, \label{Lg-2}
\end{aligned}
\end{equation}
where, $\phi = (H+ i A)/\sqrt{2}$. The charged scalar $h^+$ contribution is ignored by choosing a negligible mixing between charged scalars $\varphi \sim 0$.
Neutral scalar contributions to anomalous magnetic moments at one-loop \cite{Leveille:1977rc} as shown in Fig.~\ref{fig:AMM} (a) is 
\begin{equation}
    \Delta a_\ell^{(1)}(\phi)=\frac{m_{\ell}^2}{32\pi^2}\Big(\left\{|Y_{\ell i}|^2+|Y_{i\ell}|^2\right\}F_\phi(x,1)\pm 2\frac{m_i}{m_\ell}\Re(Y_{\ell i}Y_{i \ell})F_\phi(x,0)\Big),
    \label{eq:a1N}
\end{equation}
where,
\begin{equation}
    \begin{aligned}
    F_\phi(x,\epsilon)=\int_{0}^{1} \frac{x^2 - \epsilon x^3}{m_\ell^2x^2+(m_i^2-m_\ell^2)x+m_\phi^2(1-x)} \,dx.
    \end{aligned}
\end{equation}
In the above expression, $+$ and $-$ correspond to $H$ and $A$, respectively, with the second part representing the chiral enhancement. The contribution from charged the Higgs $H^+$ from Fig.~\ref{fig:AMM} (b) is
\begin{equation}
   \Delta a_\ell^{(1)}(H^+)= \frac{m_\ell^2}{16\pi^2}|Y_{i\ell}|^2\int_{0}^{1} \frac{x^3-x^2}{m_\ell^2x^2+(m_{H^+}^2-m_\ell^2)x}\,dx.\label{eq:charged}
\end{equation}
The analytical expressions in the limit $m_i \ll m_\phi(m_{H^+})$ for Eqs.~\eqref{eq:a1N} and \eqref{eq:charged} are given in Appendix \ref{app_1loop}.

There are also two-loop Barr-Zee \cite{Barr:1990vd,Bjorken:1977vt} diagrams arising from the neutral scalars and charged lepton loop \cite{Ilisie:2015tra,Frank:2020smf}, as shown in Fig.~\ref{fig:AMM} (c), contributing to the AMM corrections. The two-loop correction is
\begin{equation}
    \Delta a_\ell^{(2)}=\dfrac{\alpha_{em}}{4\pi^3}\dfrac{m_\ell}{m_i}\dfrac{z}{2}\Big(-C_{S_\ell}^\phi C_{S_i}^\phi G\left(x,z,1\right)+C_{P_\ell}^\phi C_{P_i}^\phi G\left(x,z,0\right)\Big),
    \label{eq:a2N}
\end{equation}
where,
\begin{equation}
    \begin{aligned}
   G(x,z,\epsilon)=\int_{0}^{1} \frac{1-2\ \epsilon\ x(1-x)}{x(1-x)-z}\log{\frac{x(1-x)}{z}}\,dx,
   \end{aligned}
   \label{eq:twoint}
\end{equation}

with $z=m_i^2/m_\phi^2$, and the coefficients may be obtained from Eq.~\eqref{Lg-2} as
\begin{equation}
    \begin{aligned}
    C_{S_i}^H=C_{P_i}^A=\frac{1}{\sqrt{2}}\text{Re}(Y_{ii}),
   \quad -C_{P_f}^H=C_{S_f}^A=\frac{i}{\sqrt{2}}\text{Im}(Y_{ii}).
    \end{aligned}
    \label{eq:twoY}
\end{equation}
The analytical expressions for the integrals of Eq.~\eqref{eq:twoint} are given in Appendix \ref{app_2loop}.

\paragraph{One-loop:}
First, we investigate the one-loop contribution to $g-2$ of muon and electron from Eq.~\eqref{eq:a1N}. As seen from Eq.~\eqref{eq:a1N}, for degenerate neutral scalars, i.e., $m_H = m_A$, the chiral enhancement vanishes, making the effective contribution to $\Delta a_\ell$ positive. When the couplings are diagonal, a positive  contribution to $\Delta a_\ell$ can also be achieved by considering $\mathcal{CP}$-odd scalar heavier than $\mathcal{CP}$-even scalar, an ideal scenario to explain $\Delta a_\mu$. However, such an assumption would contradict with $\Delta a_e$, for which an overall negative correction would require a lighter pseudoscalar. It is worth mentioning that the negative contribution from the charged scalar $H^+$ can partially cancel out the non-chiral part of the neutral scalar corrections. However, the mass of the charged scalar in our choice of the parameter space is always much larger than the mass of the neutral scalar making this effect negligible. Hence, the real Yukawa couplings $Y_{ii}\ (i = e, \mu)$ cannot explain both the AMMs simultaneously from one-loop corrections. Note that if the lepton mediator is different from the external line $Y_{ij} (i \neq j)$, Yukawa couplings with opposite signs can explain the both the AMMs, subject to constraints from LFV (cf. Sec.~\ref{sec:flavor} and Sec.~\ref{sec:constraints}).
However, concurrent solutions do exist for $Y_{\mu\mu}\in\mathbb{R}\cap Y_{ee}\in\mathbb{I}$ for $m_A>m_H$ and vice versa for $m_A<m_H$. Note that any nonzero phase in the Yukawa couplings would be constrained by  EDMs (cf. Sec.~\ref{sec:EDM}). The corrections to $a_\ell$ can appear from one or more of the following Yukawa couplings as shown:
\begin{equation}
\Delta a_{\mu}\Rightarrow\begin{pmatrix}
       . & Y_{e\mu} & .\\
   Y_{\mu e} &Y_{\mu\mu} & Y_{\mu\tau}\\
    . &Y_{\tau\mu} & .
    \end{pmatrix}+\text{h.c}.,\qquad
\Delta a_e\Rightarrow    \begin{pmatrix}
       Y_{ee} & Y_{e\mu} & Y_{e\tau}\\
   Y_{\mu e} &. & .\\
   Y_{\tau e} &. & .
    \end{pmatrix}+\text{h.c}..
    \label{eq:1looptext}
\end{equation}

\paragraph{Two-loop:} 
There are several contributions to corrections for AMMs arising at two-loop, which are studied in detail in \cite{Ilisie:2015tra, Cherchiglia:2016eui, Cherchiglia:2017uwv,Frank:2020smf}. A typical but most relevant two-loop Barr-Zee diagram is shown in Fig.~\ref{fig:AMM} (c). The contribution from the $Z$ boson line in the figure is typically suppressed by a factor of $\mathcal{O}(10^{-2})$ compared to the photon line. This is partly due to the massive $Z$ boson propagator and the smallness of its leptonic vector coupling, $\frac{g}{\cos\theta_W}(-\frac{1}{4}+\sin^2\theta_W)\sim -0.015$ \cite{Chang:2000ii}. Similar diagrams with charged scalars replacing the lepton loop also exist. We suppress, for simplicity, the contribution from this type by considering a small value of quartic coupling $\lambda_7$ ($H_2^\dagger H_2 H_1^\dagger H_2$) in the Higgs potential. Note that this quartic coupling does not contribute to the scalar boson masses. Other diagrams that involve $W^+W^-H_1$ and $H_1 H_2 H_2$ couplings also vanish in the alignment limit, $\widetilde{\alpha} =0$ (see Fig.~3, 5, 6 in Ref.~\cite{Ilisie:2015tra}). Moreover, there is no $W^+W^-H_2$ and $H_2 H_2 H_2$ vertex in the Higgs basis where $H_2$ does not get VEV. 
The only other non-vanishing diagram, with charged scalars and $W^\pm$ respectively replacing the neutral scalar and photon lines, involves neutrinos. In this case, the loop factor, which is a function of masses of the internal particles, becomes negligible. Interestingly, this diagram becomes considerably dominant if vector-like fermions are involved~\cite{Frank:2020smf} instead of neutrinos.

Hence, Fig.~\ref{fig:AMM}~(c) is the only two-loop diagram where diagonal Yukawa couplings $Y_{ii}$ are relevant to AMM corrections, as seen from Eq.~\eqref{eq:a2N} and Eq.~\eqref{eq:twoY}. Moreover, since there is a chiral enhancement in the lepton loop, $\mathbf{Y_{\tau\tau}}$ contribution becomes relevant. It should be noted that taking $Y_{\tau \tau} \sim 0$, one-loop contribution always dominates two-loop, failing to give concurrent solutions for both AMMs. Here we take $m_H < m_A$ and choose the same sign for Yukawa couplings such that the scalar contribution dominates the pseudoscalar allowing for an overall negative correction to AMMs, thereby providing the right sign to explain $\Delta a_e$.

\paragraph{$\boldsymbol{\Delta a_\mu>0}$ and $\boldsymbol{\Delta a_e <0}$:} 
The two-loop corrections, even though suppressed by $\alpha_{em}/\pi$ in comparison to one-loop, have a factor of $m_i/m_l$ from chiral enhancement, as can be seen from Eq.~\eqref{eq:a2N}. This plays an important role in providing the right signs to the AMMs. In the case of $\Delta a_e$, the factor $m_i/m_e$ enhances the two-loop contribution over the corresponding one-loop correction. Therefore, with a choice of heavier pseudoscalar, the overall correction to $\Delta a_e$ can be made negative. In doing so, one also gets a negative contribution for $\Delta a_\mu$ enhanced by $m_i/m_\mu$ from two-loop, which can become comparable to a one-loop contribution. However, by choosing $Y_{\tau \tau}$ small enough, one can suppress the two-loop contribution and get the correct sign for $\Delta a_\mu$ from one-loop (see Sec.~\ref{sec:caseIfit} for more details).  

As aforementioned, one-loop by itself can explain both AMMs without the necessity to go to the two-loop level by taking the diagonal couplings $Y_{ii} \sim 0$. In this scenario, one of the Yukawa couplings can be chosen to be negative so that, for a heavier pseudoscalar, the non-chiral part would explain $\Delta a_\mu$ while the chiral part, enhanced by $\log\left( m_i/ m_\phi\right)$ (c.f.~\eqref{a1N}), explains $\Delta a_e$.

\subsection{Electric Dipole Moments\label{sec:EDM}}
The electric dipole moment of leptons places stringent constraints on the imaginary part of the Yukawa couplings of the scalar field $\phi$. We study these constraints by turning on the relevant couplings such that the two-loop contribution becomes subdominant. These constraints are only significant when there is a chirality flip in the fermion line inside the loop, depicted in Fig.~\ref{fig:AMM} (a). In such a scenario lepton EDM is given by~\cite{Ecker:1983dj}
\begin{equation}
    d_\ell(\phi)=\frac{\mp q_im_i}{16\pi^2 m_\phi^2} \frac{\text{Im}(Y^*_{i\ell}Y^*_{\ell i})}{2}\,I\left(m_i^2/m_\phi^2,m_\ell^2/m_\phi^2\right),\label{edm}
\end{equation}
with $+(-)$ corresponding to $A\ (H)$, and \begin{equation}
    I(r,s)= \int_{0}^{1} \frac{x^2}{1-x+rx-sx(1-x)} \,dx .
\end{equation}
A tiny complex phase of $\mathcal{O}(10^{-5})$ is required to satisfy the current limits on eEDM $|d_e| \leq 1.1 \times 10^{-29}$ e-cm from ACME~\cite{ACME:2018yjb} while also satisfying $\Delta a_e$. We can simply avoid the electron EDM limit by taking all the relevant couplings that give rise to $\Delta a_e$ real. However, for the case of muon EDM, there exist regions of parameter space where the phase of the Yukawa couplings can be significant and provide enough correction to satisfy $\Delta a_\mu$ while also remaining compatible with the current upper limits from eEDM measurements $|d_\mu| \leq 1.9\times 10^{-19}$ e-cm \cite{Muong-2:2008ebm} and can potentially be measurable in future experiments \cite{Abe:2019thb,Sato:2021aor,Adelmann:2021udj}. 

\begin{figure}[!t]
    \centering
    \includegraphics[scale=0.6]{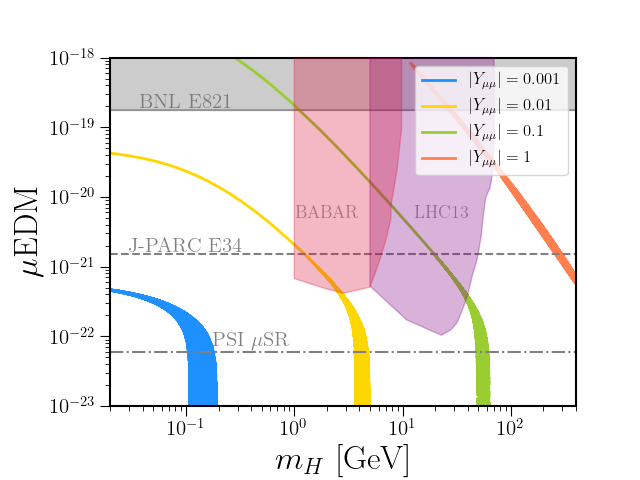}
    \caption{$\mu$EDM for different values of Yukawa couplings satisfying $\Delta a_\mu$ at $1\sigma$. The phase of the coupling $Y_{\mu\mu}=|Y_{\mu\mu}|e^{i\theta}$ is taken arbitrarily within $[0,\pi]$. The collider bounds on $Y_{\mu\mu}$ from $e^+e^-\to \mu^+\mu^- H$ searches obtained from BABAR~\cite{BaBar:2016sci}(pink) and LHC~\cite{CMS:2018yxg}(purple) experiments are projected to the maximum EDM that can be obtained for the coupling. The gray band shows the current experimental bound on $\mu$EDM, while the dashed and dotted-dashed gray lines provide the sensitivity reach from future experiments~\cite{Abe:2019thb,Sato:2021aor,Adelmann:2021udj}.}
    \label{fig:edm}
\end{figure}

Fig.~\ref{fig:edm} shows the $\mu$EDM ranges that can be probed at the near future experiments and satisfy $\Delta a_\mu$ within $1\sigma$ for different values of $Y_{\mu\mu}=|Y_{\mu\mu}|e^{i\theta}$. Here we set all the other Yukawa couplings small for simplicity and to satisfy the flavor constraints. Different color bands (red, green, yellow, blue) represent different choices of Yukawa couplings $|Y_{\mu\mu}|$ $(1, 0.1, 0.01, 0.001)$ by allowing the phase to take arbitrary values. The pink and purple shaded regions are excluded from $e^+e^-\to \mu^+\mu^- H$ searches obtained from BABAR~\cite{BaBar:2016sci} and LHC~\cite{CMS:2018yxg}, respectively. The bounds were obtained in the $Y_{\mu\mu}-m_H$ plane, which was then projected onto the maximum values of $\mu$EDM, obtained at $\theta=\pi/4$. The gray region is excluded from current experiments, and the gray dashed \cite{Sato:2021aor} and dotted-dashed lines \cite{Abe:2019thb, Adelmann:2021udj} are the projected sensitivities from various proposed experiments. 

It is worth noting that $\Delta a_\mu$ cannot be satisfied for $\theta \in [\frac{\pi}{4},\frac{3\pi}{4}]$ since the dominant chirally enhanced term in Eq.~\eqref{eq:a1N} is $\propto \cos2\theta$ which is $\leq 0$ for the said region of complex phases regardless of the value of $Y_{\mu\mu}$ and $m_H$.

\section{Non-standard Neutrino Interaction\label{sec:NSI}}
In the Zee model, the charged scalars $\eta^+$ and $H_2^+$ can induce charged-current NSI at tree level ~\cite{Babu:2019mfe}. Since the model can have leptophilic Yukawa couplings $Y_{ie}\ (i = e, \mu, \tau)$ of order unity, significant NSI can be generated. As mentioned before, $f\ll 1$ due to strong constraints from LFV as well as to correctly reproduce the neutrino oscillation parameters (see Sec.~\ref{sec:Fit}). Thus the contributions from $f$ Yukawa couplings are heavily suppressed. Using the dimension-6 operators for NSI \cite{Wolfenstein:1977ue}, the effective NSI parameters in the model can be expressed as 
\begin{equation}
\varepsilon_{i j} \equiv \varepsilon_{i j}^{\left(h^{+}\right)}+\varepsilon_{i j}^{\left(H^{+}\right)}=\frac{1}{4 \sqrt{2} G_{F}} Y_{i e} Y_{j e}^{*}\left(\frac{\sin ^{2} \varphi}{m_{h^{+}}^{2}}+\frac{\cos ^{2} \varphi}{m_{H^{+}}^{2}}\right),\label{NSI}
\end{equation}
where $h^+$ and $H^+$ are the physical masses of the charged scalars, $\varphi$ is the mixing angle between the scalars and is define in Eq.~\eqref{eq:chargMix} and Eq.~\eqref{eq:mixcharged}. Since we wish to make the neutral scalar light to explain AMM of muon and electron, we take a limit when doublet charged scalar is lighter than singlet charged scalar field. In this limit, the contribution from $H^+$ dominates, as can be seen, from Eq.~\eqref{NSI}. Note that there is a strong constraint from the T-parameter on the choice of neutral and charged scalar masses and mixing among scalars (see Sec.~\ref{sec:Tparam} for details). Due to the strong constraints from LFV (cf. Sec.~\ref{sec:LFV} and Sec.~\ref{sec:trilep}), one cannot get sizable off-diagonal NSI $\varepsilon_{ij}\ (i\neq j)$ in the model. However, sizable diagonal NSI $\varepsilon_{ii}$ can potentially be generated by the matrix elements $(Y_{ee},Y_{\mu e},Y_{\tau e})$, contingent on satisfying $\Delta a_\ell$ and reproducing neutrino oscillation parameters, which is discussed in the following section.     

\section{Low-energy Constraints\label{sec:constraints}}
In this section, we summarize various relevant low-energy constraints on the parameter space that can potentially explain the observables being explored here. We can safely ignore charged lepton flavor violation (cLFV) involving the $f_{ij}$ couplings as they are tiny to satisfy the neutrino mass constraint. On the other hand, we require $Y \sim\mathcal{O}(1)$ to explain both AMMs and to induce maximum NSI, which leads to various flavor violating processes that are severely constrained by experimental data. 

\subsection{\texorpdfstring{$l_1\to l_2 \gamma$}{l1tol2gamma}}\label{sec:LFV}
\begin{figure}
    \centering
    \includegraphics[scale=0.33]{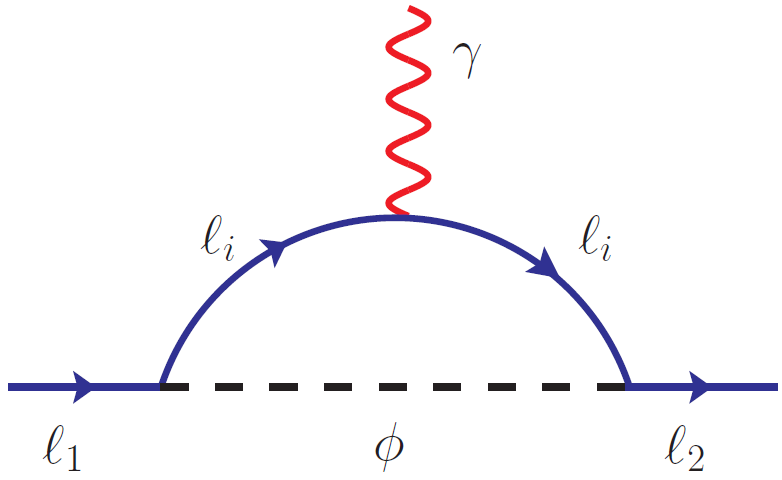}~~~~
    \includegraphics[scale=0.35]{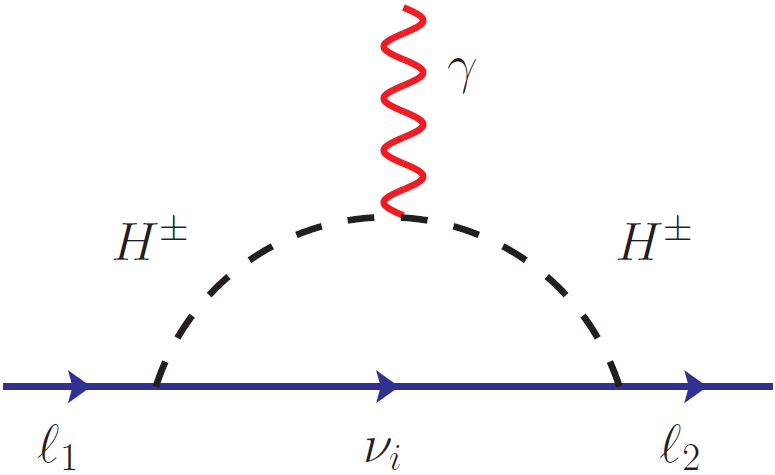}
    (a)~~~~~~~~~~~~~~~~~~~~~~~~~~~~~~~~~~~~~~~~~~~~~~~(b)
    \caption{Feynman diagrams inducing radiative decays of charged leptons. $\phi$ represents both the neutral scalar~($H$) and pseudoscalar~($A$).}
    \label{fig:radiative}
\end{figure}
The $l_1\to l_2 \gamma$ decays are induced radiatively at one-loop diagrams as shown in Fig.~\ref{fig:radiative}, and the general expression for such decays involving the neutral scalar field $\phi$ reads as \cite{Lavoura:2003xp} 

\begin{equation}
\begin{aligned}
    \Gamma_{\phi}=&\frac{\alpha_{em}}{144\,(16\pi^2)^2}\frac{m_1^5}{16m_{\phi}^4}\Big(\big(\left|Y_{2f}Y_{1f}^*\right|^2+\left|Y_{f1}Y_{f2}^*\right|^2\big)\mathcal{F}^2_1(t)\\
    &\qquad \qquad +\frac{9m^2_f}{m^2_1}\big(\left|Y_{1f}^*Y_{f2}^*\right|^2+\left|Y_{2f}Y_{f1}\right|^2\big)\mathcal{F}^2_2(t)\Big),
    \end{aligned}\label{radiative1}
\end{equation}
where, for $t=m_f^2/m_\phi^2$,
\begin{equation}
    \begin{aligned}
    \mathcal{F}_1(t)=&\frac{2+3t-6t^2+t^3+6t\log{t}}{(t-1)^4},\\
    \mathcal{F}_2(t)=&\frac{3-4t+t^2+2\log{t}}{(t-1)^3}.
    \end{aligned}\label{radiative2}
\end{equation}
The second term in Eq.~\eqref{radiative1} appears from the chirally enhanced radiative diagrams, whereas the first term has no chirality flip in the fermion line inside the loop. The bounds on the Yukawa couplings as a function of the mediator masses are shown in Table \ref{tab:radiative1}. Table \ref{tab:radiative1} has two rows showing the constraints arising from the chiral enhancement of charged lepton mediator in addition to the diagram without the enhancement. Similarly, the decay rate for $\ell_1 \to \ell_2 \gamma$ from charged scalar $H^+$ in the Zee model can be expressed as  
\begin{equation}
\begin{aligned}
    \Gamma_{H^-}=&\frac{\alpha_{em}}{144(16\pi^2)^2}\frac{m_1^5}{4m_{H^-}^4}\left|Y_{f1}Y^*_{f2}\right|^2.
\end{aligned}
\end{equation}
Note that there are no chirally enhanced contributions in these decays. The bounds on the $Y$ Yukawa couplings as a function of the mediator masses are shown in Table \ref{tab:radiative2}.
\begin{table}[t!]
    \footnotesize
    	{\renewcommand{\arraystretch}{1.5}%
    	\begin{center}
    \begin{tabular}{|c|c|c|}
    \hline
    \textbf{Process}&\textbf{Exp. Bound} \cite{ParticleDataGroup:2020ssz} &\textbf{Constraints}\\
    \hline
    \hline
        \multirow{2}*{$\mu\to e \gamma$} &\multirow{2}*{BR $<4.2\times10^{-13}$ \cite{MEG:2016leq}}&$\left|Y_{\mu f}Y_{e f}\right|^2+\left|Y_{f\mu}Y_{fe}\right|^2<1.89\times 10^{-9}\left(\frac{m_{\phi}}{100\text{GeV}}\right)^4$ \\
        \cline{3-3}
        & &$(\left|Y_{e f}Y_{f\mu}\right|^2+\left|Y_{\mu f}Y_{fe}\right|^2)\ \mathcal{C} <5.84\times 10^{-13}\ \left(\frac{m_\phi}{100\ \text{GeV}} \right)^4 \left(\frac{1\ \text{GeV}}{m_f} \right)^2 $ \\
        \hline
         \multirow{2}*{$\tau\to e \gamma$} &\multirow{2}*{BR $<3.3\times10^{-8}$ \cite{BaBar:2009hkt}}  &$\left|Y_{\tau f}Y_{e f}\right|^2+\left|Y_{f\tau}Y_{fe}\right|^2<8.31\times 10^{-4}\left(\frac{m_{\phi}}{100\text{GeV}}\right)^4$ \\
        \cline{3-3}
        & &$(\left|Y_{e f}Y_{f\tau}\right|^2+\left|Y_{\tau f}Y_{fe}\right|^2)\ \mathcal{C} <7.29\times 10^{-5}\ \left(\frac{m_\phi}{100\ \text{GeV}} \right)^4 \left(\frac{1\ \text{GeV}}{m_f} \right)^2 $ \\
         \hline
         \multirow{2}*{$\tau\to \mu \gamma$} &\multirow{2}*{BR $<4.4\times10^{-8}$ \cite{BaBar:2009hkt}}&$\left|Y_{\tau f}Y_{\mu f}\right|^2+\left|Y_{f\tau}Y_{f\mu}\right|^2<1.11\times 10^{-3}\left(\frac{m_{\phi}}{100\text{GeV}}\right)^4$ \\
        \cline{3-3}
        & &$(\left|Y_{\mu f}Y_{f\tau}\right|^2+\left|Y_{\tau f}Y_{f\mu}\right|^2)\ \mathcal{C} <9.72\times 10^{-5}\ \left(\frac{m_\phi}{100\ \text{GeV}} \right)^4 \left(\frac{1\ \text{GeV}}{m_f} \right)^2 $ \\
        \hline
    \end{tabular}
    \caption{Constraints on Yukawa couplings from radiative decay of charged leptons mediated by neutral scalars, $\phi$. Here, $\mathcal{C}=\left(\frac{3}{2}+\log{\frac{m_f^2}{m_\phi^2}}\right)^2$. For each process, there are diagrams with and without chiral enhancement from different charged lepton mediators, $f$. The first row in the "Constraints" column shows the bound on the diagram with no chiral enhancement, whereas the second row gives the bound on the chirally enhanced diagram contributed by the charged lepton mediator.}
    \label{tab:radiative1}
    \end{center}}
\end{table}

\begin{table}[t!]
    \footnotesize
    	{\renewcommand{\arraystretch}{1.5}%
    	\begin{center}
    \begin{tabular}{|c|c|c|}
    \hline
    \textbf{Process}&\textbf{Exp. Bound}~\cite{ParticleDataGroup:2020ssz}&\textbf{Constraints}\\
    \hline
    \hline
        $\mu\to e \gamma$ &BR $<4.2\times10^{-13}$ \cite{MEG:2016leq}&$\left|Y_{f\mu}Y_{fe}\right|^2<1.89\times 10^{-9}\left(\frac{m_{H^-}}{100\text{GeV}}\right)^4$ \\
        \hline
         $\tau\to e \gamma$ &BR $<3.3\times10^{-8}$ \cite{BaBar:2009hkt}&$\left|Y_{f\tau}Y_{fe}\right|^2<8.31\times 10^{-4}\left(\frac{m_{H^-}}{100\text{GeV}}\right)^4$ \\
         \hline
         $\tau\to \mu \gamma$ &BR $<4.4\times10^{-8}$ \cite{BaBar:2009hkt}&$\left|Y_{f\tau}Y_{f\mu}\right|^2<1.11\times 10^{-3}\left(\frac{m_{H^-}}{100\text{GeV}}\right)^4$ \\
        \hline
    \end{tabular}
    \caption{Constraints on Yukawa couplings from radiative decay of charged leptons mediated by charged scalar, $H^-$. The constraints here are similar to that of the neutral scalars with no chiral enhancement, given in Table \ref{tab:radiative1}.}
    \label{tab:radiative2}
    \end{center}}
\end{table}

\subsection{Trilepton Decays}\label{sec:trilep}
The flavor-changing nature of the new scalar bosons allows for the processes of form $l_{i}\to l_{k} \bar{l}_{j} l_{l}$ to realize at the tree level, and it imparts one of the most stringent constraints on the model that precludes simultaneous explanation of AMMs, reducing the flavor structure down to just two textures of Eq.~\eqref{eq:textures}. The partial rates for such trilepton decays are obtained in the limit when the masses of the decay products are neglected. The decay rate can be read as \cite{Cai:2017jrq}:
\begin{equation}
    \Gamma =\frac{1}{6144 \pi^3}\frac{m_i^5}{4m_\phi^4}\left(\frac{1}{(1+\delta_{lk})}(\left|Y^*_{ik}Y^*_{jl}\right|^2+\left|Y_{ki}Y_{lj}\right|^2)+\left|Y^*_{ik}Y_{lj}\right|^2+\left|Y_{ki}Y^*_{jl}\right|^2\right).
\end{equation}
$\delta_{lk}$ is the symmetry factor that takes care of identical particles in the final state. This expression is relevant for both $\mathcal{CP}$-even and $\mathcal{CP}$-odd neutral scalar mediated decays. Using the total muon and tau decay widths, $\Gamma_\mu^{\rm tot} = 3.00 \times 10^{-19}$ GeV and $\Gamma_\tau^{\rm tot} = 2.27 \times 10^{-12}$ GeV, we calculate the branching ratios for various processes and summarize the constraints on the Yukawa couplings as a function of the neutral scalar masses in Table  \ref{tab:trilepton decay}.

\begin{table}[t!]
        \footnotesize
    	{\renewcommand{\arraystretch}{2}%
    	\begin{center}
    \begin{tabular}{|c|c|c|}
    \hline
     \textbf{Process}    & \textbf{Exp. Bound}~\cite{ParticleDataGroup:2020ssz,HFLAV:2016hnz} &\textbf{Constraints} \\
     \hline\hline
      $\mu^-\to e^-e^+e^-$   & BR $<1.0\times 10^{-12}$~\cite{SINDRUM:1987nra}& $|Y_{ee}|^2(|Y_{e\mu}|^2+|Y_{\mu e}|^2)<1.16\times 10^{-12}\left(\frac{m_{\phi}}{100\text{ GeV}}\right)^4$\\
      \hline
      $\tau^-\to e^-e^+e^-$& BR $<2.7\times 10^{-8}$~\cite{Hayasaka:2010np}&$|Y_{ee}|^2(Y_{e\tau}|^2+|Y_{\tau e}|^2 ) <1.76\times 10^{-7}\left(\frac{m_{\phi}}{100\text{ GeV}}\right)^4$\\
      \hline
      $\tau^-\to \mu^-e^+e^-$ & BR $<1.8\times 10^{-8}$~\cite{Hayasaka:2010np} & $|Y_{ee}|^2(Y_{\mu\tau}|^2+|Y_{\tau\mu}|^2) <8.78\times 10^{-8}\left(\frac{m_{\phi}}{100\text{ GeV}}\right)^4$\\
      \hline
      $\tau^-\to \mu^-\mu^+\mu^-$ & BR $<2.1\times 10^{-8}$~\cite{Hayasaka:2010np} & $|Y_{\mu \mu}|^2(|Y_{\mu\tau}|^2+|Y_{\tau\mu}|^2) <1.37\times 10^{-7}\left(\frac{m_{\phi}}{100\text{ GeV}}\right)^4$\\
      \hline
      $\tau^-\to e^-\mu^+\mu^- $ &BR $<2.7\times 10^{-8}$~\cite{Hayasaka:2010np} &$ |Y_{\mu\mu}|^2(|Y_{e\tau}|^2+|Y_{\tau e}|^2)  <1.32\times 10^{-7}\left(\frac{m_{\phi}}{100\text{ GeV}}\right)^4$\\
      \hline
      $\tau^-\to e^-\mu^+e^-$ &BR $<1.5\times 10^{-8}$~\cite{Hayasaka:2010np} & $|Y_{e \mu}|^2(|Y_{e \tau}|^2+2|Y_{\tau e}|^2)+ (\mu \leftrightarrow e) <2.93\times 10^{-7}\left(\frac{m_{\phi}}{100\text{ GeV}}\right)^4$\\
      \hline
      $ \tau^-\to \mu^-e^+\mu^-$ & BR $<1.7\times 10^{-8}$~\cite{Hayasaka:2010np} & $ |Y_{e \mu}|^2(|Y_{ \tau\mu}|^2+2|Y_{\mu\tau}|^2)+ (\mu \leftrightarrow e) <3.32\times 10^{-7}\left(\frac{m_{\phi}}{100\text{ GeV}}\right)^4$\\
      \hline
    \end{tabular}
    \caption{Constraints on Yukawa couplings as a function of neutral scalar mass from trilepton decays $l_{i}\to l_{k} \bar{l}_{j} l_{l}$ of charged leptons. Slightly weaker constraints on tau lepton decays from BABAR, ATLAS and LHC can be found in refs.~\cite{BaBar:2010axs,ATLAS:2016jts,LHCb:2014kws}, respectively.}
    \label{tab:trilepton decay}
    \end{center}}
\end{table}

\subsection{T-parameter Constraints}\label{sec:Tparam}
\begin{figure}
    \centering
    \includegraphics[scale=0.6]{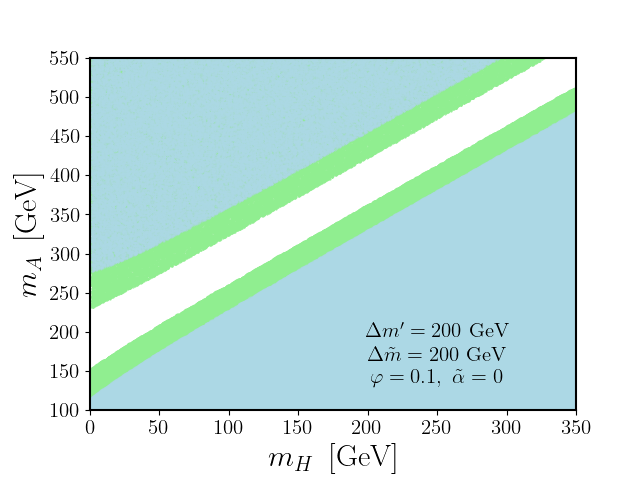}
    \caption{Allowed regions of scalar boson masses from T-parameter constraint. The green and blue regions are $1\sigma$ and $2\sigma$ exclusion from T$=0.03\pm0.12$~\cite{ParticleDataGroup:2020ssz}, with $\Delta \tilde{m} = m_{h^+}-m_H$ and $\Delta m' = m_{H^+}-m_{H}$ .}
    \label{fig:Tp}
\end{figure}

The oblique parameters S, T, and U quantify the deviation of a new physics model from the SM through radiative corrections arising from shifts in gauge boson self energies \cite{Peskin:1990zt, Peskin:1991sw, Funk:2011ad}. Out of these observables, the T-parameter imposes the most stringent constraint. In the decoupling limit $\tilde{\alpha} = 0$, T-parameter in the Zee model can be written as \cite{Grimus:2008nb}
\begin{equation}
    \begin{aligned}
    T=&\frac{1}{16 \pi^{2} \alpha_{em} \nu^2}\left\{\cos ^{2} \varphi\left[ \mathcal{F}\left(m_{h^{+}}^{2}, m_{H}^{2}\right)+\mathcal{F}\left(m_{h^{+}}^{2}, m_{A}^{2}\right)\right]- \mathcal{F}\left(m_{H}^{2}, m_{A}^{2}\right)\right.\\
  & +\sin ^{2} \varphi\left[ \mathcal{F}\left(m_{H^{+}}^{2}, m_{H}^{2}\right)+\mathcal{F}\left(m_{H^{+}}^{2}, m_{A}^{2}\right)\right] -2 \sin ^{2} \varphi \cos ^{2} \varphi \mathcal{F}\left(m_{h^{+}}^{2}, m_{H^{+}}^{2}\right) \}
\end{aligned}
\end{equation}
where, 
\begin{equation}
\mathcal{F}\left(m_{1}^{2}, m_{2}^{2}\right)=\mathcal{F}\left(m_{2}^{2}, m_{1}^{2}\right) \equiv \frac{1}{2}\left(m_{1}^{2}+m_{2}^{2}\right)-\frac{m_{1}^{2} m_{2}^{2}}{m_{1}^{2}-m_{2}^{2}} \log \left(\frac{m_{1}^{2}}{m_{2}^{2}}\right).
\end{equation}
The allowed regions of scalar masses $m_A$ and $m_H$ for the choice of mixing angle between charged scalars $\sin\varphi=0.1$ under the alignment limit are shown in Fig.~\ref{fig:Tp}, where the green and blue regions are excluded at $1\sigma$ and $2\sigma$ from T$=0.03\pm0.12$~\cite{ParticleDataGroup:2020ssz}. Here we fix the charged scalar mass heavier than neutral scalar mass by choosing $m_{h^+} = m_H + 200$ GeV and $m_{H^+} = m_H + 200$ GeV such that collider constraints on light charged scalars are easily satisfied \cite{Babu:2019mfe}. Notice that the choice of $\varphi$ is arbitrary and can be made small. Such a choice only makes the overall scale of neutrino mass smaller and does not alter the phenomenology for $\Delta a_\ell$ and NSI, as both can be incorporated with just the second Higgs doublet.  
\subsection{Muonium-antimuonium oscillations}
\label{sec:MAO}
The nontrivial mixing between bound states of muonium ($M: e^-\mu^+$) and antimuonium $\bar{M}: e^+\mu^-$ implies a non-vanishing LFV amplitude for $e^- \mu^+ \to e^+ \mu^-$ \cite{Pontecorvo:1957cp, Willmann:1998gd, Jentschura:1997tv, Jentschura:1998vkm, Ginzburg:1998df, Clark:2003tv, Harnik:2012pb, Dev:2017ftk}. These oscillation probabilities were measured by the {\tt PSI} Collaboration, with $P(M\leftrightarrow\bar{M}) < 8.3 \times 10^{-11}$ at $95\%$ C.L. \cite{Willmann:1998gd}, while {\tt MACE} Collaboration at CSNS attempts to improve the sensitivity at the level of $\mathcal{O}(10^{-13})$ \cite{mace}. These oscillations place a stringent constraint on the product of Yukawa couplings $Y_{e\mu}$ and $Y_{\mu e}$, thereby excluding a significant portion of parameter space as shown in Fig.~\ref{fig:caseII} of {\tt TX-II}. The muonium-antimuonium oscillation probability is given by \cite{Cvetic:2005gx, Han:2021nod} 
\begin{equation}
    P(M\to \bar{M}) = \frac{64 \alpha^6 m_{red}^6 \tau_\mu^2}{\pi^2} G_{M\bar{M}}^2 \simeq 1.95 \times 10^5\ G_{M\bar{M}}^2
    \label{eq:probmm}
\end{equation}
where, $m_{red} = m_e m_\mu/(m_e + m_\mu)$ is the reduced mass between a muon and an electron, $\alpha$ is the QED fine structure constant, and $\tau_\mu$ is muon lifetime. $G_{M\bar{M}}$ is the Wilson coefficient \cite{Conlin:2020veq, Fukuyama:2021iyw} associated with the dimension-six four fermion operator in the effective Hamiltonian density 
\begin{equation}
    \mathcal{H}_{eff} = \frac{G_{M\bar{M}}}{\sqrt{2}}\ [\bar{\mu} \gamma_\mu (1+\gamma_5) e] [\bar{\mu} \gamma^\mu (1-\gamma_5) e] \, .
    \label{eq:hamil}
\end{equation}
From Eq.~\eqref{eq:probmm}, one can obtain a limit on the Wilson coefficient from {\tt PSI} as $G_{M\bar{M}} \leq 1.77 \times 10^{-3}$, which translates to the bound on the Yukawa couplings as $Y_{e\mu} Y_{\mu e} \leq 2.37 \times 10^{-7} (m_H/\text{GeV})^2$. Similarly, the bound from the {\tt MACE} experiment is expected to improve the sensitivity by at least two orders $\mathcal{O}(10^{-13})$ which corresponds to $Y_{e\mu} Y_{\mu e} \leq 8.11 \times 10^{-9} (m_H/\text{GeV})^2$. 
\subsection{Direct experimental constraints}
In this section, we analyze various direct experimental constraints on the neutral scalar with mass in the range of 100 MeV to 300 GeV range that explains $\Delta a_\ell$. There are various experimental constraints one needs to consider, such as dark photon searches, rare $Z-$decay constraints, and LEP and LHC constraints. The lack of observation of dark photon $A_d$ in searches through the $e^+ e^- \to \gamma A_d$, with $A_d \to e^+ e^-$ channel at KLOE \cite{Anastasi:2015qla, Alves:2017avw} and BaBar \cite{BaBar:2014zli, Knapen:2017xzo} sets strong constraints on the couplings. By recasting the results from BaBar and KLOE, one can put a bound on the mass of the light scalars and the corresponding Yukawa couplings, as depicted by the brown and blue shaded region in Fig.~\ref{fig:case1fig}. Similarly, for the scalar mass $m_H > 200$ MeV, the dark-boson searches at the BaBar \cite{BaBar:2016sci} can be recast to limit on mass and the Yukawa couplings via the process $e^+ e^- \to \mu^+ \mu^- H$ \cite{Batell:2016ove, Batell:2017kty}, shown as the pink shaded region in Fig.~\ref{fig:case1fig}. We observe that the searches at BaBar exclude $Y_{\mu\mu} \gtrsim 6 \times 10^{-3}$~($1.5\times 10^{-2}$) at $m_{H} = 0.212~$GeV~(10~GeV). In the relatively heavier mass regime, $m_{H} \gtrsim \mathcal{O}(10)~$GeV, $Y_{\mu\mu}$ is constrained by LHC search limits. For example, direct $Z^{\prime}$ searches at the LHC in the  $pp \to \mu^{+}\mu^{-}(Z^{\prime} \to \mu^{+}\mu^{-})$ channel leads to upper limits on the production cross-section times branching ratio~\cite{CMS:2018yxg}. These upper limits could, in turn, be recast to upper bounds on the Yukawa couplings of our interest. We illustrate its implication on our parameter space as a purple shaded region in the $\{Y_{\mu\mu},m_{H}\}$ plane in Fig.~\ref{fig:case1fig}~(left panel).

The Yukawa couplings of the leptophilic Higgs boson $H$ are also susceptible to constraints from searches at the LEP. Direct searches in contact interaction processes $e^{+}e^{-} \to \ell^{+}\ell^{-}$~($\ell = e,\mu$) and Higgs production in association with leptons $e^{+}e^{-} \to \ell^{+}\ell^{-}(H \to \ell^{+}\ell^{-})$ can potentially constrain the Yukawa couplings in both \texttt{TX-I} and \texttt{TX-II}. In the heavy mass limit, LEP data exclude an effective cutoff scale $\Lambda \simeq m_H/Y_{ij}$ \cite{Electroweak:2003ram}. The LEP contact interaction constraints on $\Lambda$ for a light neutral scalar H are no longer applicable. However, due to the $t-$channel contribution of $H/A$ that interferes with the SM process, the cross-section of $e^+ e^- \to \ell_\alpha^+ \ell_\alpha^-$ can still be modified. By implementing the model file in the {\tt FeynRules} package \cite{Christensen:2008py}, we computed the cross-sections using \texttt{MadGraph5\_aMC@NLO}~\cite{Alwall:2014hca}, which are then compared with the measured cross-sections \cite{Electroweak:2003ram, OPAL:2003kcu} to get a limit on Yukawa coupling as a function of scalar mass; $Y_{ee} < 0.8$ and $Y_{e\mu} < 0.74$ for the benchmark value of $m_H = 130$ GeV~\cite{Babu:2019mfe}. This limit is slightly weaker than the bounds from searches in the $e^{+}e^{-} \to \ell^{+}\ell^{-}(H \to \ell^{+}\ell^{-})$ channel, discussed later in great detail.  

The most constraining Higgs production processes for $Y_{ee}$ and $Y_{e\mu}$ are $e^{+}e^{-} \to e^{+}e^{-}(H \to e^{+}e^{-})$ and $e^{+}\mu^{-}(H \to e^{+}\mu^{-})$, respectively. Likewise, $Y_{\mu\mu}$ would be most sensitive to the $e^{+}e^{-} \to \mu^{+}\mu^{-}(H \to \mu^{+}\mu^{-})$ process. However, the production cross-section of the latter at LEP is roughly an order of magnitude smaller than its former two counterparts. Therefore, we ignore the constraints on $Y_{\mu\mu}$ from direct searches in the $e^{+}e^{-} \to \mu^{+}\mu^{-}(H \to \mu^{+}\mu^{-})$ channel for the present analysis.
In order to correctly estimate the LEP bounds on $Y_{ee}$~($Y_{e\mu}$), we generate signal events in the $e^{+}e^{-} \to e^{+}e^{-}(H \to e^{+}e^{-})$~($e^{+}e^{-} \to e^{+}\mu^{-}(H \to e^{+}\mu^{-})$) channel assuming $\sqrt{s}=207~$GeV for several values of $m_{H}$ at the leading order~(LO) using the \texttt{MadGraph5\_aMC@NLO}~\cite{Alwall:2014hca} package. We also simulate the respective dominant background processes: $4e$~($2e2\mu$) in the same framework. We reconstruct the Higgs boson by identifying the $e^{+}e^{-}$~($e^{+}\mu^{-}$) pair with the smallest $\Delta R = \sqrt{\Delta \eta^{2} + \Delta \phi^{2}}$ separation, where $\Delta \phi$ is the azimuthal angle difference between the opposite sign lepton pairs. Consequently, we compute the signal and background efficiencies and translate them into upper bounds on $Y_{ee}$~($Y_{e\mu}$) as a function of $m_{H}$. We illustrate the current upper limits from LEP on $Y_{ee}$ and $Y_{e\mu}$ as grey shaded regions in the right panel of Fig.~\ref{fig:case1fig} and Fig.~\ref{fig:caseII}, respectively. We would like to note that we do not include any LEP detector effects in these studies, and therefore our ``LEP estimations'' must be treated cautiously. As such, these upper limits are rather conservative estimations. 

In principle, the Higgs boson could have been reconstructed by identifying the opposite sign lepton pairs with invariant mass closest to $m_{H}$. However, such a reconstruction strategy leads to an almost $100\%$ signal efficiency for all choices of $m_{H}$ since we are restricted to a simplistic truth level analysis without accounting for detector effects. This could potentially entail considerably stronger upper limits on the Yukawa couplings than a realistic scenario leading to misleading implications. We note that we have adopted this reconstruction strategy in Sec.~\ref{sec:collider}, where we perform a detailed cut-based collider analysis at the detector level to study the future sensitivity on $Y_{ee},~Y_{e\mu}$ and $Y_{\mu\mu}$ from future lepton colliders.

For completeness, we would like to note that the charged scalar in the model can be pair produced directly at colliders via $s$-channel off-shell photon or Z exchange with $H^\pm$ further decaying into $\ell \nu$. The corresponding LEP bound is $m_{H^\pm} \gtrsim 80$ GeV \cite{ALEPH:2013htx} for $\tau \nu$ final state. Moreover, these leptonic final states $\ell \nu$ mimic slepton searches in supersymmetric models that can be recast for our scenario and provide a  lower bound on its mass $m_{H^\pm} \gtrsim 100$ GeV \cite{Babu:2019mfe}. Similarly at the LHC, the $s$-channel Drell-Yan process $pp \to \gamma^*/Z^* \to H^+ H^-$ provides a somewhat stronger limit of $m_{H^\pm} \gtrsim 200$ GeV \cite{Babu:2019mfe} for BR$(H^\pm \to e \nu) =1$. Note that the LHC limits can be evaded by lowering the branching ratio, which can be always be done with more than one Yukawa coupling.

\section{Flavor Structures\label{sec:flavor}}
This section explores the texture of the Yukawa coupling matrix $Y$ for a unified explanation of electron and muon $g-2$ while correctly reproducing neutrino oscillation data. The Yukawa coupling $f$ appearing in the neutrino mass formula given in Eq.~\eqref{eq:numass} is taken to be arbitrary and small such that it automatically satisfies all the constraints and generates the correct order of the neutrino masses. However, the Yukawa matrix $Y$ has non-trivial structures and its various components explain $\Delta a_{e/\mu}$ as previously discussed in Sec.~\ref{sec:AMM}, for instance, Eq.~\eqref{eq:1looptext} for one-loop texture. It turns out that the Yukawa couplings $Y_{i e} (i= e, \mu, \tau)$, which is a part of the texture to explain $\Delta a_\ell$, also induce NSI at some level. 
A cursory glance at the analysis of LFV processes in Sec.~\ref{sec:constraints} reveals two types of Yukawa textures to accommodate both AMMs in conjunction with neutrino observables, which are:
\begin{equation}
 \textbf{TX-I:} \hspace{0.4cm}  Y= \begin{pmatrix}
    \encircled{Y_{ee}} & 0 & 0\\
    0 & \BG{Y_{\mu\mu}} & \times\\
    0 & \times & \MDB{Y_{\tau\tau}}\\
    \end{pmatrix}\, , \hspace{10mm}
   \textbf{TX-II:} \hspace{0.4cm}  Y= \begin{pmatrix}
    0 & \MDB{Y_{e \mu}} & 0\\
   \twincircled{Y_{\mu e}}& 0 & \times\\
    0 & \times & \times\\
    \end{pmatrix}.
    \label{eq:textures}
\end{equation}
These textures are studied in detail in Sec.~\ref{sec:Fit}. Here the green, blue, and red color-coded entries respectively represent $\Delta a_\mu,\ \Delta a_e$, and NSI. The same couplings that can explain more than one observables are encircled with the corresponding colors. The $\times$'s denote the Yukawa couplings required to satisfy the five neutrino oscillation observables ($\Delta m_{21}^2, \Delta m_{31}^2, \sin^2 \theta_{13}, \sin^2 \theta_{23}, \sin^2 \theta_{12}$) while satisfying the flavor constraints such as $\ell_i \to \ell_j \gamma$ and trilepton decay (c.f. Sec.~\ref{sec:constraints}). For more details, see Sec.~\ref{sec:Fit}. We also note that the zeros in the matrices of Eq.~\eqref{eq:textures} need not be exactly zero, but they need to be sufficiently small so that the flavor-changing processes remain under control (cf. Sec.~\ref{sec:constraints}). Note that the flavor structures are proposed under the assumption that $m_H$ is smaller than all other new scalar bosons.

\paragraph{{\tt \bf TX-I:}}
Here, we examine the texture where both $\Delta a_{e}$ and $\Delta a_{\mu}$ could be explained from the diagonal Yukawa couplings $Y_{ii}$. For the choice of real Yukawa couplings, $Y_{\mu \mu}$ can explain $\Delta a_\mu$ at one-loop order parametrized by the chirally enhanced term of Eq.~\eqref{eq:a1N}. The one-loop contribution from $Y_{ee}$ alone always gives the wrong sign to $\Delta a_e$. However, the inclusion of a third Yukawa coupling $Y_{\tau\tau}$, appearing from the two-loop Barr-Zee diagram, can generate the correct sign for $\Delta a_e$ \cite{Jana:2020pxx}. Furthermore, the same Yukawa coupling $Y_{ee}$ induces NSI, as previously discussed in Sec.~\ref{sec:NSI}. On the other hand, for complex Yukawa couplings, concurrent solutions do exist, for instance, $Y_{\mu\mu}\in\mathbb{R}\cap Y_{ee}\in\mathbb{I}$ for $m_A>m_H$ and vice versa for $m_A<m_H$. Note that any nonzero phase in the Yukawa couplings would be constrained by  EDMs (cf. Sec.\ref{sec:EDM}).

\paragraph{{\tt \bf TX-II:}}
The only other flavor structure to incorporate $\Delta a_{e}$ and $\Delta a_{\mu}$ while satisfying neutrino oscillation data is given in Eq.~\eqref{eq:textures} as {\tt TX-II} with Yukawa couplings $Y_{e\mu}$ and $Y_{\mu e}$. With non-zero $Y_{e \mu} (e \leftrightarrow \mu)$, the diagonal couplings $Y_{ii} (i=e, \mu)$ are highly constrained from $\mu \to e \gamma$ (cf. Sec.~\ref{sec:LFV}). Thus, the two-loop contributions are highly suppressed and can be safely ignored. To explain both the anomalies,  $Y_{e\mu}$ and $Y_{\mu e}$ can take opposite signs to get $\Delta a_e$ negative. Moreover, considering a hierarchy among the two Yukawa couplings, one can explain $\Delta a_\mu$ from the non-chiral part of Eq.~\eqref{eq:a1N}. Details on the choice of the Yukawa couplings as a function of the mass of scalar field is given in Fig.~\ref{fig:caseII} in Sec.~\ref{sec:Fit}. Note that $Y_{\mu e} \sim \mathcal{O}(1)$ can induce NSI. However, such a choice necessarily requires $|Y_{\mu e}| > |Y_{e \mu}|$ to get the correct order for $\Delta a_\ell$, which turns out to be not in favor with the normal hierarchy (NH) solution to the neutrino oscillation data (cf. Sec.~\ref{sec:Fit}).    

For the sake of completeness, we also point out that $Y_{\tau e}$ can induce large NSI, $\epsilon_{\tau\tau}\sim 9.3\%$~\cite{Babu:2019mfe}, and provide correction to $a_e$ from tau mass chiral enhancement involving the coupling $Y_{e\tau}$. As was the case before, the $\Delta a_e$ emerges from one-loop diagrams alone. However, there exists no choice of Yukawa couplings that can incorporate $\Delta a_\mu$ in this scenario; LFV processes highly suppress all the couplings that provide corrections $a_\mu$. 


\section{Collider Analysis\label{sec:collider}}

The Yukawa structure of the new Higgs boson $H$ required to simultaneously explain $\Delta a_{\ell}$, NSI as well as neutrino oscillation data entails concurrent non-zero values of all three diagonal elements $\{Y_{ee},Y_{\mu\mu},Y_{\tau\tau}\}$ or the first and second generation off-diagonal entries $\{Y_{e\mu},Y_{\mu e}\}$, as discussed in Sec.~\ref{sec:flavor}. At lepton colliders, these couplings could be directly accessed via searches in the $e^{+}e^{-}\to \ell^{+}\ell^{-}\left(H \to \ell^{+}\ell^{-}\right)$ channel, where $\ell = e,\mu$.  In this section, we explore the sensitivity of the aforesaid channels to probe $Y_{ee}$ and $Y_{e\mu}$ at the projected ILC configuration: $\sqrt{s}=1~\text{TeV}$ with an integrated luminosity $\mathcal{L}=500~\text{fb}^{-1}$~\cite{Behnke:2013xla,Baer:2013cma,Adolphsen:2013jya,Adolphsen:2013kya,Behnke:2013lya}. We also study the projected sensitivity on $Y_{\mu\mu}$ from direct searches in the $\mu^{+}\mu^{-}\to \mu^{+}\mu^{-}\left(H \to \mu^{+}\mu^{-}\right)$ channel at the future muon collider~(MuC) with configuration: $\sqrt{s}=3~\text{TeV}$, $\mathcal{L}=1~\text{ab}^{-1}$~\cite{Delahaye:2019omf,Shiltsev:2019rfl}.

We generate signal and background events at leading order~(LO) using the \texttt{MadGraph5\_aMC@NLO}~\cite{Alwall:2014hca} framework. Showering and hadronization is performed with \texttt{Pythia-8}~\cite{Sjostrand:2006za,Sjostrand:2014zea} while fast detector response is simulated with \texttt{Delphes-3.5.0}~\cite{deFavereau:2013fsa}. We utilize the default \texttt{ILCgen}~\cite{ILCgen} and \texttt{MuonCollider}~\cite{MuonCollider} detector cards to simulate the response of the ILC and the muon collider. 

\subsection{$Y_{ee}$ at ILC}
\label{sec:yee_ILC}

\begin{figure}
    \centering
    \includegraphics[scale=0.25]{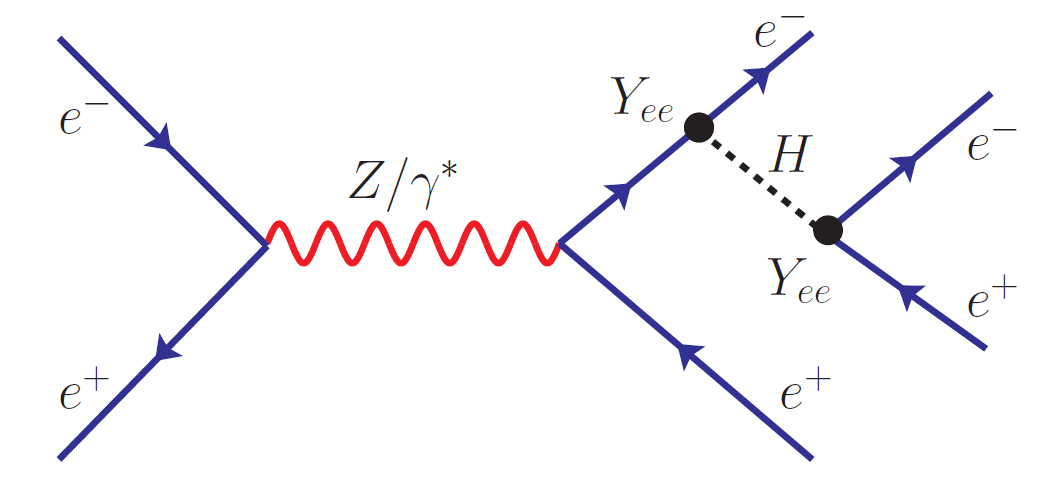}
    \includegraphics[scale=0.25]{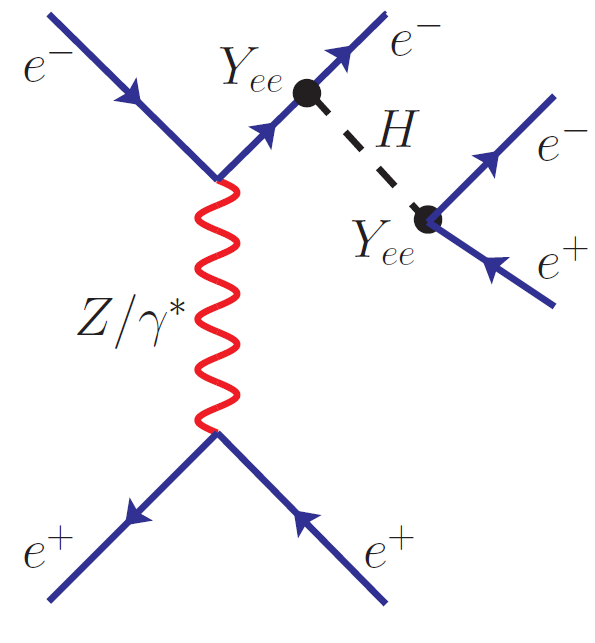}
   \includegraphics[scale=0.25]{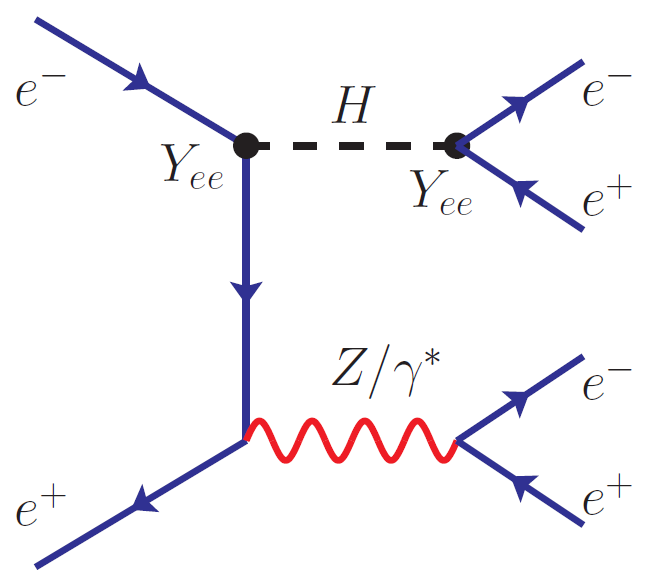}
    \caption{Some illustrative LO Feynman diagrams for the signal process: $e^{+}e^{-} \to e^{+}e^{-}\left(H_{2} \to e^{+}e^{-}\right)$. }
    \label{fig:yee_feynman_diag}
\end{figure}

We study the projected sensitivity on $Y_{ee}$ at the ILC using the channel
\begin{equation}
    e^{+}e^{-} \to e^{+}e^{-}H \to e^{+}e^{-}\left(H \to e^{+}e^{-}\right).
    \label{eqn:process_yee}
\end{equation} 
A few typical leading order~(LO) Feynman diagrams of this process are illustrated in Fig.~\ref{fig:yee_feynman_diag}.  The dominant SM background is $e^{+}e^{-} \to e^{+}e^{-}e^{+}e^{-}$. 

\begin{figure}[!htb]
    \centering
    \includegraphics[scale=0.25]{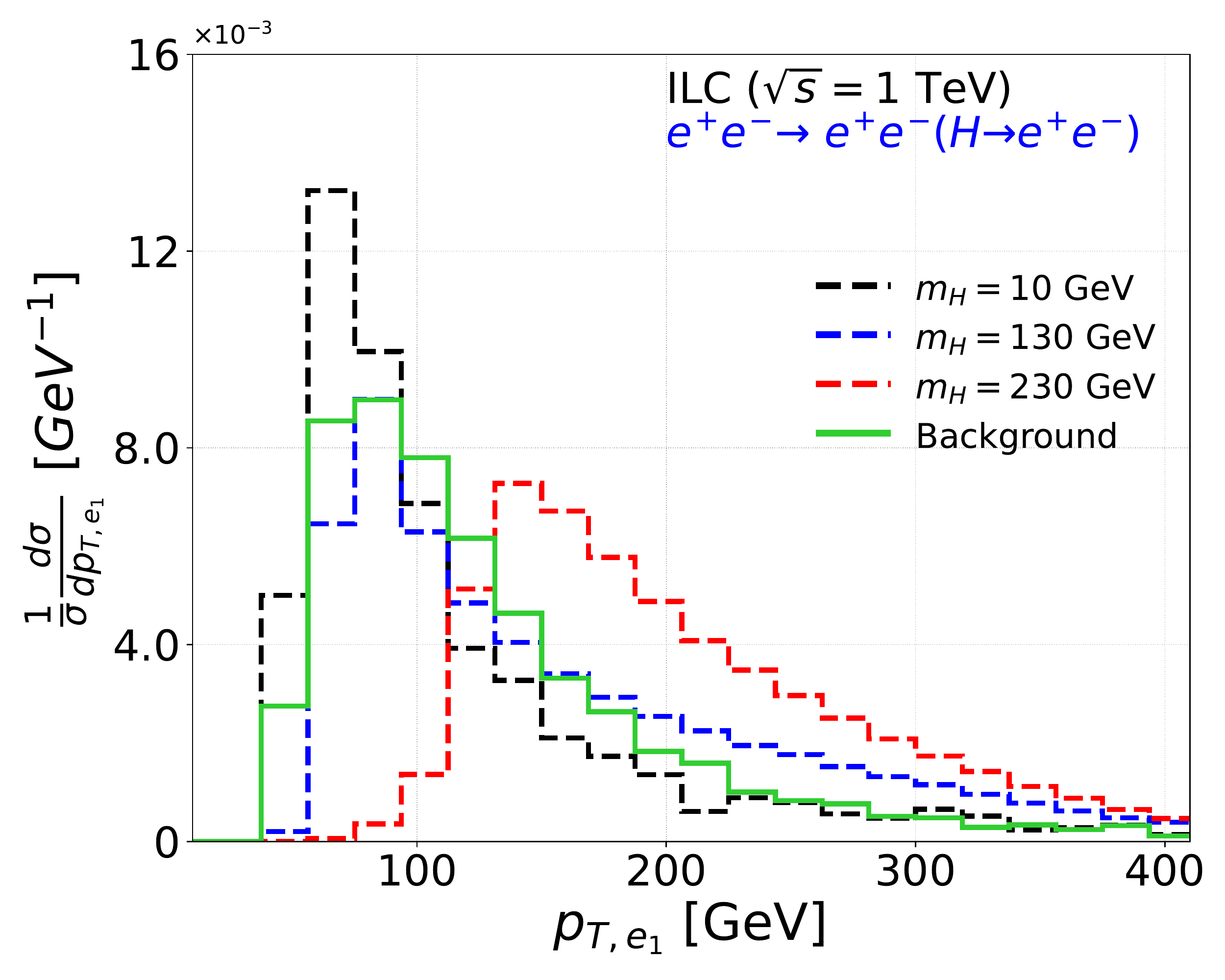}\hspace{1.0cm}\includegraphics[scale=0.25]{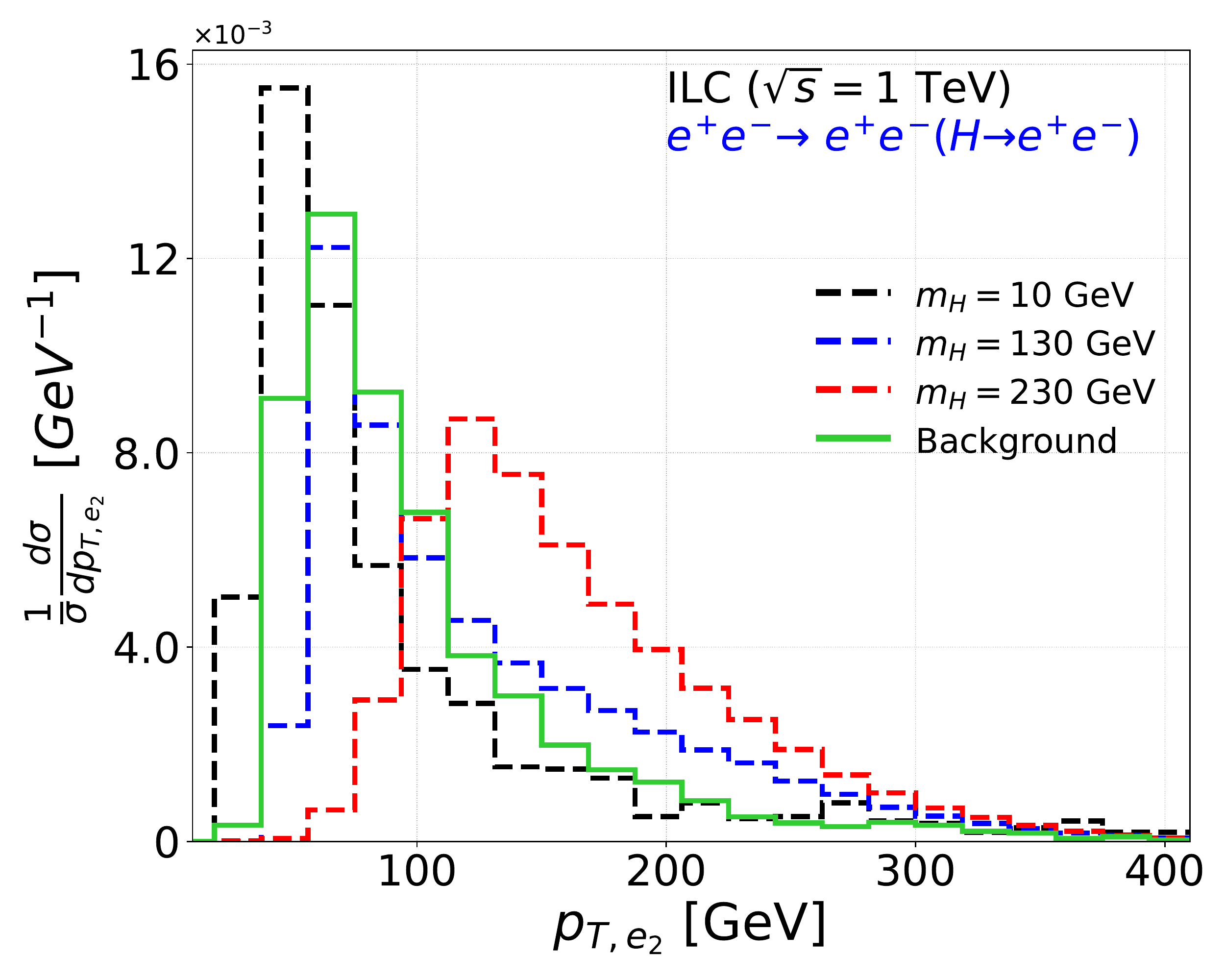}\\
    \includegraphics[scale=0.25]{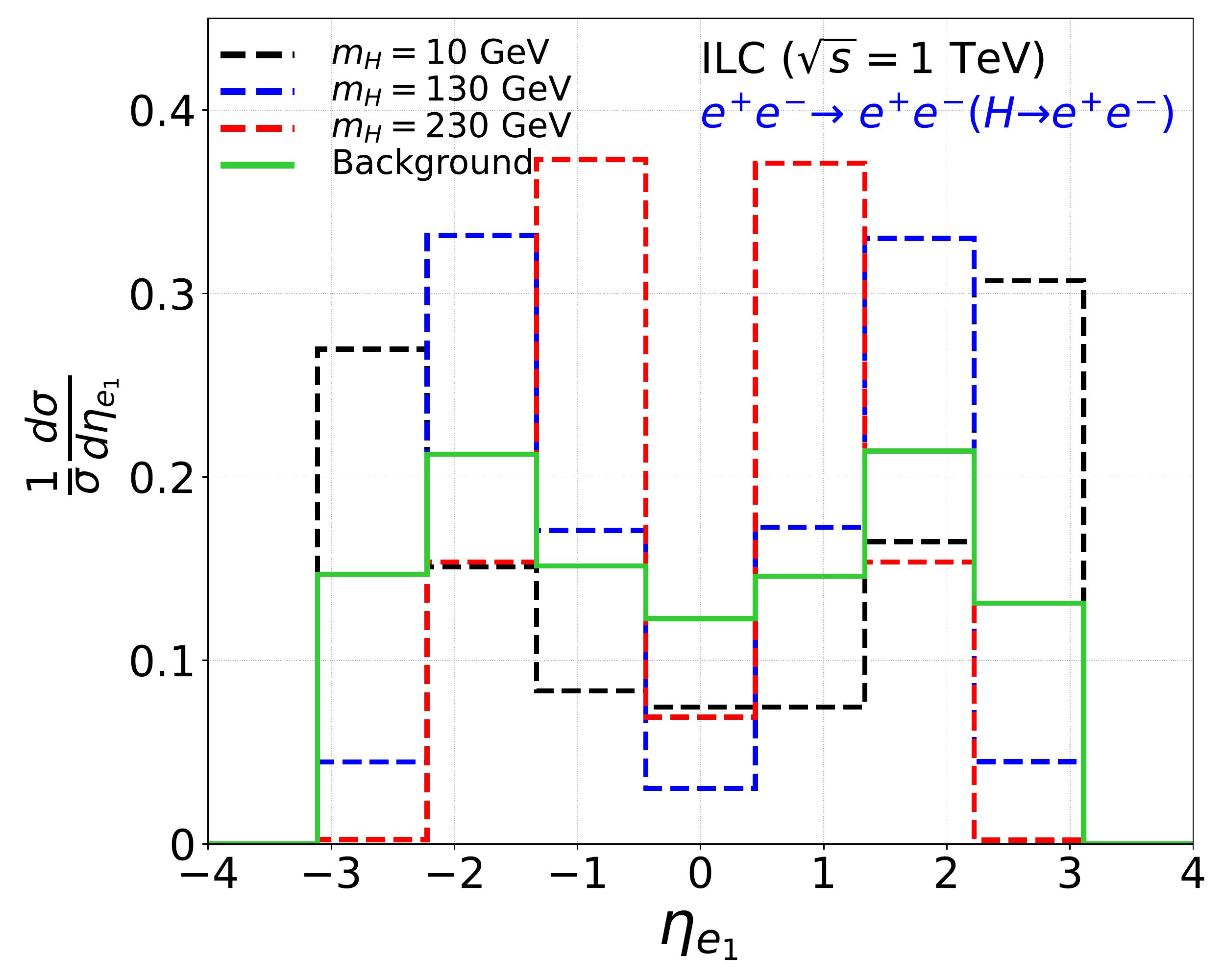}\hspace{1.0cm}\includegraphics[scale=0.25]{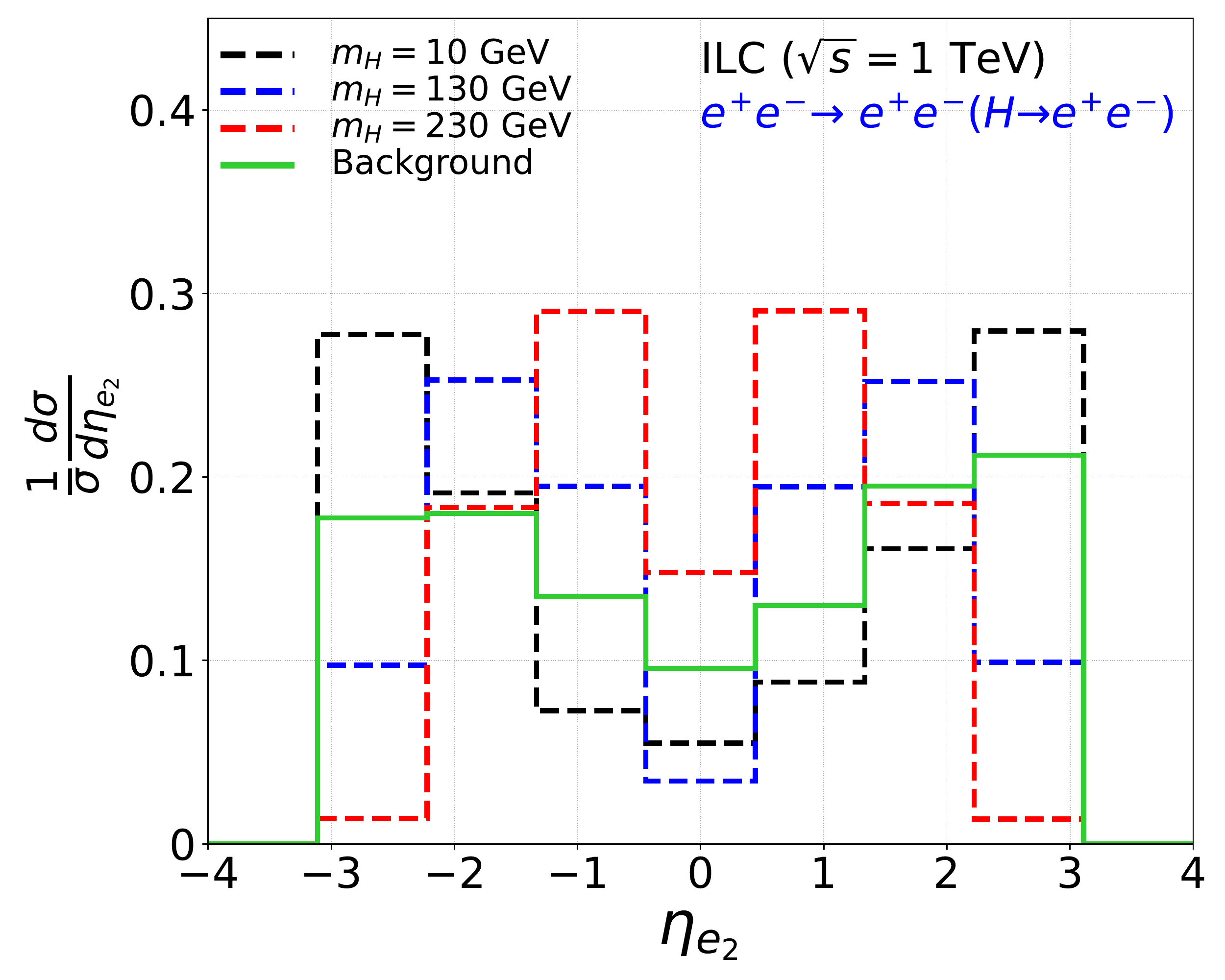}\\
    \caption{$p_{T}$ and $\eta$ distributions of the final state electrons with the highest and $2^{nd}$ highest $p_{T}$, $e_{1}$ and $e_{2}$, respectively, in the $e^{+}e^{-} \to e^{+}e^{-}(H \to e^{+}e^{-})$ channel at the ILC with $\sqrt{s}=1~$TeV. The black, blue and red dashed lines represent the distributions for signal benchmarks $m_{H} =$~10,~130 and 230~GeV, respectively. The background distribution is presented as green solid line.}
    \label{fig:yee_ILC_pt_eta}
\end{figure}

We select events containing exactly two isolated electrons and two isolated positrons in the final state with $p_{T} > 2~$GeV and $|\eta| < 3.0$. We further impose $p_{T, e_{1,2}} > 10~$GeV and $p_{T,e_{3,4}} > 5~$GeV, where $e_{1}$ and $e_{4}$ are the highest and lowest $p_{T}$ leptons, respectively. In Fig.~\ref{fig:yee_ILC_pt_eta}, we illustrate the $p_{T}$ distribution of the highest and second-highest $p_{T}$ leptons, $e_1$ and $e_2$, respectively, for three signal benchmarks, $M_{H} = 10,~130$ and 230~GeV, and the $4e$ background process. In the signal process, the Higgs boson can recoil against an electron or a positron as seen in the first two diagrams in Fig.~\ref{fig:yee_feynman_diag}. This leads to an overall improvement in the $p_{T}$ of the recoiling electron with increasing $M_{H}$. At relatively large $m_{H}$, this recoiling electron becomes the dominant constituent in $e_{1}$. Correspondingly, we expect to see an upward shift in the peak of $p_{T,e_{1}}$ distribution with increasing $m_{H}$. We observe this behaviour in Fig.~\ref{fig:yee_ILC_pt_eta} where $p_{T,e_{1}}$ peaks at $\sim 60~$GeV in the $M_{H}=10~$GeV scenario and the peak position shifts to $p_{T,e_{1}} \sim 90$ and $\sim 130~$GeV in the $m_{H} = 130~$ and 230~GeV scenarios, respectively. Furthermore, we observe that the overall distribution gets flatter with increasing $m_{H}$. The background $p_{T}$ distribution of $e_{1}$ and $e_{2}$ peaks at $s\sim 90~$ and $\sim 60~$ GeV, respectively, which roughly coincides with the peak in the $M_{H} = 130~$GeV scenario. The $4e$ background process also includes diagrams where an on-shell $Z$ boson recoils against an electron or positron leading to the aforesaid similarity in peak positions. Next, let us focus on the pseudorapidity distributions of the final state leptons. We present the $\eta_{e_{1}}$ and $\eta_{e_{2}}$ distributions in Fig.~\ref{fig:yee_ILC_pt_eta}. We observe that $e_{1}$ and $e_{2}$ in the background are mostly produced in the forwards regions of the detector due to back-to-back production of electron-positron pairs. On the contrary, the leading and sub-leading $p_{T}$ leptons in signal benchmarks with large $m_{H}$~($\sim 230~$GeV) are mostly produced in the central regions $|\eta| \lesssim 1.0$ by virtue of their larger transverse momenta. At relatively smaller Higgs masses $M_{H}\sim 10$ and $130~$GeV, the peaks in pseudorapidity distribution roughly coincides with that of background.

\begin{table}[!htb]
    \centering\scalebox{0.8}{
    \begin{tabular}{||c|c|c|c|c|c||} \hline
         $m_{H}$ & \multicolumn{3}{c|}{Optimized cuts} & \multirow{2}{*}{Signal eff.} & \multirow{2}{*}{Bkg eff.} \\ \cline{2-4} 
         $[\text{GeV}]$ & $P_{T,e_{1}}>$~[GeV] &  $|\eta_{e_{1}}| <$ & $|\eta_{e_{2}}| <$ &  & \\ \hline
         10 & 40 & 3.0 & 3.0 & 0.002 & 0.004 \\
         40 & 30 & 2.7 & 2.7 & 0.066 & 0.005  \\
         70 & 30 & 3.0 & 3.0 & 0.207 & 0.007  \\
         100 & 40 & 2.6 & 2.9 & 0.328 & 0.010  \\
         130 & 60 & 2.4 & 2.5 & 0.385 & 0.003  \\
         160 & 70 & 2.1 & 2.3 & 0.415 & 0.002  \\
         190 & 90 & 2.0 & 2.2 & 0.455 & 0.002 \\
         230 & 110 & 1.7 & 2.0 & 0.457 & 0.001  \\
         270 & 140 & 1.6 & 1.9 & 0.475 & $9 \times 10^{-4}$  \\
         310 & 150 & 1.6 & 2.2 & 0.511 & $8 \times 10^{-4}$  \\
         350 & 180 & 1.3 & 2.5 & 0.476 & $5 \times 10^{-4}$  \\ \hline
    \end{tabular}}
    \caption{Optimized selection cuts on $p_{T,e_{1}}$, $\eta_{e_{1}}$ and $\eta_{e_{2}}$, signal efficiency and background efficiency, from cut-based analysis in the $e^{+}e^{-} \to e^{+}e^{-}(H \to e^{+}e^{-})$ channel at $\sqrt{s} = 1~$TeV ILC for several signal benchmark points.}
    \label{tab:ILC_yee_cutflow}
\end{table}

\begin{figure}
    \centering
     \includegraphics[scale=0.29]{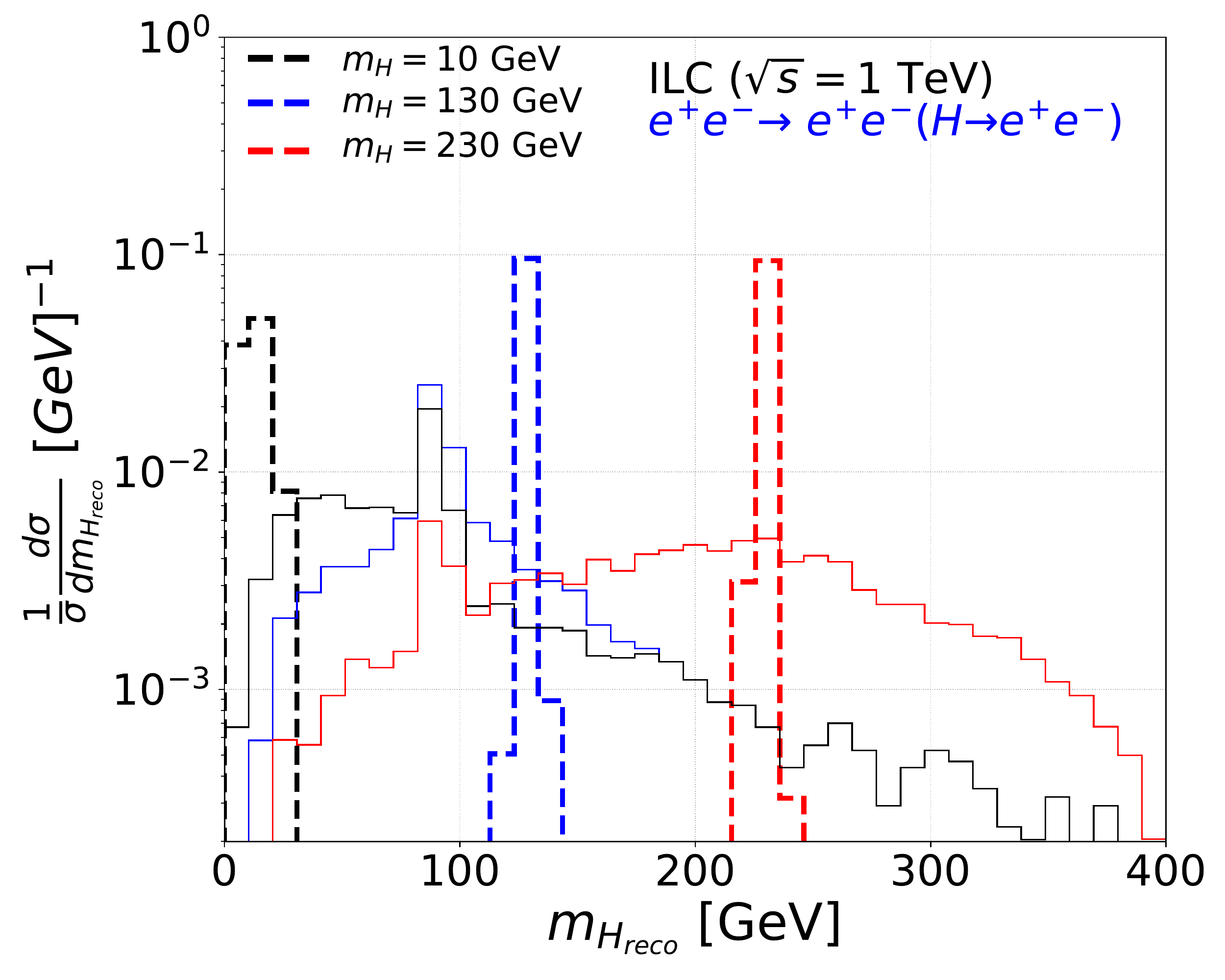}\includegraphics[scale=0.32]{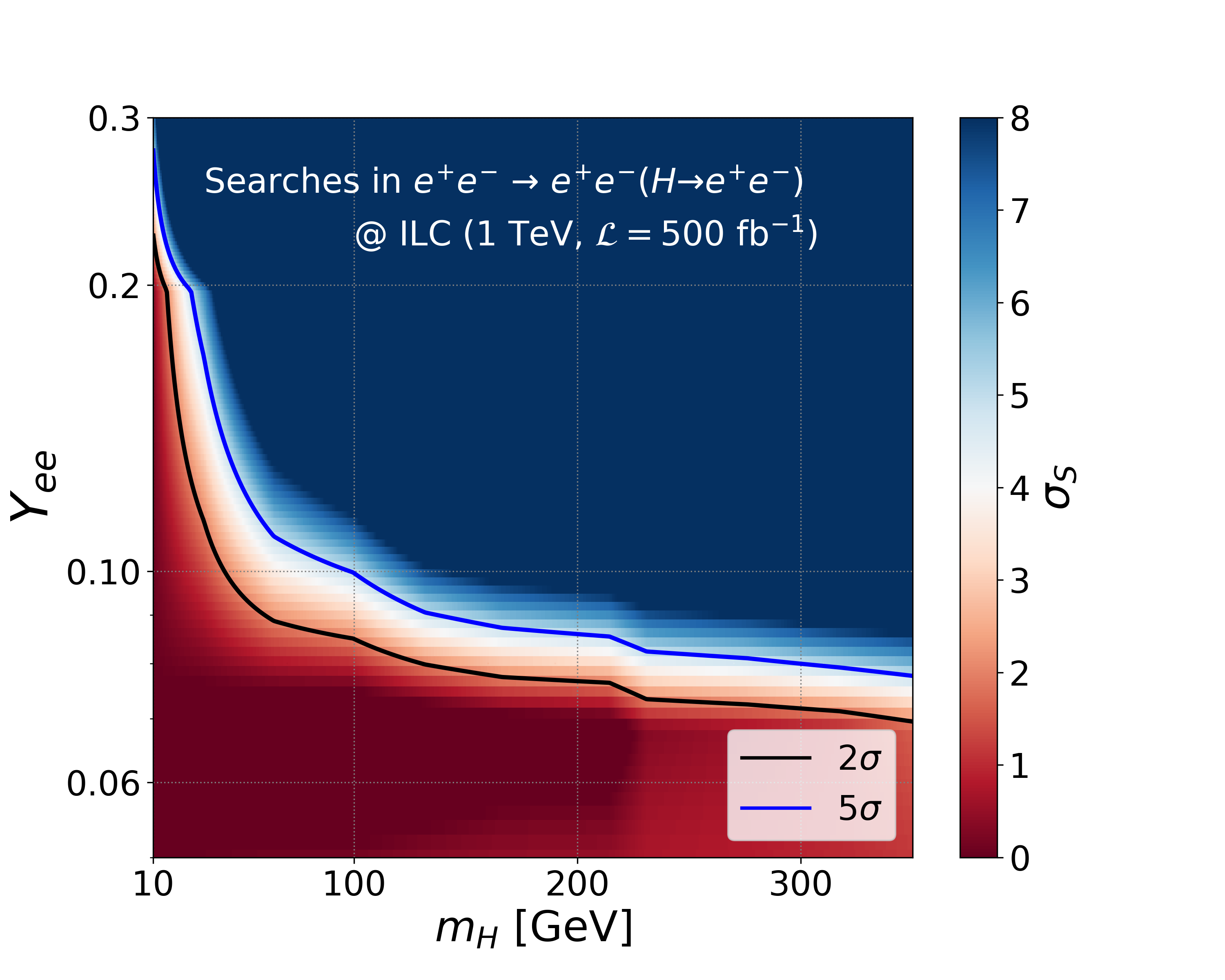}
    \caption{\textit{Left:} Invariant mass distribution of the reconstructed Higgs boson for three signal benchmarks $m_{H} = 10$~(black dashed),~130~(blue dashed) and 230~GeV~(red dashed). The corresponding solid lines represent the respective $4e$ background distributions. \textit{Right:} Projected upper limits on $Y_{ee}$ from direct searches in the $e^{+}e^{-} \to e^{+}e^{-}\left(H\to e^{+}e^{-}\right)$ channel at the 1~TeV ILC assuming $\mathcal{L}=500~{\rm fb^{-1}}$. The black solid and blue solid lines represent the projected upper limits at $2\sigma$ and $5\sigma$, respectively. The color palette in the z-axis represents the signal significance.}
    \label{fig:ILC_yee_proj}
\end{figure}

The four final state leptons can be classified into four opposite sign~(OS) lepton pairs, one of which is produced from the decay of $H$ in the signal process. The electron-positron pair associated with $H$ is identified by minimizing $\left(m_{e^{+}e^{-}}^{2} - m_{H}^{2}\right)$, where $m_{e^{+}e^{-}}$ is the invariant mass of an OS electron pair and $m_{H}$ is the mass of $H$ in the signal benchmark under consideration. We present the invariant mass distribution of the reconstructed Higgs boson $H_{rec}$ in Fig.~\ref{fig:ILC_yee_proj} for three signal benchmark scenarios $m_{H} = 10,~130$ and 230~GeV as dashed black, blue and red lines, respectively. The respective background distributions are presented in solid lines. 

Taking cognizance of the $m_{H_{rec}}$ distributions in Fig.~\ref{fig:ILC_yee_proj}, we require signal and background events to satisfy $m_{H} \pm 10~$GeV. In addition, we also optimize the selection cuts on $p_{T,e_{1}}$,$\eta_{e_{1}}$ and $\eta_{e_{2}}$ for several signal benchmarks in order to maximize the signal significance $\sigma_{S} = S /\sqrt{S+B}$, where $S$ and $B$ are the signal and background yields. In Table~\ref{tab:ILC_yee_cutflow}, we present the optimized selection cuts, signal and background yields, along with the signal significance, for several signal benchmarks within the range $10~\text{GeV} \leq M_{H} \leq 350~\text{GeV}$. We utilize the optimized signal and background efficiencies obtained in the previous step to derive the projected sensitivity on $Y_{ee}$ as a function of $M_{H}$ at the ILC $1~\text{TeV}$ machine. We present our results as upper limit projections at $2\sigma$ and $5\sigma$ in the $\{Y_{ee},M_{H}\}$ plane in Fig.~\ref{fig:ILC_yee_proj}.

We observe that the projected upper limits are weaker at small $m_{H} \sim 10~\text{GeV}$, and improves by a factor of $\sim 3$ until $m_{H} \sim 50~$GeV. Afterwards, it gradually improves with increasing $m_{H}$. At a particular $\{Y_{ee},m_{H}\}$, the projected sensitivity is determined by three factors: the signal production cross-section $\propto$ $Y_{ee}$ and $m_{H}$, signal efficiency $\propto$ $m_{H}$ and background efficiency. At the ILC 1~TeV run, the LO production cross-section of the signal process in Eq.~\ref{eqn:process_yee}, $\sigma_{Y_{ee}}$, peaks at roughly $\sim 100~\text{fb}$ with $m_{H} \sim 220-240~\text{GeV}$ for $Y_{ee} = 1.0$, and falls down to $\sim 7~\text{fb}$~($\sim 70~\text{fb}$) at $m_{H} \sim 10$~GeV~($350~\text{GeV}$), respectively. On the other hand, we observe that the signal efficiency in Table~\ref{tab:ILC_yee_cutflow} improves with increasing $m_{H}$ until around 310~GeV after which it falls down marginally at $m_{H} = 350~$GeV. The signal efficiency improves by $\mathcal{O}(100)$ from $\sim 0.002$ at $m_{H}=10~$GeV to around $0.207$ at $m_{H}=70~$GeV. Afterwards, it registers a gradual rise to $\sim 0.511$ at $m_{H}=310~$GeV. At $m_{H}=350~$GeV, it falls down to 0.476. The background efficiency, on the other hand, exhibits maximal value near $m_{H} \sim m_{Z}$. It improves from $\sim 0.004$ at $m_{H} = 10~$GeV to $\sim 0.01$ at $m_{H} = 100~$GeV beyond which it gradually falls down by a factor of $\sim 20$ to $\sim 5 \times 10^{-4}$ at $m_{H}=350~$GeV. At small $m_{H}$~($\lesssim 70~$GeV), the small signal production cross-section coupled with a relatively small signal efficiency leads to weaker sensitivity on $Y_{ee}$. In the $70~\text{GeV} \lesssim m_{H} \lesssim 230~{\rm GeV}$ region, both signal production cross-section and efficiency improves while the background efficiency deteriorates. All these factors contribute towards improving the projected sensitivity. As we move to signal benchmarks with heavier $m_{H}$, the signal production cross-section registers a decrement. However, the signal efficiency continues to increase until $m_{H} = 310~$GeV and the background efficiency continues to plummet. The later two counters the decrements in signal cross-section, and the projected upper limits continue improving. At $m_{H}=350~$GeV, the signal efficiency is $\mathcal{O}(8\%)$ smaller than its $m_{H}=310~$GeV counterpart. However, this decrement in signal efficiency is not reflected in the projected upper limits in Fig.~\ref{fig:ILC_yee_proj} since the background efficiency falls down by a relatively larger rate $\mathcal{O}(40\%)$. Overall we observe that the 1~TeV ILC machine would be able to probe $Y_{ee}$ up to $Y_{ee} \gtrsim 0.085$~(at $2\sigma$) at $m_{H} = 100~$GeV via direct searches in the $e^{+}e^{-} \to e^{+}e^{-}(H \to e^{+}e^{-})$ channel.

\subsection{$Y_{e\mu}$ at ILC}\label{sec:yeu_ILC}
To explore the projected sensitivity to $Y_{e\mu}$, we consider the channel
\begin{equation}
    e^{+}e^{-} \to e^{\pm}\mu^{\mp} H \to  e^{\pm}\mu^{\mp}(H \to e^{\pm}\mu^{\mp}).
\end{equation}

\begin{figure}[!htb]
    \centering
    \includegraphics[scale=0.25]{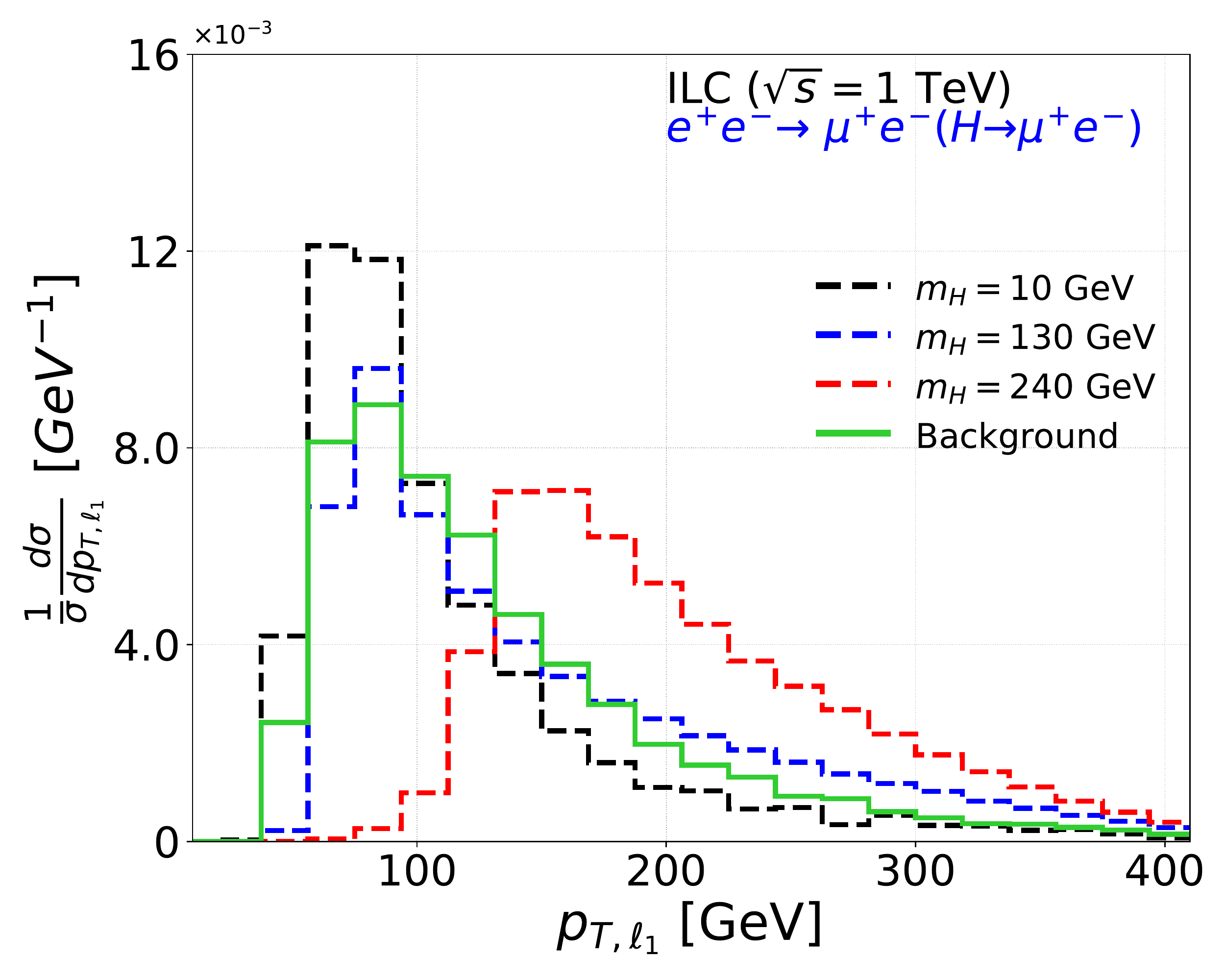}\hspace{1.0cm}\includegraphics[scale=0.25]{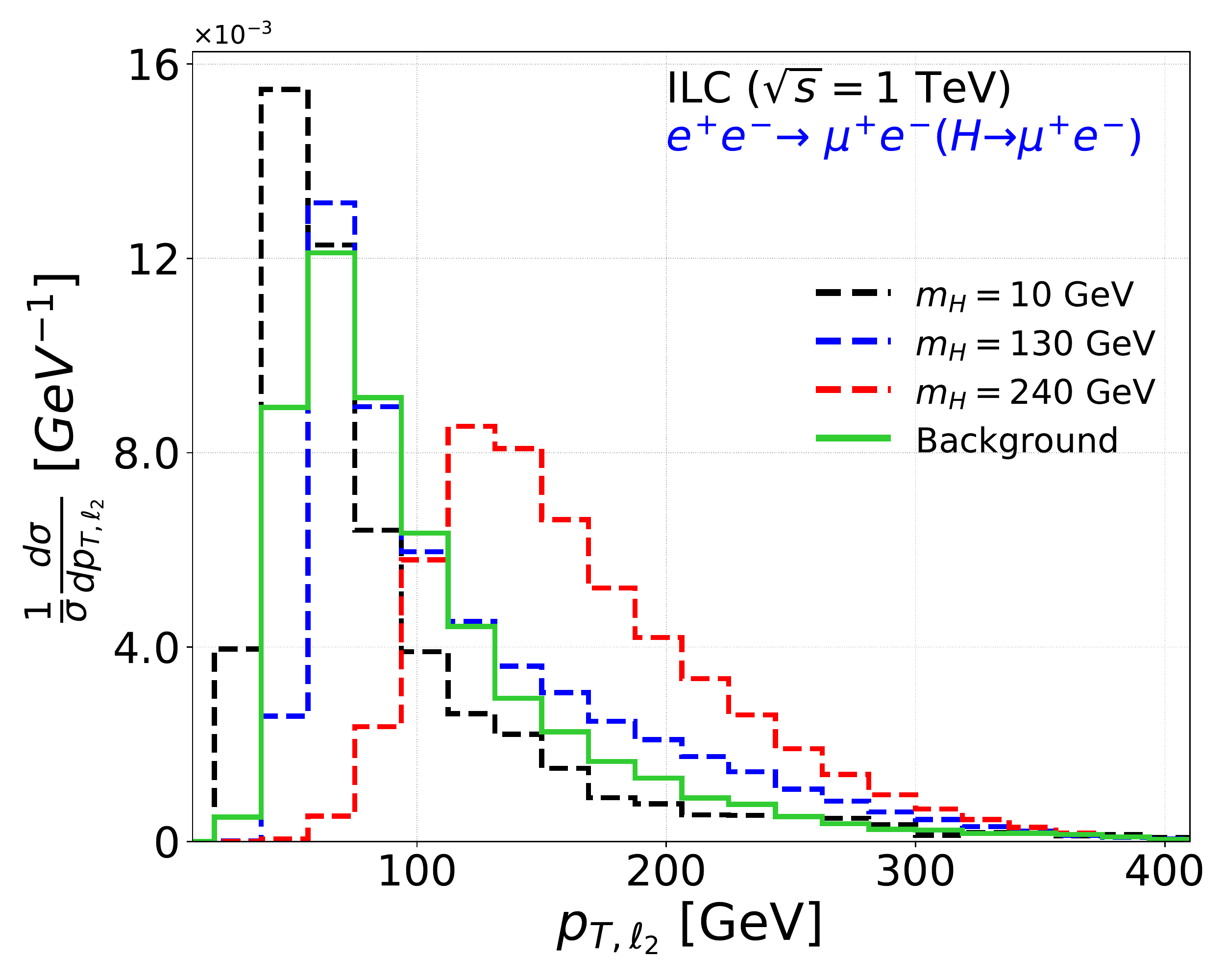}\\
    \includegraphics[scale=0.25]{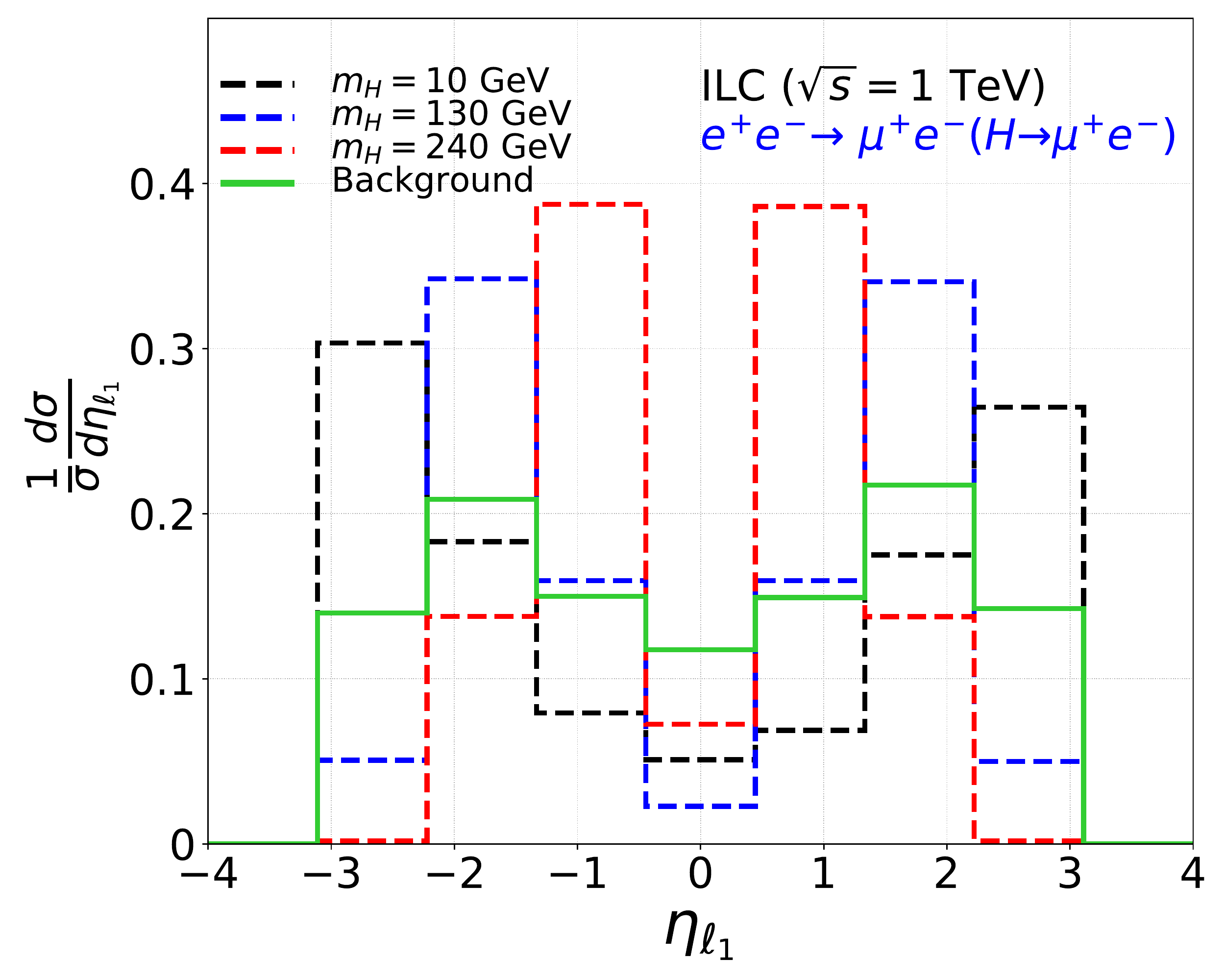}\hspace{1.0cm}\includegraphics[scale=0.25]{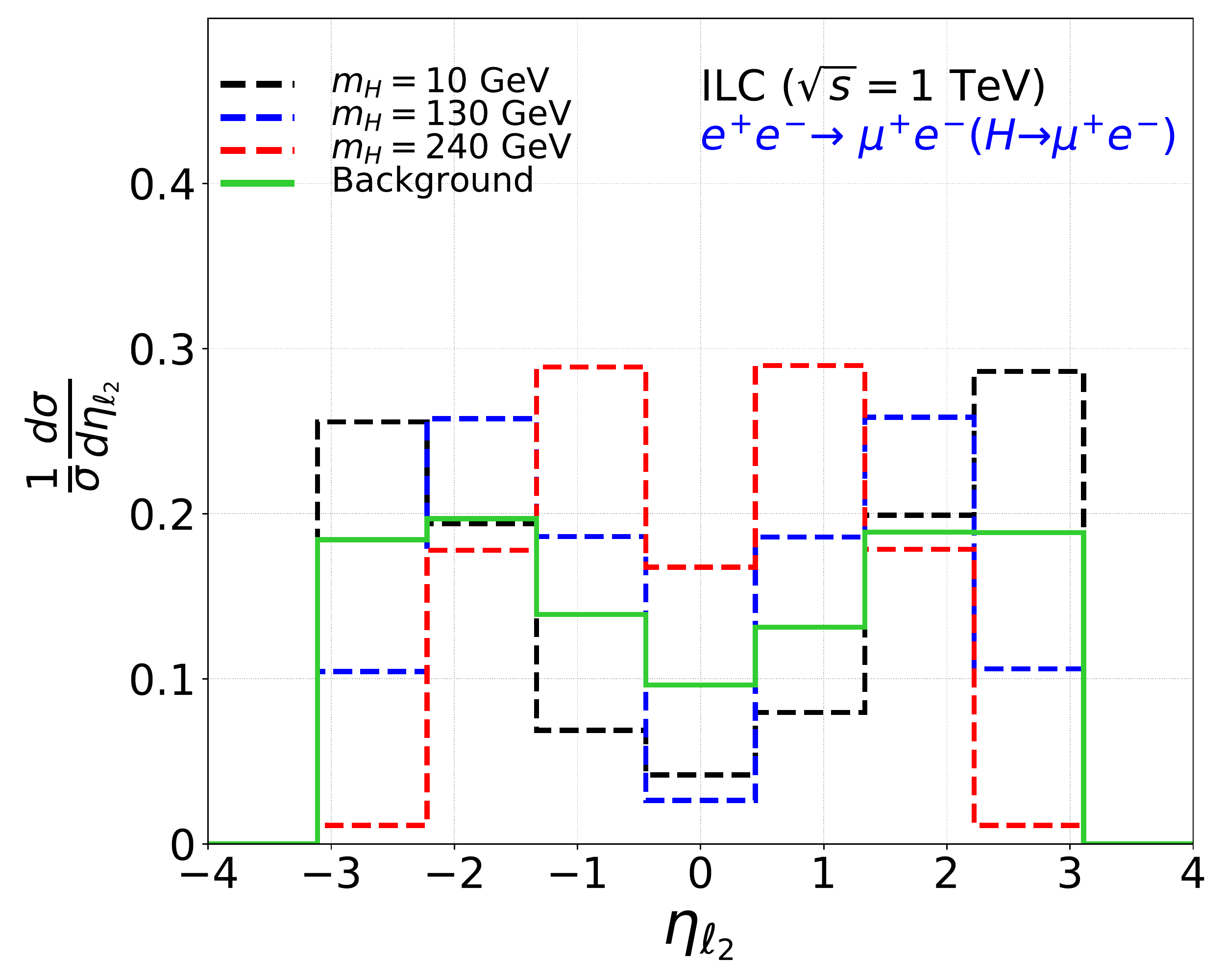}
    \caption{$p_{T}$ and $\eta$ distributions of the final state leptons with the highest and $2^{nd}$ highest $p_{T}$, $\ell_{1}$ and $\ell_{2}$, respectively, in the $e^{+}e^{-} \to \mu^{+}e^{-}(H \to \mu^{+}e^{-})$ channel at the ILC with $\sqrt{s}=1~$TeV. The black, blue and red dashed lines represent the distributions for signal benchmarks $m_{H} =$~10,~130 and 240~GeV, respectively. The background distribution is presented as green solid line.}
    \label{fig:yemu_pt_eta}
\end{figure}

We consider events containing an OS electron pair and an OS muon pair. All four leptons~($\ell = e,\mu$) are required to have $p_{T} > 2~$GeV and $|\eta| < 3.0$. The leptons are also required to satisfy $p_{T_{\ell_{1,2}}} > 10~$GeV and $p_{T_{\ell_{3,4}}} > 5~$GeV, where $\ell_{1}$ and $\ell_{4}$ are the highest and lowest $p_{T}$ leptons, respectively. The kinematic behavior of the final state leptons are similar to that exhibited in the $e^{+}e^{-} \to e^{+}e^{-}(H \to e^{+}e^{-})$ channel. One major difference however is the absence of $t$-channel electron exchange diagrams, shown in the right panel of Fig.~\ref{fig:yee_feynman_diag}, since the $H$ couples to $e^{\pm}\mu^{\mp}$ pair instead of $e^{+}e^{-}$ pair. In Fig.~\ref{fig:yemu_pt_eta}, we present the $p_{T}$ and $\eta$ distributions of $l_{1}$ and $l_{2}$ for three signal benchmarks $m_{H} = 10,~130$ and 240~GeV, and the $2e2\mu$ background. We observe that the overall features are roughly similar to that in Sec.~\ref{sec:yee_ILC}. Here as well, the peak of the $p_{T}$ distributions shifts to larger values with increasing $m_{H}$. Similarly, the relative concentration of $\ell_{1}$ and $\ell_{2}$ in the central regions of the detector improves with $m_{H}$.

\begin{table}[!htb]
    \centering\scalebox{0.8}{
    \begin{tabular}{||c|c|c|c|c|c||} \hline
         $m_{H}$ & \multicolumn{3}{c|}{Optimized cuts} & \multirow{2}{*}{Signal eff.} & \multirow{2}{*}{Bkg eff.} \\ \cline{2-4} 
         $[\text{GeV}]$ & $P_{T,e_{1}}>$~[GeV] &  $|\eta_{e_{1}}| <$ & $|\eta_{e_{2}}| <$ &  & \\ \hline
         10 & 30 & 3.0 & 3.0 & 0.007 & 0.002 \\
         40 & 20 & 3.0 & 3.0 & 0.113 & 0.003 \\
         70 & 30 & 3.0 & 3.0 & 0.281 & 0.004 \\
         100 & 40 & 2.7 & 2.9 & 0.428 & 0.006 \\
         130 & 40 & 2.5 & 2.7 & 0.505 & 0.002 \\
         190 & 90 & 2.1 & 2.3 & 0.584 & 0.001 \\
         240 & 110 & 1.8 & 2.2 & 0.618 & $7 \times 10^{-4}$ \\
         270 & 130 & 1.7 & 2.1 & 0.625 & $6 \times 10^{-4}$ \\
         310 & 140 & 1.6 & 2.1 & 0.639 & $4 \times 10^{-4}$ \\
         350 & 160 & 1.4 & 2.0 & 0.627 & $3 \times 10^{-4}$ \\ \hline
    \end{tabular}}
    \caption{Optimized selection cuts on $p_{T,e_{1}}$, $\eta_{e_{1}}$ and $\eta_{e_{2}}$, signal efficiency and background efficiency, from cut-based analysis in the $e^{+}e^{-} \to \mu^{+}e^{-}(H \to \mu^{+}e^{-})$ channel at $\sqrt{s} = 1~$TeV ILC for several signal benchmark points.}
    \label{tab:ILC_yemu_cutflow}
\end{table}

\begin{figure}[!htb]
    \centering
     \includegraphics[scale=0.29]{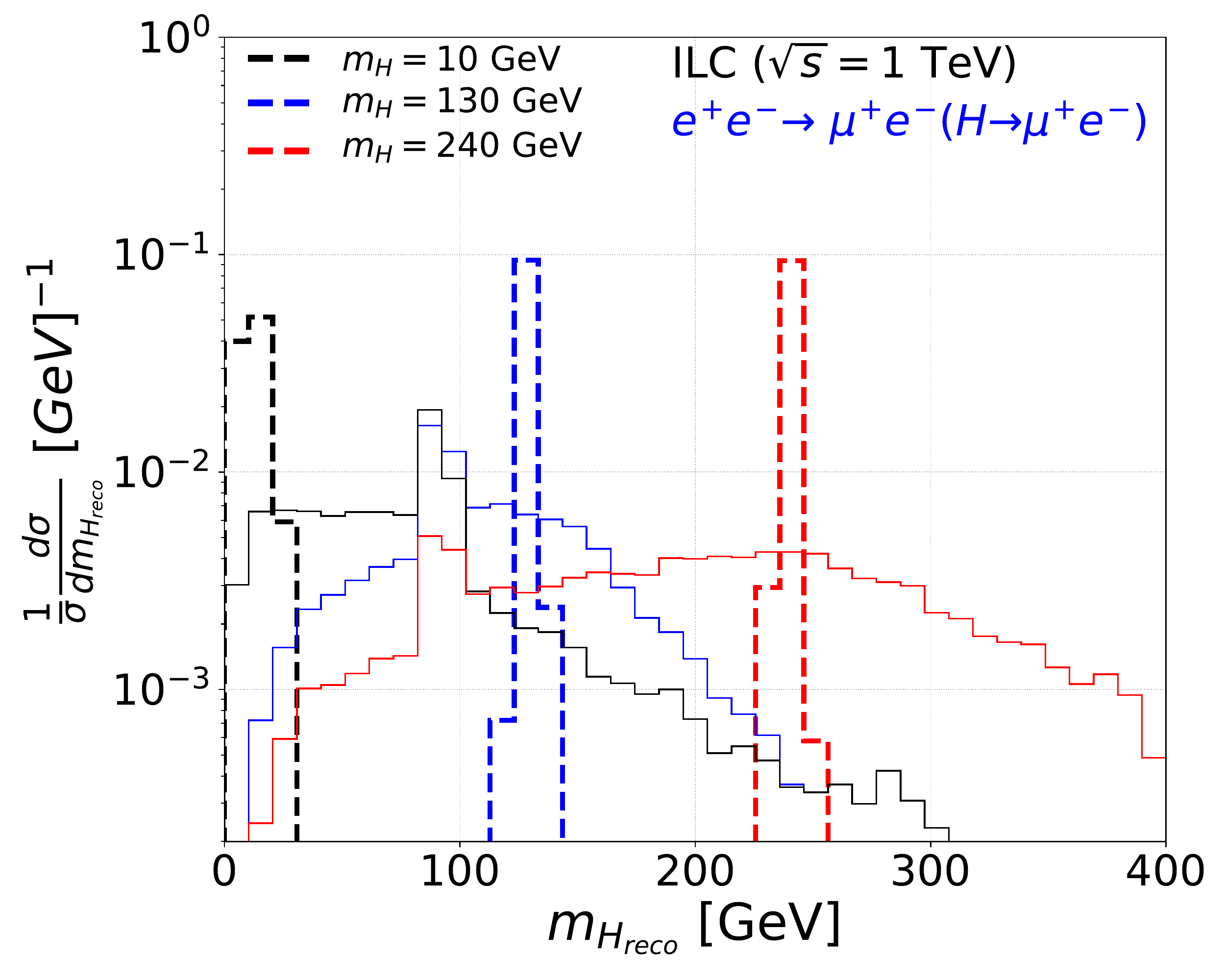}\includegraphics[scale=0.32]{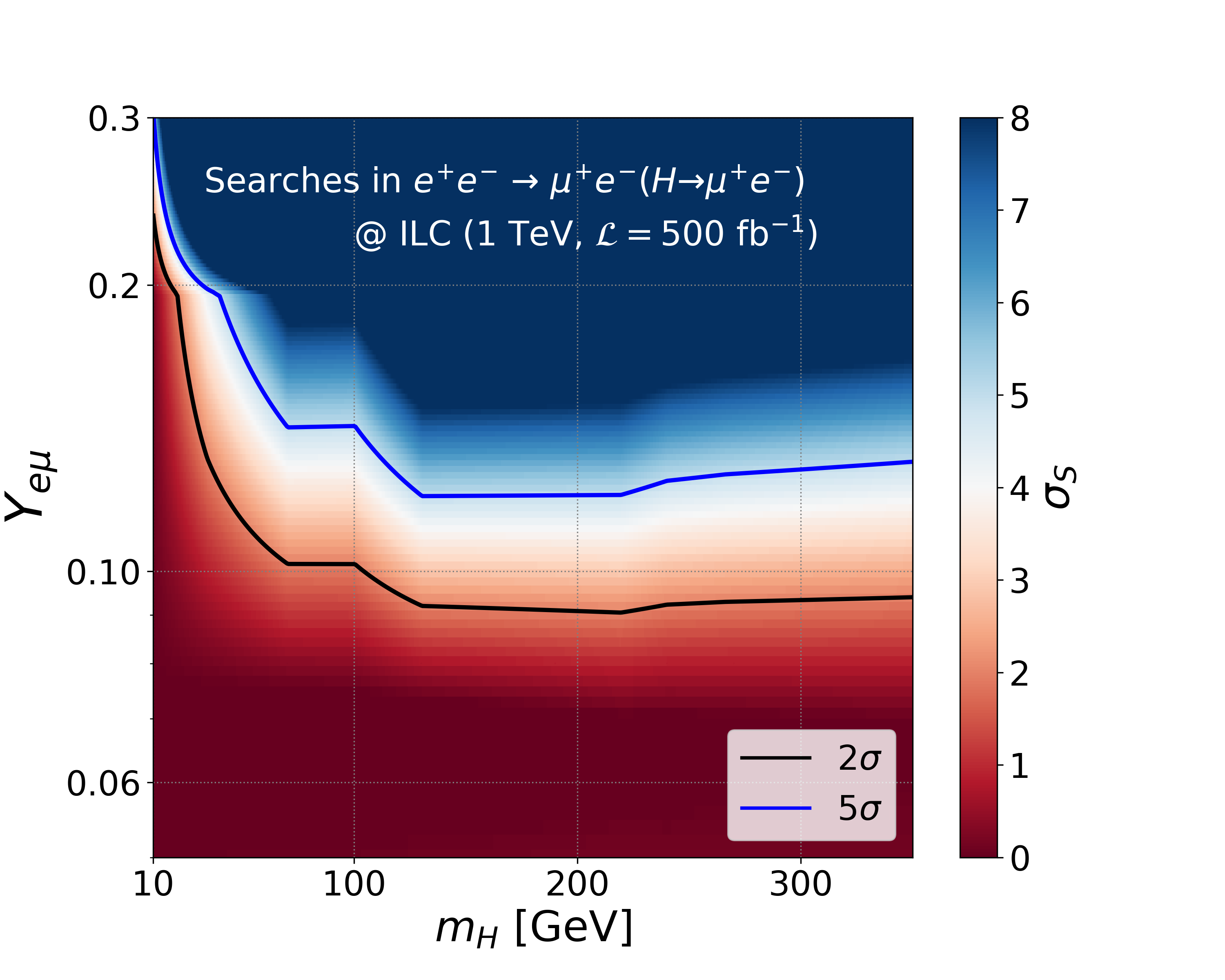}
    \caption{\textit{Left:} Invariant mass distribution of the reconstructed Higgs boson for three signal benchmarks $m_{H} = 10$~(black dashed),~130~(blue dashed) and 240~GeV~(red dashed). The corresponding solid lines represent the respective $2e2\mu$ background distributions. \textit{Right:} Projected upper limits on $Y_{e\mu}$ from direct searches in the $e^{+}e^{-} \to \mu^{+}e^{-}\left(H\to \mu^{+}e^{-}\right)$ channel at the 1~TeV ILC assuming $\mathcal{L}=500~{\rm fb^{-1}}$. The black solid and blue solid lines represent the projected upper limits at $2\sigma$ and $5\sigma$, respectively. The color palette in the z-axis represents the signal significance.}
    \label{fig:ILC_yemu_proj}
\end{figure}

In order to reconstruct the Higgs boson, we perform mass minimization similar to that in Sec.~\ref{sec:yee_ILC}. We associate an OS electron-muon pair with $H$ that minimizes $(m_{e^{\pm}\mu^{\mp}}^{2} - m_{H}^{2})$. In Fig.~\ref{fig:ILC_yemu_proj}, we present the invariant mass distribution of the reconstructed Higgs boson for several signal benchmarks and the $2e2\mu$ background. We impose $|m_{H_{reco}} \pm 10~\text{GeV}|$ to filter the signal from the $2e2\mu$ background continuum. Additionally, we also optimize selection cuts on $p_{T,\ell_{1}}$, $\eta_{\ell_{1}}$ and $\eta_{\ell_{2}}$. We present the optimized signal and background efficiency for several choices of $m_{H}$ along with the respective set of optimized cuts in Table~\ref{tab:ILC_yemu_cutflow}. Both signal and background efficiencies follow a similar behaviour to that in Sec.~\ref{sec:yee_ILC}. We observe that the signal efficiency improves by a factor of $\sim 40$ from $\sim 0.007$ at $m_{H}=10~$GeV to $\sim 0.281$ at $m_{H}=70~$GeV. On further increasing $m_{H}$, it reaches to $\sim 0.639$ at $m_{H}=310~$GeV before falling down to $\sim 0.627$ at $m_{H} = 350~$GeV. We translate these efficiencies into upper limit projections in $\{Y_{e\mu},m_{H}\}$ plane, shown in Fig.~\ref{fig:ILC_yemu_proj}. We observe that the 1~TeV ILC machine would be able to probe $Y_{e\mu}$ up to $\sim 0.1$ and $\sim 0.09$ at $m_{H} = 100$ and 200~GeV, respectively, at $2\sigma$. We note that the projected sensitivity on $Y_{e\mu}$ is relatively weaker compared to $Y_{ee}$~(see Sec.~\ref{sec:yee_ILC}) due to relatively smaller signal production cross-section in the present channel.

\subsection{$Y_{\mu\mu}$ at future muon collider}
\label{sec:ILC_yemu}
The $e^{+}e^{-} \to \mu^{+}\mu^{-}(H \to \mu^{+}\mu^{-})$ channel offers a direct probe to the $Y_{\mu\mu}$ coupling at electron-positron colliders. However, the production cross-section of this process is at least an order of magnitude smaller relative to $e^{+}e^{-} \to e^{+}e^{-}(H \to e^{+}e^{-})$ process due to the absence of $t$-channel $Z/\gamma^{*}$ exchange diagrams, leading to weaker sensitivity. The issue of smaller production cross-section could be, however, circumvented at a muon collider~(MuC). Muons produce less synchrotron radiation relative to the electrons, and therefore, a MuC machine could be operated at a much higher center of mass energies compared to an electron-positron collider. Consequently, the most stringent sensitivity on muon Yukawa coupling via direct searches in the $\ell^{+}\ell^{-}H$ channel could perhaps be achieved at a future MuC. Taking this into consideration, in the present section, we explore the potential capability of a muon collider with configuration $\{\sqrt{s}=3~$TeV, $\mathcal{L}=1~\text{ab}^{-1}\}$ to probe $Y_{\mu\mu}$ via direct searches in the $\mu^{+}\mu^{-} \to \mu^{+}\mu^{-}(H \to \mu^{+}\mu^{-})$ channel. In order to simulate detector response, we utilize the default \texttt{Delphes} card for a muon collider~\cite{MuonCollider}.

\begin{table}[!htb]
    \centering\scalebox{1.0}{
    \begin{tabular}{||c|c|c||} \hline
         $m_{H}$ & \multirow{2}{*}{Signal eff.} & \multirow{2}{*}{Bkg eff.} \\ 
         $[\text{GeV}]$ &  & \\ \hline
         10 & 0.006 & 0.005 \\
         40 & 0.151 & 0.005 \\
         70 & 0.267 & 0.005 \\
         100 & 0.291 & 0.022 \\
         150 & 0.292 & 0.003 \\
         200 & 0.297 & 0.003 \\
         250 & 0.321 & 0.002 \\
         300 & 0.369 & 0.002 \\
         350 & 0.395 & 0.002 \\ \hline 
    \end{tabular}}
    \caption{Signal and background efficiencies, from cut-based analysis in the $\mu^{+}\mu^{-} \to \mu^{+}\mu^{-}(H \to \mu^{+}\mu^{-})$ channel at $\sqrt{s} = 3~$TeV MuC for several signal benchmark points.}
    \label{tab:ILC_ymm_cutflow}
\end{table}

\begin{figure}
    \centering
    \includegraphics[scale=0.32]{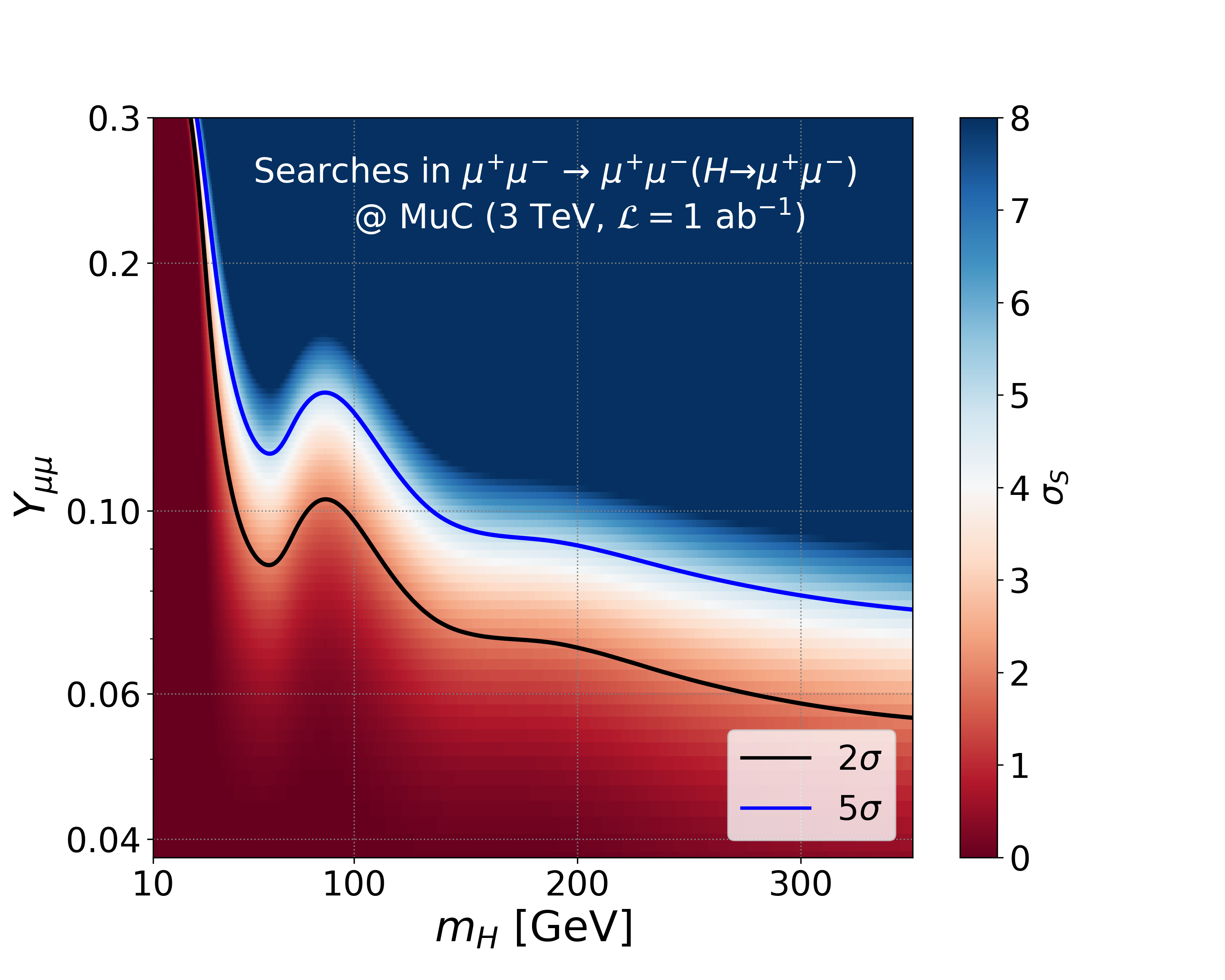}
    \caption{Projected upper limits on $Y_{\mu\mu}$ from direct searches in the $\mu^{+}\mu^{-} \to \mu^{+}\mu^{-}\left(H\to \mu^{+}\mu^{-}\right)$ channel at the 3~TeV MuC assuming $\mathcal{L}=1~{\rm ab^{-1}}$. The black solid and blue solid lines represent the projected upper limits at $2\sigma$ and $5\sigma$, respectively. The color palette in the z-axis represents the signal significance as a function of $\{Y_{\mu\mu},m_{H}\}$.}
    \label{fig:muC_ymm_projections}
\end{figure}

We require events to contain exactly four isolated muons in the final state. We assume detector coverage up to $|\eta| < 2.5$, and also impose $p_{T,\mu} > 10~$GeV. The final state muons exhibit kinematic features that are similar to that of the final state electrons in Sec.~\ref{sec:yee_ILC}. Consequently, we adopt the mass minimization strategy from Sec.~\ref{sec:yee_ILC} to reconstruct the Higgs boson. We do not present the $m_{H_{reco}}$ distribution in the present scenario due to its close similarity with Fig.~\ref{fig:ILC_yee_proj}. The dominant background source is the $\mu^{+}\mu^{-} \to 4\mu$ process. We impose additional selection cuts on the $p_{T}$ of the final state muons in order to improve signal-background discrimination: 
\begin{equation}
    p_{T,\mu_{1}} > 90~\text{GeV}\quad p_{T,\mu_{2}} > 20~\text{GeV}, \quad p_{T,\mu_{3}} > 15~\text{GeV}, 
\end{equation}
where, $\mu_{1}$ represents the highest $p_{T}$ muon.  

We tabulate the signal and background efficiencies for various signal benchmarks corresponding to different values of $m_{H}$ in Table.~\ref{tab:ILC_ymm_cutflow}. In the Higgs mass range of our interest $10~\text{GeV} < m_{H} < 350~\text{GeV}$, we observe that the signal efficiency improves with increasing $m_{H}$. The background efficiency, on the other hand, falls down with $m_{H}$ except for $m_{H}$ close to $m_{Z}$ where the leptons from $Z$ resonance fill into the $m_{H_{reco}}$ distribution. We utilize these efficiencies to derive projected upper limits on $Y_{\mu\mu}$ as a function of $m_{H}$. We illustrate the $2\sigma$ and $5\sigma$ projections in Fig.~\ref{fig:muC_ymm_projections}. The signal efficiency improves by more than one order of magnitude from $\sim 0.006$ at $m_{H} = 10~$GeV to $\sim 0.151$ at $m_{H}=40~$GeV, while the signal production cross-section improves by a factor of $\sim 6$. The background efficiency, on the other hand, remains almost unchanged. This leads to an order of magnitude improvement in the projected upper limits on $Y_{\mu\mu}$. At $m_{H} = 100~$GeV, the background efficiency is roughly $4$ times higher than at its neighbour signal benchmark points in Table.~\ref{tab:ILC_ymm_cutflow} due to its closeness to the $Z$ resonance. This translates into a weakening in the projected upper limits in the vicinity of $m_{Z}$ as seen in Fig.~\ref{fig:muC_ymm_projections}. Above $m_{Z}$, we observe that the signal efficiency continues to grow while the background efficiency keeps falling gently with increasing $m_{H}$. The signal production cross-section also continues to rise. All these factors lead to a gradual strengthening in the projected upper limits on $Y_{\mu\mu}$ with increasing $m_{H}$. At $m_{H} = 300~$GeV, we observe that MuC would be able to probe $Y_{\mu\mu}$ up to $\sim 0.06$ at $2\sigma$.

\section{Results and Discussions\label{sec:Fit}} 
In this section, we present numerical analysis for the model parameter space and reconcile electron and muon $g-2$ within their $1\sigma$ measured values while being consistent with all the low-energy, LHC, and LEP constraints discussed in the previous sections. After exhausting all the possibilities, we find two minimum textures discussed in  Eq.~\eqref{eq:textures} of Sec.~\ref{sec:flavor} to incorporate both of these anomalies and have a consistent neutrino oscillation fit ($\Delta m_{21}^2, \Delta m_{31}^2, \sin^2 \theta_{13}, \sin^2 \theta_{23}, \sin^2 \theta_{12}$). The neutrino mass matrix given in Eq.~\eqref{eq:numass} is diagonalized by a unitary transformation\
\begin{equation}
    U^T M_\nu U = \widehat{M}_\nu \, , 
\end{equation}
where $\widehat{M}_\nu$ is the diagonal mass matrix, and $U$ is the $3\times 3$ PMNS lepton mixing matrix. We diagonalize the mass matrix numerically by scanning over the input parameters while being consistent with $\Delta a_\ell$ and LFV constraints. For the ease of satisfying the flavor constraints on the Yukawa coupling $f$, we factor out $f_{\mu \tau}$ into the overall factor and define $a_0 = \kappa f_{\mu \tau}$, where $\kappa$ is the one-loop factor given in Eq.~\eqref{eq:loopfactor}. Moreover, we perform a constrained minimization where the observables are confined to $3\sigma$ of their experimental measured values. The fits to the two textures discussed in Eq.~\eqref{eq:textures} are shown in the subsequent sections. It is beyond the scope of this work to explore the entire parameter space of the theory; instead, we find benchmark points to show that the model is consistent with neutrino oscillation data while explaining the anomalies for both textures. As was eluded to earlier, we choose the masses of the scalar bosons such that the major contributions to the AMMs appear from the $\mathcal{CP}$-even Higgs $H$, taking all the other scalars heavier, i.e., fixing $\Delta m (\equiv m_A - m_H) = 200$ GeV and $\Delta m' (\equiv m_{H^+}- m_{H} ) = 200$ GeV.     

\subsection{Fit to {\tt TX-I}}\label{sec:caseIfit}
\begin{figure}[!t]
    \includegraphics[scale=0.48]{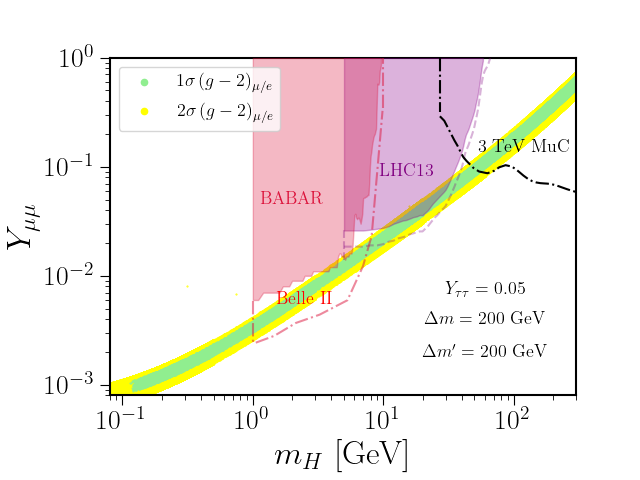}
    \includegraphics[scale=0.48]{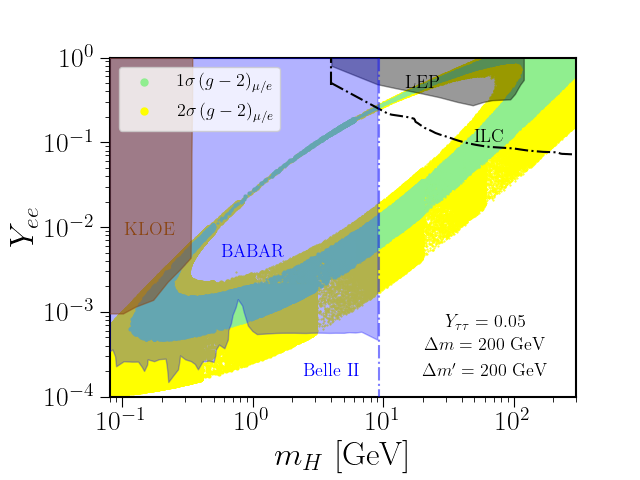}
    \caption{The parameter space of Yukawa couplings~$(Y_{\ell\ell})$ vs. scalar boson mass~($m_H$) satisfying both the AMMs. The green and yellow bands correspond to $1\sigma$ and $2\sigma$ regions allowed by $\Delta a_\ell$. The shaded regions in the plots indicate the excluded parameter space by different experiments: pink and purple regions are obtained from $e^+e^-\to \mu^+\mu^- H$ searches at BABAR~\cite{BaBar:2016sci} and LHC~\cite{CMS:2018yxg}, respectively; blue and brown regions from dark photon searches through $e^+e^-\to\gamma H$ at BABAR~\cite{BaBar:2016sci} and KLOE~\cite{Anastasi:2015qla}. Black shaded region is excluded from $e^+e^-\to e^+e^- H$ searches at LEP~\cite{OPAL:2003kcu,Electroweak:2003ram}. Blue and pink dash-dotted are the projected sensitivities from dark-photon searches at Belle-II \cite{Belle-II:2010dht, Belle-II:2018jsg} through $e^+ e^- \to \gamma H$ and $e^+ e^- \to \mu^+ \mu^- H$ searches, respectively. The black dash-dotted line on the left plot shows the projected sensitivity reach on $Y_{\mu\mu}$ from direct searches in the $\mu^{+}\mu^{-} \to \mu^{+}\mu^{-}\left(H\to \mu^{+}\mu^{-}\right)$ channel at the 3~TeV Muon Collider assuming $\mathcal{L}=1~{\rm ab^{-1}}$ (cf. Fig~\ref{fig:muC_ymm_projections}), and the one on the right plot shows the projected upper limits on $Y_{ee}$ from direct searches in the $e^{+}e^{-} \to e^{+}e^{-}\left(H\to e^{+}e^{-}\right)$ channel at the 1~TeV ILC assuming $\mathcal{L}=500~{\rm fb^{-1}}$ (cf. Fig.~\ref{fig:ILC_yee_proj}). The pink (purple) shaded region on the left plot assumes the BR$(H\to\mu\mu)=1 \, (0.5)$ with the purple dotted line  BR$(H\to\mu\mu)=1$. Here, $\Delta m=m_A-m_H$ and $\Delta m'=m_{H^+}-m_H$. }
    \label{fig:case1fig}
\end{figure}
The allowed parameter space of the flavor structure of {\tt TX-I} in Eq.~\eqref{eq:textures}  is explored here, as shown in Fig.~\ref{fig:case1fig}. The green and yellow bands in both plots correspond to $1\sigma$ and $2\sigma$ regions that can simultaneously explain both the AMMs. Here we fix $Y_{\tau\tau} = 0.05$ such that it provides the required sign for $\Delta a_e$ from the Barr-Zee diagram. Note, a smaller value of the $Y_{\tau\tau}$ requires a larger $Y_{ii} (i\neq\tau)$ to satisfy AMMs, excluding more parameter ranges. On the other hand, making $Y_{\tau\tau}$ larger does allow wider parameter space as the plot in Fig.~\ref{fig:case1fig} would shift downwards. However, it conflicts with the fit to neutrino oscillation data as there would be a large hierarchy in the elements of the neutrino mass matrix. 
The various shaded regions in Fig.~\ref{fig:case1fig} are excluded from various experimental constraints. The pink and purple shaded regions in Fig.~\ref{fig:case1fig} are excluded from $e^+e^-\to \mu^+\mu^- H$ searches at BaBar~\cite{BaBar:2016sci} and LHC~\cite{CMS:2018yxg}. Here we considered BR$(H\to\mu\mu)$=1 (dotted line) and 0.5 (purple shaded region) to show that more parameter space is allowed for $BR <1$. Blue and brown regions in the right plot are exclusion regions from the dark photon searches through $e^+e^-\to\gamma A_d$ channel at BABAR~\cite{BaBar:2016sci} and KLOE~\cite{Anastasi:2015qla}. The combination of these constraints on Yukawa couplings $Y_{ee}$ and $Y_{\mu \mu}$ would exclude light scalar mass below 10 GeV in the parameter space of our interest. The black region is obtained from LEP~\cite{Electroweak:2003ram} constraints through $e^+e^-\to e^+e^- (H \to e^+ e^-)$ searches, and the black dash-dotted line shows the projected sensitivity at 1 TeV ILC machine (3 TeV Muon Collider), as discussed in Sec.~\ref{sec:yee_ILC} (Sec.~\ref{sec:ILC_yemu}). As it can be seen from the figure that though LEP bound does not constrain the parameter space much, ILC would be able to probe a substantial parameter space. For instance, at $m_H =100$ GeV, the ILC would be sensitive to $Y_{ee} \gtrsim 0.085$ at $2\sigma$.  Our results for the fit to the {\rm TX-I} of Eq.~\eqref{eq:textures} is shown below: 
\paragraph{Fit ({\rm TX-I}):} With $a_0 = \kappa\ f_{\mu \tau} = 2.95 \times 10^{-7}  $ and $m_H = 85$ GeV, 
\begin{equation}
f  = f_{\mu \tau}  \begin{pmatrix}
0 & 2.14 \times 10^{-3} & 1.18 \times 10^{-4} \\
 & 0 & 1 \\
&  & 0 
\end{pmatrix} , \hspace{5mm}
y  = \begin{pmatrix}
0.31 & 0 & 0 \\
0 & 0.169 & -4.4 \times 10^{-4} \\
0 & 2.6\times 10^{-5} & 0.01
\end{pmatrix}\, . 
\label{eq:FitI}
\end{equation} 
For the Yukawa texture above, the corresponding fit for the neutrino oscillation parameters and $\Delta a_\ell$ are shown in Table~\ref{tab:fit} as Model {\tt Fit I}. Here the diagonal entries $Y_{ii}$ explain AMMs, while the rest of the Yukawa couplings are required to fit the neutrino oscillation data. It is important to point out that for the texture given in Eq.~\eqref{eq:FitI}, $Y_{\mu\mu} \simeq Y_{\tau \tau} m_\tau/m_\mu $ is required to get a NH solution while being consistent with the LFV from $\tau \to \mu \gamma$ and $\tau \to 3\mu$. Thus, for a larger value of $Y_{\tau \tau}$, say $0.05$, $Y_{\mu \mu} \simeq 0.8$ is required to get NH solution. For such a large Yukawa coupling, AMM can be satisfied by increasing the scalar mass to about 300 GeV. As noted in Sec.~\ref{sec:NSI}, order one coupling $Y_{e e}$ can also induce NSI; for the benchmark point $Y_{ee}=0.31, \varphi=0.1$, and $m_{H^+}=285\, \text{GeV} \, (\Delta m'= 200\, \text{GeV})$, we find $\varepsilon_{ee} = 4.2 \%$. This can be improved up to $8\%$ \cite{Babu:2019mfe} with a proper choice of parameters and is limited by direct experimental searches TEXONO \cite{TEXONO:2010tnr}.    

\begin{table}[!t]
	\centering
\begin{tabular}{|c|c|c|c|}
\hline \hline
\textbf{Oscillation} & \textbf{3\ $\sigma$ allowed range}  & \textbf{Model} & \textbf{Model}  \\
\textbf{parameters} &  \textbf{from NuFit5.1}~\cite{Esteban:2020cvm} & \textbf{{\tt Fit I}} & \textbf{{\tt Fit II}}  \\ \hline \hline
$\sin^2\theta_{12}$ & 0.269 -- 0.343 & 0.314 & 0.315  \\ \hline
$\sin^2\theta_{13}$ & 0.02032 -- 0.02410 & 0.02244 & 0.0219  \\ \hline
$\sin^2\theta_{23}$ & 0.415 -- 0.616 & 0.527 & 0.578 \\ \hline
$\Delta m_{21}^2$ $( 10^{-5} \ {\rm eV}^2)$ & 6.82 -- 8.04 & 7.42 & 7.35 \\ \hline
$\Delta m_{23}^2$ $( 10^{-3} \ {\rm eV}^2)$ & 2.435 -- 2.598 & 2.52 & 2.52  \\ \hline
\textbf{Observable} & \textbf{$1 \, \sigma$ allowed range} &  &    \\ \hline \hline
$\Delta a_e \, \, (10^{-14})$ & $-88 \pm 36$ & $-56$ & $-62$\\ \hline
$\Delta a_\mu \, \, (10^{-10})$  & $25 \pm 6$ & $29$ & $21$  \\ \hline \hline
\end{tabular}
\caption{Fits to the neutrino oscillation parameters in the model with normal hierarchy, along with $\Delta a_e$ and $\Delta a_\mu$ for the two benchmark fits given in Eq.~\eqref{eq:FitI} and Eq.~\eqref{eq:FitII}. For comparison, the $3\sigma$ allowed range for the oscillation parameters and $1\sigma$ for $\Delta a_\ell$ are also given.  }
 \label{tab:fit}
\end{table}

\subsection{Fit to {\tt TX-II}}
\begin{figure}[!t]
    \centering
    \includegraphics[scale=0.45]{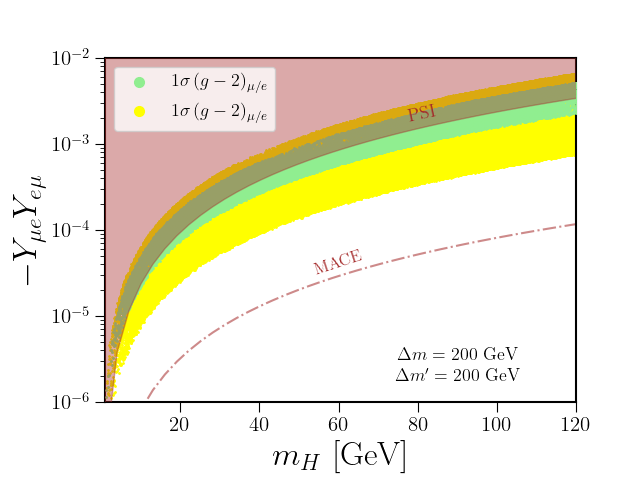}
    \includegraphics[scale=0.45]{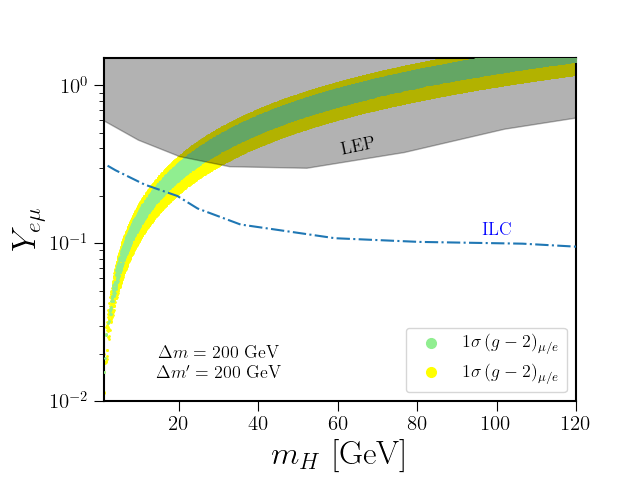}
    \caption{The parameter space of the Yukawa couplings $Y_{e\mu} Y_{\mu e}$ (left) and $Y_{e \mu}$ (right) as a function of scalar mass ($m_H$) satisfying both AMMs. The green and yellow band in both the figures respectively represents $1\sigma$ and $2\sigma$ allowed region. The region shaded in gray is excluded by the muonium oscillation PSI experiment \cite{Willmann:1998gd}, whereas the dash-dotted line represents the future sensitivity bound from the MACE experiment \cite{mace}. Black shaded region is excluded from $e^+e^-\to e^\pm \mu^\mp H$ searches at LEP~\cite{OPAL:2003kcu,Electroweak:2003ram}, whereas blue dash-dotted line is the projected sensitivity from ILC with $\sqrt{s} = 1$ TeV with integrated luminosity $\mathcal{L} = 500$ fb$^-1$.}
    \label{fig:caseII}
\end{figure}

The parameter space explored in {\tt TX-II} of Eq.~\eqref{eq:textures} is provided in this section. As aforementioned, off-diagonal couplings $Y_{e\mu}$ and $Y_{\mu e}$ need to take opposite signs to get a negative $\Delta a_e$ from the chirally-enhanced part of Eq.~\eqref{eq:a1N}. Furthermore, one of the two couplings needs to be large to provide the required correction to $\Delta a_\mu$ via the non-chiral part of Eq.~\eqref{eq:a1N}. The allowed parameter space as a function of Yukawa couplings and scalar mass is shown in Fig.~\ref{fig:caseII}, where the green and yellow bands correspond to $1\sigma$ and $2\sigma$ regions of $\Delta a_{\mu /e}$. The same couplings that give rise to $\Delta a_{e/\mu}$ also induce muonium oscillations, and the probability of these oscillations at the PSI experiment \cite{Willmann:1998gd} excludes a considerable region of the allowed parameter space, with a lower bound on the scalar mass of 8 (1) GeV at $1\sigma$ ($2\sigma$). These bounds are expected to improve at the MACE experiment \cite{mace} that can probe/exclude all the allowed parameter space. These couplings are directly accessible at lepton colliders via searches in the $e^+ e^- \to e^\pm \mu^\mp (H \to e^\pm \mu^\mp)$ channel and can be used to obtain the bound from LEP \cite{Electroweak:2003ram} as shown by the black shaded region in Fig.~\ref{fig:caseII}. The LEP bound imposes an upper limit of about 30 GeV on the scalar mass with the simultaneous explanation of $\Delta a_{e/\mu}$. Furthermore, the projected sensitivity of Yukawa coupling $Y_{e\mu}$ from ILC is shown in the dash-dotted line, which can probe $Y_{e\mu}$ up to $\sim 0.1$ at $m_H = 100$ GeV and explore the parameter space for AMMs as low as 20 GeV scalar mass. Our results for the fit to the {\tt TX-II} of Eq.~\eqref{eq:textures} is shown below:

\paragraph{Fit II ({\tt TX-II}):} With $a_0 = \kappa\ f_{\mu \tau} = 1.80 \times 10^{-6}$ and $m_H = 22$ GeV, 
\begin{equation}
f \ = f_{\mu \tau} \ \left(\begin{array}{ccc}
0~~ & 0.119 &~~ -0.198 \\
 & 0 &~~ 1.0 \\
 &  &~~ 0 
\end{array}\right), \hspace{10mm}
y \ = \ \left(\begin{array}{ccc}
0 & 0.3 & 0 \\
-3.5\times 10^{-4} & 0 & 5.56 \times 10^{-5} \\
1.18 \times 10^{-5} & 0 & 5.19 \times 10^{-6}
\end{array}\right) \, . 
\label{eq:FitII}
\end{equation} 
The corresponding fit for the neutrino oscillation and $\Delta a_\ell$ are shown in Table~\ref{tab:fit} as Model {\tt Fit II}. The fits are in excellent agreement with the observed experimental values. Here the non-zero Yukawa couplings are introduced to get the neutrino oscillation fit, whereas $0$'s in the texture are to satisfy LFV such as $\mu \to e \gamma$ and $\tau \to e \mu \mu$. It is worth pointing out that although it is ideal to choose $Y_{\mu e}\sim\mathcal{O}(1)$ instead of $Y_{e \mu}$ such that it can induce diagonal NSI, $\varepsilon_{\mu \mu}$, there is no solution to neutrino oscillation data while simultaneously explaining $\Delta a_\ell$ and satisfying flavor constraints.

\section{Conclusion\label{sec:conclusion}}

In conclusion, we have proposed a novel scenario in a radiative neutrino mass model, namely the Zee Model, that solves the intriguing deviations observed in muon and electron AMMs simultaneously within $1\sigma$ uncertainty, while being consistent with neutrino oscillation data, as well as all the relevant flavor and collider constraints. The neutral scalar in the second Higgs doublet is responsible for explaining both of these anomalies via one-loop and two-loop contributions. The two-loop Barr-Zee diagram is essential for getting the negative correction to $a_e$. Furthermore, by exhausting all the possibilities, we find two minimum textures in the model that can incorporate both of these anomalies and have a consistent neutrino oscillation fit. We observe that the currently allowed parameter space accommodates the scalar mass in the range of roughly 10-300 GeV in {\tt TX-I} and 1-30 GeV in {\tt TX-II}.  In {\tt TX-I}, for $m_{H} > 200~$GeV~($m_{H} > 40~$GeV), direct searches in the $e^{+}e^{-} \to e^{+}e^{-}(H \to e^{+}e^{-})$ channel at the 1 TeV ILC would be able to probe all allowed parameter space points with $Y_{ee} \gtrsim 0.07$~($Y_{ee} > 0.1$) at $2\sigma$ uncertainty. Similarly, through searches in the $\mu^{+}\mu^{-} \to \mu^{+}\mu^{-}(H \to \mu^{+}\mu^{-})$ chanel, the 3 TeV MuC collider would be able to probe the entire region with $m_{H} \gtrsim 40~$GeV irrespective of $Y_{\mu\mu}$. In {\tt TX-II}, we observe that all allowed parameter space points with $m_{H} > 20~$GeV fall within the projected exclusion reach of 1 TeV ILC via searches in the $e^{+}e^{-} \to \mu^{+}e^{-}(H \to \mu^{+}e^{-})$ channel. The model also features charged scalars required for generating neutrino masses which induce diagonal NSI $\varepsilon_{ee}$ as large as $8 \%$. Furthermore, the model can also give a sizable EDM of muon that can potentially be measured in future experiments. This task of connecting both the AMMs and finding textures to get neutrino oscillation fit is a highly non-trivial. As we have shown in detail, the allowed parameter space in the model can be probed in future experiments, which will either discover NP or rule out the possibility of explaining both AMMs within the context of the Zee Model. 

\section*{Acknowledgement}
We would like to thank K.S. Babu, Ahmed Ismail and Julian Heeck for useful discussion. RKB and RD thank the U.S.~Department of Energy for the financial support, under grant number DE-SC 0016013. Some computing for this project was performed at the High Performance Computing Center at Oklahoma State University, supported in part through the National Science Foundation grant OAC-1531128.

\appendix
\section{AMM expressions}
\subsection{One-loop\label{app_1loop}}
In the limit $m_i \ll m_\phi$ the expression for one-loop contribution to AMM \cite{Lindner:2016bgg} from neutral scalars simplifies to
\begin{equation}
    \Delta a_\ell^{(1)}(\phi)=\frac{1}{16\pi^2}\frac{m_\ell^2}{m_{\phi}^2}\Bigg(\frac{|Y_{\ell i}|^2+|Y_{i\ell}|^2}{6}\mp 2\frac{m_i}{m_\ell}\left(\frac{3}{4}+\log{\left(\frac{m_i}{m_\phi}\right)}\right)\Re(Y_{\ell i}Y_{i \ell})\Bigg),
    \label{a1N}
\end{equation} 
In the expression above, $-$ and $+$ correspond to $H$ and $A$, respectively. Similarly, the charged scalar contribution is simply
\begin{equation}
    \Delta a_\ell^{(1)}(H^+)=\frac{-1}{96\pi^2}\frac{m_\ell^2}{ m^2_{H^+}}|Y_{i\ell}|^2.
\end{equation}

\subsection{Two-loop\label{app_2loop}}

The loop functions are 
\begin{align}
    G(z,1)&=\begin{cases}
    2 \Big[ (2+\log{z})+\frac{1-2z}{\sqrt{1-4z}}\big\{Li_2(\frac{x_-}{x_+})-Li_2\left(\frac{x_+}{x_-}\right)\big\} \Big] \quad  \text{ for } z< \frac{1}{4}, \\[5pt]
    4  \quad \text{ for } z=\frac{1}{4},             \\[5pt]
  2(2+\log{z}) - \frac{2i(2z-1)}{\sqrt{4z-1}} \Big[ \log{z}\log{\frac{y_-}{y_+}} - Li_2 \left( \frac{2i}{y_+}\right) + Li_2 \left( \frac{-2i}{y_-}\right) \Big] \,   \text{ for } z> \frac{1}{4} 
    \end{cases} \\[15pt] 
%
    G(z,0)&=\begin{cases}
    \dfrac{2}{\sqrt{1-4z}}\Big[Li_2\left( \frac{x_-}{x_+} \right)-Li_2\left( \frac{x_+}{x_-} \right)\Big] \quad \text{ for } z<\frac{1}{4}, \\[10pt]
    2\ \log{16} \quad \text{ for } z=\frac{1}{4} \, ,   \\[5pt]
    \dfrac{2i}{\sqrt{4z-1}}\Big[\log{z}\log{(\frac{y_-}{y_+})}+Li_2\left(\frac{-2i}{y_-}\right)-Li_2\left(\frac{2i}{y_+}\right)\Big] \quad \text{ for } z> \frac{1}{4} \, ,
    \end{cases}
\end{align}
where, $x_{\pm}=\sqrt{1-4z} \pm 1$, $y_{\pm}= \sqrt{4z-1} \pm i$ and $z=m_i^2/m_\phi^2$.
\end{sloppypar}

\bibliographystyle{utphys}
\bibliography{references}

\providecommand{\href}[2]{#2}\begingroup\raggedright\begin{thebibliography}{100}

\bibitem{ParticleDataGroup:2020ssz}
{\bfseries Particle Data Group} Collaboration, P.~A. Zyla {\em et~al.},
  ``{Review of Particle Physics},''
  \href{http://dx.doi.org/10.1093/ptep/ptaa104}{{\em PTEP} {\bfseries 2020}
  no.~8, (2020) 083C01}.

\bibitem{Weinberg:1979sa}
S.~Weinberg, ``{Baryon and Lepton Nonconserving Processes},''
  \href{http://dx.doi.org/10.1103/PhysRevLett.43.1566}{{\em Phys. Rev. Lett.}
  {\bfseries 43} (1979) 1566--1570}.

\bibitem{Minkowski:1977sc}
P.~Minkowski, ``{$\mu \to e\gamma$ at a Rate of One Out of $10^{9}$ Muon
  Decays?},'' \href{http://dx.doi.org/10.1016/0370-2693(77)90435-X}{{\em Phys.
  Lett. B} {\bfseries 67} (1977) 421--428}.

\bibitem{Mohapatra:1979ia}
R.~N. Mohapatra and G.~Senjanovic, ``{Neutrino Mass and Spontaneous Parity
  Nonconservation},'' \href{http://dx.doi.org/10.1103/PhysRevLett.44.912}{{\em
  Phys. Rev. Lett.} {\bfseries 44} (1980) 912}.

\bibitem{Yanagida:1980xy}
T.~Yanagida, ``{Horizontal Symmetry and Masses of Neutrinos},''
  \href{http://dx.doi.org/10.1143/PTP.64.1103}{{\em Prog. Theor. Phys.}
  {\bfseries 64} (1980) 1103}.

\bibitem{Gell-Mann:1979vob}
M.~Gell-Mann, P.~Ramond, and R.~Slansky, ``{Complex Spinors and Unified
  Theories},'' {\em Conf. Proc. C} {\bfseries 790927} (1979) 315--321,
  \href{http://arxiv.org/abs/1306.4669}{{\ttfamily arXiv:1306.4669 [hep-th]}}.

\bibitem{Glashow:1979nm}
S.~L. Glashow, ``{The Future of Elementary Particle Physics},''
  \href{http://dx.doi.org/10.1007/978-1-4684-7197-7_15}{{\em NATO Sci. Ser. B}
  {\bfseries 61} (1980) 687}.

\bibitem{Schechter:1980gr}
J.~Schechter and J.~W.~F. Valle, ``{Neutrino Masses in SU(2) x U(1)
  Theories},'' \href{http://dx.doi.org/10.1103/PhysRevD.22.2227}{{\em Phys.
  Rev. D} {\bfseries 22} (1980) 2227}.

\bibitem{Cheng:1980qt}
T.~P. Cheng and L.-F. Li, ``{Neutrino Masses, Mixings and Oscillations in SU(2)
  x U(1) Models of Electroweak Interactions},''
  \href{http://dx.doi.org/10.1103/PhysRevD.22.2860}{{\em Phys. Rev. D}
  {\bfseries 22} (1980) 2860}.

\bibitem{Mohapatra:1980yp}
R.~N. Mohapatra and G.~Senjanovic, ``{Neutrino Masses and Mixings in Gauge
  Models with Spontaneous Parity Violation},''
  \href{http://dx.doi.org/10.1103/PhysRevD.23.165}{{\em Phys. Rev. D}
  {\bfseries 23} (1981) 165}.

\bibitem{Lazarides:1980nt}
G.~Lazarides, Q.~Shafi, and C.~Wetterich, ``{Proton Lifetime and Fermion Masses
  in an SO(10) Model},''
  \href{http://dx.doi.org/10.1016/0550-3213(81)90354-0}{{\em Nucl. Phys. B}
  {\bfseries 181} (1981) 287--300}.

\bibitem{Foot:1988aq}
R.~Foot, H.~Lew, X.~G. He, and G.~C. Joshi, ``{Seesaw Neutrino Masses Induced
  by a Triplet of Leptons},'' \href{http://dx.doi.org/10.1007/BF01415558}{{\em
  Z. Phys. C} {\bfseries 44} (1989) 441}.

\bibitem{Zee:1980ai}
A.~Zee, ``{A Theory of Lepton Number Violation, Neutrino Majorana Mass, and
  Oscillation},'' \href{http://dx.doi.org/10.1016/0370-2693(80)90349-4}{{\em
  Phys. Lett. B} {\bfseries 93} (1980) 389}. [Erratum: Phys.Lett.B 95, 461
  (1980)].

\bibitem{Zee:1985id}
A.~Zee, ``{Quantum Numbers of Majorana Neutrino Masses},''
  \href{http://dx.doi.org/10.1016/0550-3213(86)90475-X}{{\em Nucl. Phys. B}
  {\bfseries 264} (1986) 99--110}.

\bibitem{Babu:1988wk}
K.~S. Babu, E.~Ma, and J.~T. Pantaleone, ``{Model of Radiative Neutrino Masses:
  Mixing and a Possible Fourth Generation},''
  \href{http://dx.doi.org/10.1016/0370-2693(89)91425-1}{{\em Phys. Lett. B}
  {\bfseries 218} (1989) 233--237}.

\bibitem{Cai:2017jrq}
Y.~Cai, J.~Herrero-Garc\'\i{}a, M.~A. Schmidt, A.~Vicente, and R.~R. Volkas,
  ``{From the trees to the forest: a review of radiative neutrino mass
  models},'' \href{http://dx.doi.org/10.3389/fphy.2017.00063}{{\em Front. in
  Phys.} {\bfseries 5} (2017) 63},
  \href{http://arxiv.org/abs/1706.08524}{{\ttfamily arXiv:1706.08524
  [hep-ph]}}.

\bibitem{Babu:2019mfe}
K.~S. Babu, P.~S.~B. Dev, S.~Jana, and A.~Thapa, ``{Non-Standard Interactions
  in Radiative Neutrino Mass Models},''
  \href{http://dx.doi.org/10.1007/JHEP03(2020)006}{{\em JHEP} {\bfseries 03}
  (2020) 006}, \href{http://arxiv.org/abs/1907.09498}{{\ttfamily
  arXiv:1907.09498 [hep-ph]}}.

\bibitem{Schwinger:1948iu}
J.~S. Schwinger, ``{On Quantum electrodynamics and the magnetic moment of the
  electron},'' \href{http://dx.doi.org/10.1103/PhysRev.73.416}{{\em Phys. Rev.}
  {\bfseries 73} (1948) 416--417}.

\bibitem{Kusch:1948mvb}
P.~Kusch and H.~M. Foley, ``{The Magnetic Moment of the Electron},''
  \href{http://dx.doi.org/10.1103/PhysRev.74.250}{{\em Phys. Rev.} {\bfseries
  74} no.~3, (1948) 250}.

\bibitem{Sommerfield:1957zz}
C.~M. Sommerfield, ``{Magnetic Dipole Moment of the Electron},''
  \href{http://dx.doi.org/10.1103/PhysRev.107.328}{{\em Phys. Rev.} {\bfseries
  107} (1957) 328--329}.

\bibitem{Petermann:1957hs}
A.~Petermann, ``{Fourth order magnetic moment of the electron},''
  \href{http://dx.doi.org/10.5169/seals-112823}{{\em Helv. Phys. Acta}
  {\bfseries 30} (1957) 407--408}.

\bibitem{Kinoshita:1981vs}
T.~Kinoshita and W.~B. Lindquist, ``{Eighth Order Anomalous Magnetic Moment of
  the electron},'' \href{http://dx.doi.org/10.1103/PhysRevLett.47.1573}{{\em
  Phys. Rev. Lett.} {\bfseries 47} (1981) 1573}.

\bibitem{Kinoshita:1990wp}
T.~Kinoshita, B.~Nizic, and Y.~Okamoto, ``{Eighth order QED contribution to the
  anomalous magnetic moment of the muon},''
  \href{http://dx.doi.org/10.1103/PhysRevD.41.593}{{\em Phys. Rev. D}
  {\bfseries 41} (1990) 593--610}.

\bibitem{Laporta:1996mq}
S.~Laporta and E.~Remiddi, ``{The Analytical value of the electron (g-2) at
  order alpha**3 in QED},''
  \href{http://dx.doi.org/10.1016/0370-2693(96)00439-X}{{\em Phys. Lett. B}
  {\bfseries 379} (1996) 283--291},
  \href{http://arxiv.org/abs/hep-ph/9602417}{{\ttfamily arXiv:hep-ph/9602417}}.

\bibitem{Degrassi:1998es}
G.~Degrassi and G.~F. Giudice, ``{QED logarithms in the electroweak corrections
  to the muon anomalous magnetic moment},''
  \href{http://dx.doi.org/10.1103/PhysRevD.58.053007}{{\em Phys. Rev. D}
  {\bfseries 58} (1998) 053007},
  \href{http://arxiv.org/abs/hep-ph/9803384}{{\ttfamily arXiv:hep-ph/9803384}}.

\bibitem{Czarnecki:1998nd}
A.~Czarnecki and W.~J. Marciano, ``{Lepton anomalous magnetic moments: A Theory
  update},'' \href{http://dx.doi.org/10.1016/S0920-5632(99)00474-0}{{\em Nucl.
  Phys. B Proc. Suppl.} {\bfseries 76} (1999) 245--252},
  \href{http://arxiv.org/abs/hep-ph/9810512}{{\ttfamily arXiv:hep-ph/9810512}}.

\bibitem{Kinoshita:2004wi}
T.~Kinoshita and M.~Nio, ``{Improved alpha**4 term of the muon anomalous
  magnetic moment},'' \href{http://dx.doi.org/10.1103/PhysRevD.70.113001}{{\em
  Phys. Rev. D} {\bfseries 70} (2004) 113001},
  \href{http://arxiv.org/abs/hep-ph/0402206}{{\ttfamily arXiv:hep-ph/0402206}}.

\bibitem{Kinoshita:2005sm}
T.~Kinoshita and M.~Nio, ``{The Tenth-order QED contribution to the lepton g-2:
  Evaluation of dominant alpha**5 terms of muon g-2},''
  \href{http://dx.doi.org/10.1103/PhysRevD.73.053007}{{\em Phys. Rev. D}
  {\bfseries 73} (2006) 053007},
  \href{http://arxiv.org/abs/hep-ph/0512330}{{\ttfamily arXiv:hep-ph/0512330}}.

\bibitem{Passera:2006gc}
M.~Passera, ``{Precise mass-dependent QED contributions to leptonic g-2 at
  order alpha**2 and alpha**3},''
  \href{http://dx.doi.org/10.1103/PhysRevD.75.013002}{{\em Phys. Rev. D}
  {\bfseries 75} (2007) 013002},
  \href{http://arxiv.org/abs/hep-ph/0606174}{{\ttfamily arXiv:hep-ph/0606174}}.

\bibitem{Kataev:2006yh}
A.~L. Kataev, ``{Reconsidered estimates of the 10th order QED contributions to
  the muon anomaly},'' \href{http://dx.doi.org/10.1103/PhysRevD.74.073011}{{\em
  Phys. Rev. D} {\bfseries 74} (2006) 073011},
  \href{http://arxiv.org/abs/hep-ph/0608120}{{\ttfamily arXiv:hep-ph/0608120}}.

\bibitem{Aoyama:2007mn}
T.~Aoyama, M.~Hayakawa, T.~Kinoshita, and M.~Nio, ``{Revised value of the
  eighth-order QED contribution to the anomalous magnetic moment of the
  electron},'' \href{http://dx.doi.org/10.1103/PhysRevD.77.053012}{{\em Phys.
  Rev. D} {\bfseries 77} (2008) 053012},
  \href{http://arxiv.org/abs/0712.2607}{{\ttfamily arXiv:0712.2607 [hep-ph]}}.

\bibitem{Karshenboim:2008zz}
S.~G. Karshenboim, ``{New recommended values of the fundamental physical
  constants (CODATA 2006)},''
  \href{http://dx.doi.org/10.3367/UFNr.0178.200810c.1057}{{\em Phys. Usp.}
  {\bfseries 51} (2008) 1019--1026}.

\bibitem{Aoyama:2012wk}
T.~Aoyama, M.~Hayakawa, T.~Kinoshita, and M.~Nio, ``{Complete Tenth-Order QED
  Contribution to the Muon g-2},''
  \href{http://dx.doi.org/10.1103/PhysRevLett.109.111808}{{\em Phys. Rev.
  Lett.} {\bfseries 109} (2012) 111808},
  \href{http://arxiv.org/abs/1205.5370}{{\ttfamily arXiv:1205.5370 [hep-ph]}}.

\bibitem{Schnetz:2017bko}
O.~Schnetz, ``{The Galois coaction on the electron anomalous magnetic
  moment},'' \href{http://dx.doi.org/10.4310/CNTP.2018.v12.n2.a4}{{\em Commun.
  Num. Theor. Phys.} {\bfseries 12} (2018) 335--354},
  \href{http://arxiv.org/abs/1711.05118}{{\ttfamily arXiv:1711.05118
  [math-ph]}}.

\bibitem{Aoyama:2017uqe}
T.~Aoyama, T.~Kinoshita, and M.~Nio, ``{Revised and Improved Value of the QED
  Tenth-Order Electron Anomalous Magnetic Moment},''
  \href{http://dx.doi.org/10.1103/PhysRevD.97.036001}{{\em Phys. Rev. D}
  {\bfseries 97} no.~3, (2018) 036001},
  \href{http://arxiv.org/abs/1712.06060}{{\ttfamily arXiv:1712.06060
  [hep-ph]}}.

\bibitem{Volkov:2017xaq}
S.~Volkov, ``{New method of computing the contributions of graphs without
  lepton loops to the electron anomalous magnetic moment in QED},''
  \href{http://dx.doi.org/10.1103/PhysRevD.96.096018}{{\em Phys. Rev. D}
  {\bfseries 96} no.~9, (2017) 096018},
  \href{http://arxiv.org/abs/1705.05800}{{\ttfamily arXiv:1705.05800
  [hep-ph]}}.

\bibitem{Volkov:2018jhy}
S.~Volkov, ``{Numerical calculation of high-order QED contributions to the
  electron anomalous magnetic moment},''
  \href{http://dx.doi.org/10.1103/PhysRevD.98.076018}{{\em Phys. Rev. D}
  {\bfseries 98} no.~7, (2018) 076018},
  \href{http://arxiv.org/abs/1807.05281}{{\ttfamily arXiv:1807.05281
  [hep-ph]}}.

\bibitem{Czarnecki:1995wq}
A.~Czarnecki, B.~Krause, and W.~J. Marciano, ``{Electroweak Fermion loop
  contributions to the muon anomalous magnetic moment},''
  \href{http://dx.doi.org/10.1103/PhysRevD.52.R2619}{{\em Phys. Rev. D}
  {\bfseries 52} (1995) R2619--R2623},
  \href{http://arxiv.org/abs/hep-ph/9506256}{{\ttfamily arXiv:hep-ph/9506256}}.

\bibitem{Czarnecki:1995sz}
A.~Czarnecki, B.~Krause, and W.~J. Marciano, ``{Electroweak corrections to the
  muon anomalous magnetic moment},''
  \href{http://dx.doi.org/10.1103/PhysRevLett.76.3267}{{\em Phys. Rev. Lett.}
  {\bfseries 76} (1996) 3267--3270},
  \href{http://arxiv.org/abs/hep-ph/9512369}{{\ttfamily arXiv:hep-ph/9512369}}.

\bibitem{Czarnecki:1996if}
A.~Czarnecki and B.~Krause, ``{Electroweak corrections to the muon anomalous
  magnetic moment},''
  \href{http://dx.doi.org/10.1016/S0920-5632(96)90019-5}{{\em Nucl. Phys. B
  Proc. Suppl.} {\bfseries 51} (1996) 148--153},
  \href{http://arxiv.org/abs/hep-ph/9606393}{{\ttfamily arXiv:hep-ph/9606393}}.

\bibitem{Czarnecki:2002nt}
A.~Czarnecki, W.~J. Marciano, and A.~Vainshtein, ``{Refinements in electroweak
  contributions to the muon anomalous magnetic moment},''
  \href{http://dx.doi.org/10.1103/PhysRevD.67.073006}{{\em Phys. Rev. D}
  {\bfseries 67} (2003) 073006},
  \href{http://arxiv.org/abs/hep-ph/0212229}{{\ttfamily arXiv:hep-ph/0212229}}.
  [Erratum: Phys.Rev.D 73, 119901 (2006)].

\bibitem{Heinemeyer:2004yq}
S.~Heinemeyer, D.~Stockinger, and G.~Weiglein, ``{Electroweak and
  supersymmetric two-loop corrections to (g-2)(mu)},''
  \href{http://dx.doi.org/10.1016/j.nuclphysb.2004.08.014}{{\em Nucl. Phys. B}
  {\bfseries 699} (2004) 103--123},
  \href{http://arxiv.org/abs/hep-ph/0405255}{{\ttfamily arXiv:hep-ph/0405255}}.

\bibitem{Gribouk:2005ee}
T.~Gribouk and A.~Czarnecki, ``{Electroweak interactions and the muon g-2:
  Bosonic two-loop effects},''
  \href{http://dx.doi.org/10.1103/PhysRevD.72.053016}{{\em Phys. Rev. D}
  {\bfseries 72} (2005) 053016},
  \href{http://arxiv.org/abs/hep-ph/0509205}{{\ttfamily arXiv:hep-ph/0509205}}.

\bibitem{Gnendiger:2013pva}
C.~Gnendiger, D.~St\"ockinger, and H.~St\"ockinger-Kim, ``{The electroweak
  contributions to $(g-2)_\mu$ after the Higgs boson mass measurement},''
  \href{http://dx.doi.org/10.1103/PhysRevD.88.053005}{{\em Phys. Rev. D}
  {\bfseries 88} (2013) 053005},
  \href{http://arxiv.org/abs/1306.5546}{{\ttfamily arXiv:1306.5546 [hep-ph]}}.

\bibitem{Jegerlehner:1985gq}
F.~Jegerlehner, ``{Hadronic Contributions to Electroweak Parameter Shifts: A
  Detailed Analysis},'' \href{http://dx.doi.org/10.1007/BF01552495}{{\em Z.
  Phys. C} {\bfseries 32} (1986) 195}.

\bibitem{Lynn:1985sq}
B.~W. Lynn, G.~Penso, and C.~Verzegnassi, ``{STRONG INTERACTION CONTRIBUTIONS
  TO ONE LOOP LEPTONIC PROCESS},''
  \href{http://dx.doi.org/10.1103/PhysRevD.35.42}{{\em Phys. Rev. D} {\bfseries
  35} (1987) 42}.

\bibitem{Swartz:1995hc}
M.~L. Swartz, ``{Reevaluation of the hadronic contribution to alpha
  (M(Z)**2)},'' \href{http://dx.doi.org/10.1103/PhysRevD.53.5268}{{\em Phys.
  Rev. D} {\bfseries 53} (1996) 5268--5282},
  \href{http://arxiv.org/abs/hep-ph/9509248}{{\ttfamily arXiv:hep-ph/9509248}}.

\bibitem{Martin:1994we}
A.~D. Martin and D.~Zeppenfeld, ``{A Determination of the QED coupling at the Z
  pole},'' \href{http://dx.doi.org/10.1016/0370-2693(94)01659-Z}{{\em Phys.
  Lett. B} {\bfseries 345} (1995) 558--563},
  \href{http://arxiv.org/abs/hep-ph/9411377}{{\ttfamily arXiv:hep-ph/9411377}}.

\bibitem{Eidelman:1998vc}
S.~Eidelman, F.~Jegerlehner, A.~L. Kataev, and O.~Veretin, ``{Testing
  nonperturbative strong interaction effects via the Adler function},''
  \href{http://dx.doi.org/10.1016/S0370-2693(99)00389-5}{{\em Phys. Lett. B}
  {\bfseries 454} (1999) 369--380},
  \href{http://arxiv.org/abs/hep-ph/9812521}{{\ttfamily arXiv:hep-ph/9812521}}.

\bibitem{Krause:1996rf}
B.~Krause, ``{Higher order hadronic contributions to the anomalous magnetic
  moment of leptons},''
  \href{http://dx.doi.org/10.1016/S0370-2693(96)01346-9}{{\em Phys. Lett. B}
  {\bfseries 390} (1997) 392--400},
  \href{http://arxiv.org/abs/hep-ph/9607259}{{\ttfamily arXiv:hep-ph/9607259}}.

\bibitem{Davier:1998si}
M.~Davier and A.~Hocker, ``{New results on the hadronic contributions to
  alpha(M(Z)**2) and to (g-2)(mu)},''
  \href{http://dx.doi.org/10.1016/S0370-2693(98)00825-9}{{\em Phys. Lett. B}
  {\bfseries 435} (1998) 427--440},
  \href{http://arxiv.org/abs/hep-ph/9805470}{{\ttfamily arXiv:hep-ph/9805470}}.

\bibitem{Jegerlehner:2003qp}
F.~Jegerlehner, ``{Theoretical precision in estimates of the hadronic
  contributions to (g-2)(mu) and alpha(QED(M(Z))},''
  \href{http://dx.doi.org/10.1016/S0920-5632(03)02352-1}{{\em Nucl. Phys. B
  Proc. Suppl.} {\bfseries 126} (2004) 325--334},
  \href{http://arxiv.org/abs/hep-ph/0310234}{{\ttfamily arXiv:hep-ph/0310234}}.

\bibitem{deTroconiz:2004yzs}
J.~F. de~Troconiz and F.~J. Yndurain, ``{The Hadronic contributions to the
  anomalous magnetic moment of the muon},''
  \href{http://dx.doi.org/10.1103/PhysRevD.71.073008}{{\em Phys. Rev. D}
  {\bfseries 71} (2005) 073008},
  \href{http://arxiv.org/abs/hep-ph/0402285}{{\ttfamily arXiv:hep-ph/0402285}}.

\bibitem{Davier:2007ua}
M.~Davier, ``{The Hadronic contribution to (g-2)(mu)},''
  \href{http://dx.doi.org/10.1016/j.nuclphysbps.2007.03.023}{{\em Nucl. Phys. B
  Proc. Suppl.} {\bfseries 169} (2007) 288--296},
  \href{http://arxiv.org/abs/hep-ph/0701163}{{\ttfamily arXiv:hep-ph/0701163}}.

\bibitem{Campanario:2019mjh}
F.~Campanario, H.~Czy\.z, J.~Gluza, T.~Jeli\'nski, G.~Rodrigo, S.~Tracz, and
  D.~Zhuridov, ``{Standard model radiative corrections in the pion form factor
  measurements do not explain the $a_\mu$ anomaly},''
  \href{http://dx.doi.org/10.1103/PhysRevD.100.076004}{{\em Phys. Rev. D}
  {\bfseries 100} no.~7, (2019) 076004},
  \href{http://arxiv.org/abs/1903.10197}{{\ttfamily arXiv:1903.10197
  [hep-ph]}}.

\bibitem{Davier:2017zfy}
M.~Davier, A.~Hoecker, B.~Malaescu, and Z.~Zhang, ``{Reevaluation of the
  hadronic vacuum polarisation contributions to the Standard Model predictions
  of the muon $g-2$ and ${\alpha (m_Z^2)}$ using newest hadronic cross-section
  data},'' \href{http://dx.doi.org/10.1140/epjc/s10052-017-5161-6}{{\em Eur.
  Phys. J. C} {\bfseries 77} no.~12, (2017) 827},
  \href{http://arxiv.org/abs/1706.09436}{{\ttfamily arXiv:1706.09436
  [hep-ph]}}.

\bibitem{Keshavarzi:2018mgv}
A.~Keshavarzi, D.~Nomura, and T.~Teubner, ``{Muon $g-2$ and $\alpha(M_Z^2)$: a
  new data-based analysis},''
  \href{http://dx.doi.org/10.1103/PhysRevD.97.114025}{{\em Phys. Rev. D}
  {\bfseries 97} no.~11, (2018) 114025},
  \href{http://arxiv.org/abs/1802.02995}{{\ttfamily arXiv:1802.02995
  [hep-ph]}}.

\bibitem{Colangelo:2018mtw}
G.~Colangelo, M.~Hoferichter, and P.~Stoffer, ``{Two-pion contribution to
  hadronic vacuum polarization},''
  \href{http://dx.doi.org/10.1007/JHEP02(2019)006}{{\em JHEP} {\bfseries 02}
  (2019) 006}, \href{http://arxiv.org/abs/1810.00007}{{\ttfamily
  arXiv:1810.00007 [hep-ph]}}.

\bibitem{Hoferichter:2019mqg}
M.~Hoferichter, B.-L. Hoid, and B.~Kubis, ``{Three-pion contribution to
  hadronic vacuum polarization},''
  \href{http://dx.doi.org/10.1007/JHEP08(2019)137}{{\em JHEP} {\bfseries 08}
  (2019) 137}, \href{http://arxiv.org/abs/1907.01556}{{\ttfamily
  arXiv:1907.01556 [hep-ph]}}.

\bibitem{Davier:2019can}
M.~Davier, A.~Hoecker, B.~Malaescu, and Z.~Zhang, ``{A new evaluation of the
  hadronic vacuum polarisation contributions to the muon anomalous magnetic
  moment and to $\mathbf{\boldsymbol\alpha(m_Z^2)}$},''
  \href{http://dx.doi.org/10.1140/epjc/s10052-020-7792-2}{{\em Eur. Phys. J. C}
  {\bfseries 80} no.~3, (2020) 241},
  \href{http://arxiv.org/abs/1908.00921}{{\ttfamily arXiv:1908.00921
  [hep-ph]}}. [Erratum: Eur.Phys.J.C 80, 410 (2020)].

\bibitem{Keshavarzi:2019abf}
A.~Keshavarzi, D.~Nomura, and T.~Teubner, ``{$g-2$ of charged leptons, $\alpha
  (M^2_Z)$ , and the hyperfine splitting of muonium},''
  \href{http://dx.doi.org/10.1103/PhysRevD.101.014029}{{\em Phys. Rev. D}
  {\bfseries 101} no.~1, (2020) 014029},
  \href{http://arxiv.org/abs/1911.00367}{{\ttfamily arXiv:1911.00367
  [hep-ph]}}.

\bibitem{Kurz:2014wya}
A.~Kurz, T.~Liu, P.~Marquard, and M.~Steinhauser, ``{Hadronic contribution to
  the muon anomalous magnetic moment to next-to-next-to-leading order},''
  \href{http://dx.doi.org/10.1016/j.physletb.2014.05.043}{{\em Phys. Lett. B}
  {\bfseries 734} (2014) 144--147},
  \href{http://arxiv.org/abs/1403.6400}{{\ttfamily arXiv:1403.6400 [hep-ph]}}.

\bibitem{Bijnens:1995xf}
J.~Bijnens, E.~Pallante, and J.~Prades, ``{Analysis of the hadronic light by
  light contributions to the muon g-2},''
  \href{http://dx.doi.org/10.1016/0550-3213(96)00288-X}{{\em Nucl. Phys. B}
  {\bfseries 474} (1996) 379--420},
  \href{http://arxiv.org/abs/hep-ph/9511388}{{\ttfamily arXiv:hep-ph/9511388}}.

\bibitem{Hayakawa:1997rq}
M.~Hayakawa and T.~Kinoshita, ``{Pseudoscalar pole terms in the hadronic light
  by light scattering contribution to muon g - 2},''
  \href{http://dx.doi.org/10.1103/PhysRevD.57.465}{{\em Phys. Rev. D}
  {\bfseries 57} (1998) 465--477},
  \href{http://arxiv.org/abs/hep-ph/9708227}{{\ttfamily arXiv:hep-ph/9708227}}.
  [Erratum: Phys.Rev.D 66, 019902 (2002)].

\bibitem{Knecht:2001qf}
M.~Knecht and A.~Nyffeler, ``{Hadronic light by light corrections to the muon
  g-2: The Pion pole contribution},''
  \href{http://dx.doi.org/10.1103/PhysRevD.65.073034}{{\em Phys. Rev. D}
  {\bfseries 65} (2002) 073034},
  \href{http://arxiv.org/abs/hep-ph/0111058}{{\ttfamily arXiv:hep-ph/0111058}}.

\bibitem{Knecht:2001qg}
M.~Knecht, A.~Nyffeler, M.~Perrottet, and E.~de~Rafael, ``{Hadronic light by
  light scattering contribution to the muon g-2: An Effective field theory
  approach},'' \href{http://dx.doi.org/10.1103/PhysRevLett.88.071802}{{\em
  Phys. Rev. Lett.} {\bfseries 88} (2002) 071802},
  \href{http://arxiv.org/abs/hep-ph/0111059}{{\ttfamily arXiv:hep-ph/0111059}}.

\bibitem{Ramsey-Musolf:2002gmi}
M.~J. Ramsey-Musolf and M.~B. Wise, ``{Hadronic light by light contribution to
  muon g-2 in chiral perturbation theory},''
  \href{http://dx.doi.org/10.1103/PhysRevLett.89.041601}{{\em Phys. Rev. Lett.}
  {\bfseries 89} (2002) 041601},
  \href{http://arxiv.org/abs/hep-ph/0201297}{{\ttfamily arXiv:hep-ph/0201297}}.

\bibitem{Melnikov:2003xd}
K.~Melnikov and A.~Vainshtein, ``{Hadronic light-by-light scattering
  contribution to the muon anomalous magnetic moment revisited},''
  \href{http://dx.doi.org/10.1103/PhysRevD.70.113006}{{\em Phys. Rev. D}
  {\bfseries 70} (2004) 113006},
  \href{http://arxiv.org/abs/hep-ph/0312226}{{\ttfamily arXiv:hep-ph/0312226}}.

\bibitem{Bijnens:2007pz}
J.~Bijnens and J.~Prades, ``{The Hadronic Light-by-Light Contribution to the
  Muon Anomalous Magnetic Moment: Where do we stand?},''
  \href{http://dx.doi.org/10.1142/S0217732307022992}{{\em Mod. Phys. Lett. A}
  {\bfseries 22} (2007) 767--782},
  \href{http://arxiv.org/abs/hep-ph/0702170}{{\ttfamily arXiv:hep-ph/0702170}}.

\bibitem{Prades:2009tw}
J.~Prades, E.~de~Rafael, and A.~Vainshtein, ``{The Hadronic Light-by-Light
  Scattering Contribution to the Muon and Electron Anomalous Magnetic
  Moments},'' \href{http://dx.doi.org/10.1142/9789814271844_0009}{{\em Adv.
  Ser. Direct. High Energy Phys.} {\bfseries 20} (2009) 303--317},
  \href{http://arxiv.org/abs/0901.0306}{{\ttfamily arXiv:0901.0306 [hep-ph]}}.

\bibitem{Kataev:2012kn}
A.~L. Kataev, ``{Analytical eighth-order light-by-light QED contributions from
  leptons with heavier masses to the anomalous magnetic moment of electron},''
  \href{http://dx.doi.org/10.1103/PhysRevD.86.013010}{{\em Phys. Rev. D}
  {\bfseries 86} (2012) 013010},
  \href{http://arxiv.org/abs/1205.6191}{{\ttfamily arXiv:1205.6191 [hep-ph]}}.

\bibitem{Masjuan:2017tvw}
P.~Masjuan and P.~Sanchez-Puertas, ``{Pseudoscalar-pole contribution to the
  $(g_{\mu}-2)$: a rational approach},''
  \href{http://dx.doi.org/10.1103/PhysRevD.95.054026}{{\em Phys. Rev. D}
  {\bfseries 95} no.~5, (2017) 054026},
  \href{http://arxiv.org/abs/1701.05829}{{\ttfamily arXiv:1701.05829
  [hep-ph]}}.

\bibitem{Colangelo:2017fiz}
G.~Colangelo, M.~Hoferichter, M.~Procura, and P.~Stoffer, ``{Dispersion
  relation for hadronic light-by-light scattering: two-pion contributions},''
  \href{http://dx.doi.org/10.1007/JHEP04(2017)161}{{\em JHEP} {\bfseries 04}
  (2017) 161}, \href{http://arxiv.org/abs/1702.07347}{{\ttfamily
  arXiv:1702.07347 [hep-ph]}}.

\bibitem{Hoferichter:2018kwz}
M.~Hoferichter, B.-L. Hoid, B.~Kubis, S.~Leupold, and S.~P. Schneider,
  ``{Dispersion relation for hadronic light-by-light scattering: pion pole},''
  \href{http://dx.doi.org/10.1007/JHEP10(2018)141}{{\em JHEP} {\bfseries 10}
  (2018) 141}, \href{http://arxiv.org/abs/1808.04823}{{\ttfamily
  arXiv:1808.04823 [hep-ph]}}.

\bibitem{Gerardin:2019vio}
A.~G\'erardin, H.~B. Meyer, and A.~Nyffeler, ``{Lattice calculation of the pion
  transition form factor with $N_f=2+1$ Wilson quarks},''
  \href{http://dx.doi.org/10.1103/PhysRevD.100.034520}{{\em Phys. Rev. D}
  {\bfseries 100} no.~3, (2019) 034520},
  \href{http://arxiv.org/abs/1903.09471}{{\ttfamily arXiv:1903.09471
  [hep-lat]}}.

\bibitem{Bijnens:2019ghy}
J.~Bijnens, N.~Hermansson-Truedsson, and A.~Rodr\'\i{}guez-S\'anchez,
  ``{Short-distance constraints for the HLbL contribution to the muon anomalous
  magnetic moment},''
  \href{http://dx.doi.org/10.1016/j.physletb.2019.134994}{{\em Phys. Lett. B}
  {\bfseries 798} (2019) 134994},
  \href{http://arxiv.org/abs/1908.03331}{{\ttfamily arXiv:1908.03331
  [hep-ph]}}.

\bibitem{Colangelo:2019uex}
G.~Colangelo, F.~Hagelstein, M.~Hoferichter, L.~Laub, and P.~Stoffer,
  ``{Longitudinal short-distance constraints for the hadronic light-by-light
  contribution to $(g-2)_\mu$ with large-$N_c$ Regge models},''
  \href{http://dx.doi.org/10.1007/JHEP03(2020)101}{{\em JHEP} {\bfseries 03}
  (2020) 101}, \href{http://arxiv.org/abs/1910.13432}{{\ttfamily
  arXiv:1910.13432 [hep-ph]}}.

\bibitem{Colangelo:2014qya}
G.~Colangelo, M.~Hoferichter, A.~Nyffeler, M.~Passera, and P.~Stoffer,
  ``{Remarks on higher-order hadronic corrections to the muon
  g\ensuremath{-}2},''
  \href{http://dx.doi.org/10.1016/j.physletb.2014.06.012}{{\em Phys. Lett. B}
  {\bfseries 735} (2014) 90--91},
  \href{http://arxiv.org/abs/1403.7512}{{\ttfamily arXiv:1403.7512 [hep-ph]}}.

\bibitem{Pauk:2014rta}
V.~Pauk and M.~Vanderhaeghen, ``{Single meson contributions to the muon`s
  anomalous magnetic moment},''
  \href{http://dx.doi.org/10.1140/epjc/s10052-014-3008-y}{{\em Eur. Phys. J. C}
  {\bfseries 74} no.~8, (2014) 3008},
  \href{http://arxiv.org/abs/1401.0832}{{\ttfamily arXiv:1401.0832 [hep-ph]}}.

\bibitem{Danilkin:2016hnh}
I.~Danilkin and M.~Vanderhaeghen, ``{Light-by-light scattering sum rules in
  light of new data},''
  \href{http://dx.doi.org/10.1103/PhysRevD.95.014019}{{\em Phys. Rev. D}
  {\bfseries 95} no.~1, (2017) 014019},
  \href{http://arxiv.org/abs/1611.04646}{{\ttfamily arXiv:1611.04646
  [hep-ph]}}.

\bibitem{Jegerlehner:2017gek}
F.~Jegerlehner, \href{http://dx.doi.org/10.1007/978-3-319-63577-4}{{\em {The
  Anomalous Magnetic Moment of the Muon}}}, vol.~274.
\newblock Springer, Cham, 2017.

\bibitem{Knecht:2018sci}
M.~Knecht, S.~Narison, A.~Rabemananjara, and D.~Rabetiarivony, ``{Scalar meson
  contributions to a \ensuremath{\mu} from hadronic light-by-light
  scattering},'' \href{http://dx.doi.org/10.1016/j.physletb.2018.10.048}{{\em
  Phys. Lett. B} {\bfseries 787} (2018) 111--123},
  \href{http://arxiv.org/abs/1808.03848}{{\ttfamily arXiv:1808.03848
  [hep-ph]}}.

\bibitem{Eichmann:2019bqf}
G.~Eichmann, C.~S. Fischer, and R.~Williams, ``{Kaon-box contribution to the
  anomalous magnetic moment of the muon},''
  \href{http://dx.doi.org/10.1103/PhysRevD.101.054015}{{\em Phys. Rev. D}
  {\bfseries 101} no.~5, (2020) 054015},
  \href{http://arxiv.org/abs/1910.06795}{{\ttfamily arXiv:1910.06795
  [hep-ph]}}.

\bibitem{Roig:2019reh}
P.~Roig and P.~Sanchez-Puertas, ``{Axial-vector exchange contribution to the
  hadronic light-by-light piece of the muon anomalous magnetic moment},''
  \href{http://dx.doi.org/10.1103/PhysRevD.101.074019}{{\em Phys. Rev. D}
  {\bfseries 101} no.~7, (2020) 074019},
  \href{http://arxiv.org/abs/1910.02881}{{\ttfamily arXiv:1910.02881
  [hep-ph]}}.

\bibitem{Blum:2019ugy}
T.~Blum, N.~Christ, M.~Hayakawa, T.~Izubuchi, L.~Jin, C.~Jung, and C.~Lehner,
  ``{Hadronic Light-by-Light Scattering Contribution to the Muon Anomalous
  Magnetic Moment from Lattice QCD},''
  \href{http://dx.doi.org/10.1103/PhysRevLett.124.132002}{{\em Phys. Rev.
  Lett.} {\bfseries 124} no.~13, (2020) 132002},
  \href{http://arxiv.org/abs/1911.08123}{{\ttfamily arXiv:1911.08123
  [hep-lat]}}.

\bibitem{Muong-2:2021ojo}
{\bfseries Muon g-2} Collaboration, B.~Abi {\em et~al.}, ``{Measurement of the
  Positive Muon Anomalous Magnetic Moment to 0.46 ppm},''
  \href{http://dx.doi.org/10.1103/PhysRevLett.126.141801}{{\em Phys. Rev.
  Lett.} {\bfseries 126} no.~14, (2021) 141801},
  \href{http://arxiv.org/abs/2104.03281}{{\ttfamily arXiv:2104.03281
  [hep-ex]}}.

\bibitem{Muong-2:2006rrc}
{\bfseries Muon g-2} Collaboration, G.~W. Bennett {\em et~al.}, ``{Final Report
  of the Muon E821 Anomalous Magnetic Moment Measurement at BNL},''
  \href{http://dx.doi.org/10.1103/PhysRevD.73.072003}{{\em Phys. Rev. D}
  {\bfseries 73} (2006) 072003},
  \href{http://arxiv.org/abs/hep-ex/0602035}{{\ttfamily arXiv:hep-ex/0602035}}.

\bibitem{Muong-2:2001kxu}
{\bfseries Muon g-2} Collaboration, H.~N. Brown {\em et~al.}, ``{Precise
  measurement of the positive muon anomalous magnetic moment},''
  \href{http://dx.doi.org/10.1103/PhysRevLett.86.2227}{{\em Phys. Rev. Lett.}
  {\bfseries 86} (2001) 2227--2231},
  \href{http://arxiv.org/abs/hep-ex/0102017}{{\ttfamily arXiv:hep-ex/0102017}}.

\bibitem{Aoyama:2020ynm}
T.~Aoyama {\em et~al.}, ``{The anomalous magnetic moment of the muon in the
  Standard Model},''
  \href{http://dx.doi.org/10.1016/j.physrep.2020.07.006}{{\em Phys. Rept.}
  {\bfseries 887} (2020) 1--166},
  \href{http://arxiv.org/abs/2006.04822}{{\ttfamily arXiv:2006.04822
  [hep-ph]}}.

\bibitem{Davier:2010nc}
M.~Davier, A.~Hoecker, B.~Malaescu, and Z.~Zhang, ``{Reevaluation of the
  Hadronic Contributions to the Muon g-2 and to alpha(MZ)},''
  \href{http://dx.doi.org/10.1140/epjc/s10052-012-1874-8}{{\em Eur. Phys. J. C}
  {\bfseries 71} (2011) 1515}, \href{http://arxiv.org/abs/1010.4180}{{\ttfamily
  arXiv:1010.4180 [hep-ph]}}. [Erratum: Eur.Phys.J.C 72, 1874 (2012)].

\bibitem{Gerardin:2020gpp}
A.~G\'erardin, ``{The anomalous magnetic moment of the muon: status of Lattice
  QCD calculations},''
  \href{http://dx.doi.org/10.1140/epja/s10050-021-00426-7}{{\em Eur. Phys. J.
  A} {\bfseries 57} no.~4, (2021) 116},
  \href{http://arxiv.org/abs/2012.03931}{{\ttfamily arXiv:2012.03931
  [hep-lat]}}.

\bibitem{Hanneke:2008tm}
D.~Hanneke, S.~Fogwell, and G.~Gabrielse, ``{New Measurement of the Electron
  Magnetic Moment and the Fine Structure Constant},''
  \href{http://dx.doi.org/10.1103/PhysRevLett.100.120801}{{\em Phys. Rev.
  Lett.} {\bfseries 100} (2008) 120801},
  \href{http://arxiv.org/abs/0801.1134}{{\ttfamily arXiv:0801.1134
  [physics.atom-ph]}}.

\bibitem{Parker:2018vye}
R.~H. Parker, C.~Yu, W.~Zhong, B.~Estey, and H.~M\"uller, ``{Measurement of the
  fine-structure constant as a test of the Standard Model},''
  \href{http://dx.doi.org/10.1126/science.aap7706}{{\em Science} {\bfseries
  360} (2018) 191}, \href{http://arxiv.org/abs/1812.04130}{{\ttfamily
  arXiv:1812.04130 [physics.atom-ph]}}.

\bibitem{Aoyama:2012wj}
T.~Aoyama, M.~Hayakawa, T.~Kinoshita, and M.~Nio, ``{Tenth-Order QED
  Contribution to the Electron g-2 and an Improved Value of the Fine Structure
  Constant},'' \href{http://dx.doi.org/10.1103/PhysRevLett.109.111807}{{\em
  Phys. Rev. Lett.} {\bfseries 109} (2012) 111807},
  \href{http://arxiv.org/abs/1205.5368}{{\ttfamily arXiv:1205.5368 [hep-ph]}}.

\bibitem{Laporta:2017okg}
S.~Laporta, ``{High-precision calculation of the 4-loop contribution to the
  electron g-2 in QED},''
  \href{http://dx.doi.org/10.1016/j.physletb.2017.06.056}{{\em Phys. Lett. B}
  {\bfseries 772} (2017) 232--238},
  \href{http://arxiv.org/abs/1704.06996}{{\ttfamily arXiv:1704.06996
  [hep-ph]}}.

\bibitem{Volkov:2019phy}
S.~Volkov, ``{Calculating the five-loop QED contribution to the electron
  anomalous magnetic moment: Graphs without lepton loops},''
  \href{http://dx.doi.org/10.1103/PhysRevD.100.096004}{{\em Phys. Rev. D}
  {\bfseries 100} no.~9, (2019) 096004},
  \href{http://arxiv.org/abs/1909.08015}{{\ttfamily arXiv:1909.08015
  [hep-ph]}}.

\bibitem{Morel:2020dww}
L.~Morel, Z.~Yao, P.~Clad\'e, and S.~Guellati-Kh\'elifa, ``{Determination of
  the fine-structure constant with an accuracy of 81 parts per trillion},''
  \href{http://dx.doi.org/10.1038/s41586-020-2964-7}{{\em Nature} {\bfseries
  588} no.~7836, (2020) 61--65}.

\bibitem{Crivellin:2018qmi}
A.~Crivellin, M.~Hoferichter, and P.~Schmidt-Wellenburg, ``{Combined
  explanations of $(g-2)_{\mu,e}$ and implications for a large muon EDM},''
  \href{http://dx.doi.org/10.1103/PhysRevD.98.113002}{{\em Phys. Rev. D}
  {\bfseries 98} no.~11, (2018) 113002},
  \href{http://arxiv.org/abs/1807.11484}{{\ttfamily arXiv:1807.11484
  [hep-ph]}}.

\bibitem{Hiller:2019mou}
G.~Hiller, C.~Hormigos-Feliu, D.~F. Litim, and T.~Steudtner, ``{Anomalous
  magnetic moments from asymptotic safety},''
  \href{http://dx.doi.org/10.1103/PhysRevD.102.071901}{{\em Phys. Rev. D}
  {\bfseries 102} no.~7, (2020) 071901},
  \href{http://arxiv.org/abs/1910.14062}{{\ttfamily arXiv:1910.14062
  [hep-ph]}}.

\bibitem{Chun:2020uzw}
E.~J. Chun and T.~Mondal, ``{Explaining $g-2$ anomalies in two Higgs doublet
  model with vector-like leptons},''
  \href{http://dx.doi.org/10.1007/JHEP11(2020)077}{{\em JHEP} {\bfseries 11}
  (2020) 077}, \href{http://arxiv.org/abs/2009.08314}{{\ttfamily
  arXiv:2009.08314 [hep-ph]}}.

\bibitem{Chen:2020tfr}
K.-F. Chen, C.-W. Chiang, and K.~Yagyu, ``{An explanation for the muon and
  electron $g-2$ anomalies and dark matter},''
  \href{http://dx.doi.org/10.1007/JHEP09(2020)119}{{\em JHEP} {\bfseries 09}
  (2020) 119}, \href{http://arxiv.org/abs/2006.07929}{{\ttfamily
  arXiv:2006.07929 [hep-ph]}}.

\bibitem{Hati:2020fzp}
C.~Hati, J.~Kriewald, J.~Orloff, and A.~M. Teixeira, ``{Anomalies in $^8$Be
  nuclear transitions and $(g-2)_{e,\mu}$: towards a minimal combined
  explanation},'' \href{http://dx.doi.org/10.1007/JHEP07(2020)235}{{\em JHEP}
  {\bfseries 07} (2020) 235}, \href{http://arxiv.org/abs/2005.00028}{{\ttfamily
  arXiv:2005.00028 [hep-ph]}}.

\bibitem{Escribano:2021css}
P.~Escribano, J.~Terol-Calvo, and A.~Vicente, ``{$\boldsymbol{(g-2)_{e,\mu}}$
  in an extended inverse type-III seesaw model},''
  \href{http://dx.doi.org/10.1103/PhysRevD.103.115018}{{\em Phys. Rev. D}
  {\bfseries 103} no.~11, (2021) 115018},
  \href{http://arxiv.org/abs/2104.03705}{{\ttfamily arXiv:2104.03705
  [hep-ph]}}.

\bibitem{Hernandez:2021tii}
A.~E.~C. Hern\'andez, S.~F. King, and H.~Lee, ``{Fermion mass hierarchies from
  vectorlike families with an extended 2HDM and a possible explanation for the
  electron and muon anomalous magnetic moments},''
  \href{http://dx.doi.org/10.1103/PhysRevD.103.115024}{{\em Phys. Rev. D}
  {\bfseries 103} no.~11, (2021) 115024},
  \href{http://arxiv.org/abs/2101.05819}{{\ttfamily arXiv:2101.05819
  [hep-ph]}}.

\bibitem{Borah:2021khc}
D.~Borah, M.~Dutta, S.~Mahapatra, and N.~Sahu, ``{Lepton Anomalous Magnetic
  Moment with Singlet-Doublet Fermion Dark Matter in Scotogenic
  $U(1)_{L_{\mu}-L_{\tau}}$ Model},''
  \href{http://arxiv.org/abs/2109.02699}{{\ttfamily arXiv:2109.02699
  [hep-ph]}}.

\bibitem{Bharadwaj:2021tgp}
H.~Bharadwaj, S.~Dutta, and A.~Goyal, ``{Leptonic g \ensuremath{-} 2 anomaly in
  an extended Higgs sector with vector-like leptons},''
  \href{http://dx.doi.org/10.1007/JHEP11(2021)056}{{\em JHEP} {\bfseries 11}
  (2021) 056}, \href{http://arxiv.org/abs/2109.02586}{{\ttfamily
  arXiv:2109.02586 [hep-ph]}}.

\bibitem{Davoudiasl:2018fbb}
H.~Davoudiasl and W.~J. Marciano, ``{Tale of two anomalies},''
  \href{http://dx.doi.org/10.1103/PhysRevD.98.075011}{{\em Phys. Rev. D}
  {\bfseries 98} no.~7, (2018) 075011},
  \href{http://arxiv.org/abs/1806.10252}{{\ttfamily arXiv:1806.10252
  [hep-ph]}}.

\bibitem{Liu:2018xkx}
J.~Liu, C.~E.~M. Wagner, and X.-P. Wang, ``{A light complex scalar for the
  electron and muon anomalous magnetic moments},''
  \href{http://dx.doi.org/10.1007/JHEP03(2019)008}{{\em JHEP} {\bfseries 03}
  (2019) 008}, \href{http://arxiv.org/abs/1810.11028}{{\ttfamily
  arXiv:1810.11028 [hep-ph]}}.

\bibitem{Han:2018znu}
X.-F. Han, T.~Li, L.~Wang, and Y.~Zhang, ``{Simple interpretations of lepton
  anomalies in the lepton-specific inert two-Higgs-doublet model},''
  \href{http://dx.doi.org/10.1103/PhysRevD.99.095034}{{\em Phys. Rev. D}
  {\bfseries 99} no.~9, (2019) 095034},
  \href{http://arxiv.org/abs/1812.02449}{{\ttfamily arXiv:1812.02449
  [hep-ph]}}.

\bibitem{Bauer:2019gfk}
M.~Bauer, M.~Neubert, S.~Renner, M.~Schnubel, and A.~Thamm, ``{Axionlike
  Particles, Lepton-Flavor Violation, and a New Explanation of $a_\mu$ and
  $a_e$},'' \href{http://dx.doi.org/10.1103/PhysRevLett.124.211803}{{\em Phys.
  Rev. Lett.} {\bfseries 124} no.~21, (2020) 211803},
  \href{http://arxiv.org/abs/1908.00008}{{\ttfamily arXiv:1908.00008
  [hep-ph]}}.

\bibitem{Cornella:2019uxs}
C.~Cornella, P.~Paradisi, and O.~Sumensari, ``{Hunting for ALPs with Lepton
  Flavor Violation},'' \href{http://dx.doi.org/10.1007/JHEP01(2020)158}{{\em
  JHEP} {\bfseries 01} (2020) 158},
  \href{http://arxiv.org/abs/1911.06279}{{\ttfamily arXiv:1911.06279
  [hep-ph]}}.

\bibitem{Dutta:2020scq}
B.~Dutta, S.~Ghosh, and T.~Li, ``{Explaining $(g-2)_{\mu,e}$, the KOTO anomaly
  and the MiniBooNE excess in an extended Higgs model with sterile
  neutrinos},'' \href{http://dx.doi.org/10.1103/PhysRevD.102.055017}{{\em Phys.
  Rev. D} {\bfseries 102} no.~5, (2020) 055017},
  \href{http://arxiv.org/abs/2006.01319}{{\ttfamily arXiv:2006.01319
  [hep-ph]}}.

\bibitem{Endo:2020mev}
M.~Endo, S.~Iguro, and T.~Kitahara, ``{Probing $e\mu$ flavor-violating ALP at
  Belle II},'' \href{http://dx.doi.org/10.1007/JHEP06(2020)040}{{\em JHEP}
  {\bfseries 06} (2020) 040}, \href{http://arxiv.org/abs/2002.05948}{{\ttfamily
  arXiv:2002.05948 [hep-ph]}}.

\bibitem{Haba:2020gkr}
N.~Haba, Y.~Shimizu, and T.~Yamada, ``{Muon and electron $g-2$ and the origin
  of the fermion mass hierarchy},''
  \href{http://dx.doi.org/10.1093/ptep/ptaa098}{{\em PTEP} {\bfseries 2020}
  no.~9, (2020) 093B05}, \href{http://arxiv.org/abs/2002.10230}{{\ttfamily
  arXiv:2002.10230 [hep-ph]}}.

\bibitem{Hernandez:2021xet}
A.~E.~C. Hern\'andez, D.~T. Huong, and I.~Schmidt, ``{Universal Inverse seesaw
  mechanism as a source of the SM fermion mass hierarchy},''
  \href{http://arxiv.org/abs/2109.12118}{{\ttfamily arXiv:2109.12118
  [hep-ph]}}.

\bibitem{Mondal:2021vou}
T.~Mondal and H.~Okada, ``{Inverse seesaw and $(g-2)$ anomalies in $B-L$
  extended two Higgs doublet model},''
  \href{http://arxiv.org/abs/2103.13149}{{\ttfamily arXiv:2103.13149
  [hep-ph]}}.

\bibitem{Adhikari:2021yvx}
R.~Adhikari, I.~A. Bhat, D.~Borah, E.~Ma, and D.~Nanda, ``{Anomalous Magnetic
  Moment and Higgs Coupling of the Muon in a Sequential U(1) Gauge Model with
  Dark Matter},'' \href{http://arxiv.org/abs/2109.05417}{{\ttfamily
  arXiv:2109.05417 [hep-ph]}}.

\bibitem{Bauer:2021mvw}
M.~Bauer, M.~Neubert, S.~Renner, M.~Schnubel, and A.~Thamm, ``{Flavor probes of
  axion-like particles},'' \href{http://arxiv.org/abs/2110.10698}{{\ttfamily
  arXiv:2110.10698 [hep-ph]}}.

\bibitem{De:2021crr}
B.~De, D.~Das, M.~Mitra, and N.~Sahoo, ``{Magnetic Moments of Leptons, Charged
  Lepton Flavor Violations and Dark Matter Phenomenology of a Minimal Radiative
  Dirac Neutrino Mass Model},''
  \href{http://arxiv.org/abs/2106.00979}{{\ttfamily arXiv:2106.00979
  [hep-ph]}}.

\bibitem{Hue:2021xzl}
L.~T. Hue, A.~E.~C. Hern\'andez, H.~N. Long, and T.~T. Hong, ``{Heavy singly
  charged Higgs bosons and inverse seesaw neutrinos as origins of large
  $(g-2)_{e,\mu}$ in two higgs doublet models},''
  \href{http://arxiv.org/abs/2110.01356}{{\ttfamily arXiv:2110.01356
  [hep-ph]}}.

\bibitem{Keung:2021rps}
W.-Y. Keung, D.~Marfatia, and P.-Y. Tseng, ``{Axion-Like Particles,
  Two-Higgs-Doublet Models, Leptoquarks, and the Electron and Muon ($g-2$)},''
  \href{http://dx.doi.org/10.31526/lhep.2021.209}{{\em LHEP} {\bfseries 2021}
  (2021) 209}, \href{http://arxiv.org/abs/2104.03341}{{\ttfamily
  arXiv:2104.03341 [hep-ph]}}.

\bibitem{Bigaran:2020jil}
I.~Bigaran and R.~R. Volkas, ``{Getting chirality right: Single scalar
  leptoquark solutions to the $(g-2)_{e,\mu}$ puzzle},''
  \href{http://dx.doi.org/10.1103/PhysRevD.102.075037}{{\em Phys. Rev. D}
  {\bfseries 102} no.~7, (2020) 075037},
  \href{http://arxiv.org/abs/2002.12544}{{\ttfamily arXiv:2002.12544
  [hep-ph]}}.

\bibitem{Dorsner:2020aaz}
I.~Dor\v{s}ner, S.~Fajfer, and S.~Saad, ``{$\mu \to e \gamma$ selecting scalar
  leptoquark solutions for the $(g-2)_{e,\mu}$ puzzles},''
  \href{http://dx.doi.org/10.1103/PhysRevD.102.075007}{{\em Phys. Rev. D}
  {\bfseries 102} no.~7, (2020) 075007},
  \href{http://arxiv.org/abs/2006.11624}{{\ttfamily arXiv:2006.11624
  [hep-ph]}}.

\bibitem{Bigaran:2021kmn}
I.~Bigaran and R.~R. Volkas, ``{Reflecting on Chirality: CP-violating
  extensions of the single scalar-leptoquark solutions for the $(g-2)_{e,\mu}$
  puzzles and their implications for lepton EDMs},''
  \href{http://arxiv.org/abs/2110.03707}{{\ttfamily arXiv:2110.03707
  [hep-ph]}}.

\bibitem{Abdullah:2019ofw}
M.~Abdullah, B.~Dutta, S.~Ghosh, and T.~Li, ``{$(g-2)_{\mu,e}$ and the ANITA
  anomalous events in a three-loop neutrino mass model},''
  \href{http://dx.doi.org/10.1103/PhysRevD.100.115006}{{\em Phys. Rev. D}
  {\bfseries 100} no.~11, (2019) 115006},
  \href{http://arxiv.org/abs/1907.08109}{{\ttfamily arXiv:1907.08109
  [hep-ph]}}.

\bibitem{CarcamoHernandez:2020pxw}
A.~E. C\'arcamo~Hern\'andez, Y.~Hidalgo~Vel\'asquez, S.~Kovalenko, H.~N. Long,
  N.~A. P\'erez-Julve, and V.~V. Vien, ``{Fermion spectrum and $g-2$ anomalies
  in a low scale 3-3-1 model},''
  \href{http://dx.doi.org/10.1140/epjc/s10052-021-08974-4}{{\em Eur. Phys. J.
  C} {\bfseries 81} no.~2, (2021) 191},
  \href{http://arxiv.org/abs/2002.07347}{{\ttfamily arXiv:2002.07347
  [hep-ph]}}.

\bibitem{CarcamoHernandez:2019ydc}
A.~E. C\'arcamo~Hern\'andez, S.~F. King, H.~Lee, and S.~J. Rowley, ``{Is it
  possible to explain the muon and electron $g-2$ in a $Z'$ model?},''
  \href{http://dx.doi.org/10.1103/PhysRevD.101.115016}{{\em Phys. Rev. D}
  {\bfseries 101} no.~11, (2020) 115016},
  \href{http://arxiv.org/abs/1910.10734}{{\ttfamily arXiv:1910.10734
  [hep-ph]}}.

\bibitem{Bodas:2021fsy}
A.~Bodas, R.~Coy, and S.~J.~D. King, ``{Solving the electron and muon $g-2$
  anomalies in $Z'$ models},''
  \href{http://arxiv.org/abs/2102.07781}{{\ttfamily arXiv:2102.07781
  [hep-ph]}}.

\bibitem{Chowdhury:2021tnm}
T.~A. Chowdhury and S.~Saad, ``{Non-Abelian vector dark matter and lepton
  g-2},'' \href{http://dx.doi.org/10.1088/1475-7516/2021/10/014}{{\em JCAP}
  {\bfseries 10} (2021) 014}, \href{http://arxiv.org/abs/2107.11863}{{\ttfamily
  arXiv:2107.11863 [hep-ph]}}.

\bibitem{Hernandez:2021iss}
A.~E.~C. Hern\'andez, S.~Kovalenko, M.~Maniatis, and I.~Schmidt, ``{Fermion
  mass hierarchy and g-2 anomalies in an extended 3HDM Model},''
  \href{http://dx.doi.org/10.1007/JHEP10(2021)036}{{\em JHEP} {\bfseries 10}
  (2021) 036}, \href{http://arxiv.org/abs/2104.07047}{{\ttfamily
  arXiv:2104.07047 [hep-ph]}}.

\bibitem{He:2019uvu}
F.~He and P.~Wang, ``{Pauli form factors of electron and muon in nonlocal
  quantum electrodynamics},''
  \href{http://dx.doi.org/10.1140/epjp/s13360-020-00151-y}{{\em Eur. Phys. J.
  Plus} {\bfseries 135} no.~2, (2020) 156},
  \href{http://arxiv.org/abs/1901.00271}{{\ttfamily arXiv:1901.00271
  [hep-ph]}}.

\bibitem{Badziak:2019gaf}
M.~Badziak and K.~Sakurai, ``{Explanation of electron and muon g \ensuremath{-}
  2 anomalies in the MSSM},''
  \href{http://dx.doi.org/10.1007/JHEP10(2019)024}{{\em JHEP} {\bfseries 10}
  (2019) 024}, \href{http://arxiv.org/abs/1908.03607}{{\ttfamily
  arXiv:1908.03607 [hep-ph]}}.

\bibitem{Endo:2019bcj}
M.~Endo and W.~Yin, ``{Explaining electron and muon $g-2$ anomaly in SUSY
  without lepton-flavor mixings},''
  \href{http://dx.doi.org/10.1007/JHEP08(2019)122}{{\em JHEP} {\bfseries 08}
  (2019) 122}, \href{http://arxiv.org/abs/1906.08768}{{\ttfamily
  arXiv:1906.08768 [hep-ph]}}.

\bibitem{Dong:2019iaf}
X.-X. Dong, S.-M. Zhao, H.-B. Zhang, and T.-F. Feng, ``{The two-loop
  corrections to lepton MDMs and EDMs in the EBLMSSM},''
  \href{http://dx.doi.org/10.1088/1361-6471/ab5f8f}{{\em J. Phys. G} {\bfseries
  47} no.~4, (2020) 045002}, \href{http://arxiv.org/abs/1901.07701}{{\ttfamily
  arXiv:1901.07701 [hep-ph]}}.

\bibitem{Yang:2020bmh}
J.-L. Yang, T.-F. Feng, and H.-B. Zhang, ``{Electron and muon $(g-2)$ in the
  B-LSSM},'' \href{http://dx.doi.org/10.1088/1361-6471/ab7986}{{\em J. Phys. G}
  {\bfseries 47} no.~5, (2020) 055004},
  \href{http://arxiv.org/abs/2003.09781}{{\ttfamily arXiv:2003.09781
  [hep-ph]}}.

\bibitem{Cao:2021lmj}
J.~Cao, Y.~He, J.~Lian, D.~Zhang, and P.~Zhu, ``{Electron and muon anomalous
  magnetic moments in the inverse seesaw extended NMSSM},''
  \href{http://dx.doi.org/10.1103/PhysRevD.104.055009}{{\em Phys. Rev. D}
  {\bfseries 104} no.~5, (2021) 055009},
  \href{http://arxiv.org/abs/2102.11355}{{\ttfamily arXiv:2102.11355
  [hep-ph]}}.

\bibitem{Frank:2021nkq}
M.~Frank, Y.~Hi\c{c}y\i{}lmaz, S.~Mondal, O.~\"Ozdal, and C.~S. \"Un,
  ``{Electron and muon magnetic moments and implications for dark matter and
  model characterisation in non-universal U(1)' supersymmetric models},''
  \href{http://dx.doi.org/10.1007/JHEP10(2021)063}{{\em JHEP} {\bfseries 10}
  (2021) 063}, \href{http://arxiv.org/abs/2107.04116}{{\ttfamily
  arXiv:2107.04116 [hep-ph]}}.

\bibitem{Li:2021koa}
S.~Li, Y.~Xiao, and J.~M. Yang, ``{Can electron and muon $g-2$ anomalies be
  jointly explained in SUSY?},''
  \href{http://arxiv.org/abs/2107.04962}{{\ttfamily arXiv:2107.04962
  [hep-ph]}}.

\bibitem{DelleRose:2020oaa}
L.~Delle~Rose, S.~Khalil, and S.~Moretti, ``{Explaining electron and muon $g$
  \ensuremath{-} 2 anomalies in an Aligned 2-Higgs Doublet Model with
  right-handed neutrinos},''
  \href{http://dx.doi.org/10.1016/j.physletb.2021.136216}{{\em Phys. Lett. B}
  {\bfseries 816} (2021) 136216},
  \href{http://arxiv.org/abs/2012.06911}{{\ttfamily arXiv:2012.06911
  [hep-ph]}}.

\bibitem{Botella:2020xzf}
F.~J. Botella, F.~Cornet-Gomez, and M.~Nebot, ``{Electron and muon $g-2$
  anomalies in general flavour conserving two Higgs doublets models},''
  \href{http://dx.doi.org/10.1103/PhysRevD.102.035023}{{\em Phys. Rev. D}
  {\bfseries 102} no.~3, (2020) 035023},
  \href{http://arxiv.org/abs/2006.01934}{{\ttfamily arXiv:2006.01934
  [hep-ph]}}.

\bibitem{Jana:2020pxx}
S.~Jana, V.~P. K., and S.~Saad, ``{Resolving electron and muon $g-2$ within the
  2HDM},'' \href{http://dx.doi.org/10.1103/PhysRevD.101.115037}{{\em Phys. Rev.
  D} {\bfseries 101} no.~11, (2020) 115037},
  \href{http://arxiv.org/abs/2003.03386}{{\ttfamily arXiv:2003.03386
  [hep-ph]}}.

\bibitem{Fajfer:2021cxa}
S.~Fajfer, J.~F. Kamenik, and M.~Tammaro, ``{Interplay of New Physics effects
  in $(g-2)_\ell$ and $h \to \ell^{+}\ell^{-}$ lessons from SMEFT},''
  \href{http://dx.doi.org/10.1007/JHEP06(2021)099}{{\em JHEP} {\bfseries 06}
  (2021) 099}, \href{http://arxiv.org/abs/2103.10859}{{\ttfamily
  arXiv:2103.10859 [hep-ph]}}.

\bibitem{ACME:2018yjb}
{\bfseries ACME} Collaboration, V.~Andreev {\em et~al.}, ``{Improved limit on
  the electric dipole moment of the electron},''
  \href{http://dx.doi.org/10.1038/s41586-018-0599-8}{{\em Nature} {\bfseries
  562} no.~7727, (2018) 355--360}.

\bibitem{Muong-2:2008ebm}
{\bfseries Muon (g-2)} Collaboration, G.~W. Bennett {\em et~al.}, ``{An
  Improved Limit on the Muon Electric Dipole Moment},''
  \href{http://dx.doi.org/10.1103/PhysRevD.80.052008}{{\em Phys. Rev. D}
  {\bfseries 80} (2009) 052008},
  \href{http://arxiv.org/abs/0811.1207}{{\ttfamily arXiv:0811.1207 [hep-ex]}}.

\bibitem{Abe:2019thb}
M.~Abe {\em et~al.}, ``{A New Approach for Measuring the Muon Anomalous
  Magnetic Moment and Electric Dipole Moment},''
  \href{http://dx.doi.org/10.1093/ptep/ptz030}{{\em PTEP} {\bfseries 2019}
  no.~5, (2019) 053C02}, \href{http://arxiv.org/abs/1901.03047}{{\ttfamily
  arXiv:1901.03047 [physics.ins-det]}}.

\bibitem{Sato:2021aor}
{\bfseries J-PARC E34} Collaboration, Y.~Sato, ``{J-PARC Muon g \ensuremath{-}
  2/EDM experiment},'' \href{http://dx.doi.org/10.7566/JPSCP.33.011110}{{\em
  JPS Conf. Proc.} {\bfseries 33} (2021) 011110}.

\bibitem{Adelmann:2021udj}
A.~Adelmann {\em et~al.}, ``{Search for a muon EDM using the frozen-spin
  technique},'' \href{http://arxiv.org/abs/2102.08838}{{\ttfamily
  arXiv:2102.08838 [hep-ex]}}.

\bibitem{Wolfenstein:1977ue}
L.~Wolfenstein, ``{Neutrino Oscillations in Matter},''
  \href{http://dx.doi.org/10.1103/PhysRevD.17.2369}{{\em Phys. Rev. D}
  {\bfseries 17} (1978) 2369--2374}.

\bibitem{Wolfenstein:1979ni}
L.~Wolfenstein, ``{Neutrino Oscillations and Stellar Collapse},''
  \href{http://dx.doi.org/10.1103/PhysRevD.20.2634}{{\em Phys. Rev. D}
  {\bfseries 20} (1979) 2634--2635}.

\bibitem{Proceedings:2019qno}
\href{http://dx.doi.org/10.21468/SciPostPhysProc.2.001}{{\em {Neutrino
  Non-Standard Interactions: A Status Report}}}, vol.~2.
\newblock 2019.
\newblock \href{http://arxiv.org/abs/1907.00991}{{\ttfamily arXiv:1907.00991
  [hep-ph]}}.

\bibitem{Behnke:2013xla}
``{The International Linear Collider Technical Design Report - Volume 1:
  Executive Summary},'' \href{http://arxiv.org/abs/1306.6327}{{\ttfamily
  arXiv:1306.6327 [physics.acc-ph]}}.

\bibitem{Baer:2013cma}
``{The International Linear Collider Technical Design Report - Volume 2:
  Physics},'' \href{http://arxiv.org/abs/1306.6352}{{\ttfamily arXiv:1306.6352
  [hep-ph]}}.

\bibitem{Adolphsen:2013jya}
``{The International Linear Collider Technical Design Report - Volume 3.I:
  Accelerator \textbackslash{}\& in the Technical Design Phase},''
  \href{http://arxiv.org/abs/1306.6353}{{\ttfamily arXiv:1306.6353
  [physics.acc-ph]}}.

\bibitem{Adolphsen:2013kya}
``{The International Linear Collider Technical Design Report - Volume 3.II:
  Accelerator Baseline Design},''
  \href{http://arxiv.org/abs/1306.6328}{{\ttfamily arXiv:1306.6328
  [physics.acc-ph]}}.

\bibitem{Behnke:2013lya}
H.~Abramowicz {\em et~al.}, ``{The International Linear Collider Technical
  Design Report - Volume 4: Detectors},''
  \href{http://arxiv.org/abs/1306.6329}{{\ttfamily arXiv:1306.6329
  [physics.ins-det]}}.

\bibitem{Delahaye:2019omf}
J.~P. Delahaye, M.~Diemoz, K.~Long, B.~Mansouli\'e, N.~Pastrone, L.~Rivkin,
  D.~Schulte, A.~Skrinsky, and A.~Wulzer, ``{Muon Colliders},''
  \href{http://arxiv.org/abs/1901.06150}{{\ttfamily arXiv:1901.06150
  [physics.acc-ph]}}.

\bibitem{Shiltsev:2019rfl}
V.~Shiltsev and F.~Zimmermann, ``{Modern and Future Colliders},''
  \href{http://dx.doi.org/10.1103/RevModPhys.93.015006}{{\em Rev. Mod. Phys.}
  {\bfseries 93} (2021) 015006},
  \href{http://arxiv.org/abs/2003.09084}{{\ttfamily arXiv:2003.09084
  [physics.acc-ph]}}.

\bibitem{Wolfenstein:1980sy}
L.~Wolfenstein, ``{A Theoretical Pattern for Neutrino Oscillations},''
  \href{http://dx.doi.org/10.1016/0550-3213(80)90004-8}{{\em Nucl. Phys. B}
  {\bfseries 175} (1980) 93--96}.

\bibitem{Lee:1973iz}
T.~D. Lee, ``{A Theory of Spontaneous T Violation},''
  \href{http://dx.doi.org/10.1103/PhysRevD.8.1226}{{\em Phys. Rev. D}
  {\bfseries 8} (1973) 1226--1239}.

\bibitem{Branco:2011iw}
G.~C. Branco, P.~M. Ferreira, L.~Lavoura, M.~N. Rebelo, M.~Sher, and J.~P.
  Silva, ``{Theory and phenomenology of two-Higgs-doublet models},''
  \href{http://dx.doi.org/10.1016/j.physrep.2012.02.002}{{\em Phys. Rept.}
  {\bfseries 516} (2012) 1--102},
  \href{http://arxiv.org/abs/1106.0034}{{\ttfamily arXiv:1106.0034 [hep-ph]}}.

\bibitem{Davidson:2005cw}
S.~Davidson and H.~E. Haber, ``{Basis-independent methods for the
  two-Higgs-doublet model},''
  \href{http://dx.doi.org/10.1103/PhysRevD.72.099902}{{\em Phys. Rev. D}
  {\bfseries 72} (2005) 035004},
  \href{http://arxiv.org/abs/hep-ph/0504050}{{\ttfamily arXiv:hep-ph/0504050}}.
  [Erratum: Phys.Rev.D 72, 099902 (2005)].

\bibitem{Gunion:2002zf}
J.~F. Gunion and H.~E. Haber, ``{The CP conserving two Higgs doublet model: The
  Approach to the decoupling limit},''
  \href{http://dx.doi.org/10.1103/PhysRevD.67.075019}{{\em Phys. Rev. D}
  {\bfseries 67} (2003) 075019},
  \href{http://arxiv.org/abs/hep-ph/0207010}{{\ttfamily arXiv:hep-ph/0207010}}.

\bibitem{Carena:2013ooa}
M.~Carena, I.~Low, N.~R. Shah, and C.~E.~M. Wagner, ``{Impersonating the
  Standard Model Higgs Boson: Alignment without Decoupling},''
  \href{http://dx.doi.org/10.1007/JHEP04(2014)015}{{\em JHEP} {\bfseries 04}
  (2014) 015}, \href{http://arxiv.org/abs/1310.2248}{{\ttfamily arXiv:1310.2248
  [hep-ph]}}.

\bibitem{BhupalDev:2014bir}
P.~S. Bhupal~Dev and A.~Pilaftsis, ``{Maximally Symmetric Two Higgs Doublet
  Model with Natural Standard Model Alignment},''
  \href{http://dx.doi.org/10.1007/JHEP12(2014)024}{{\em JHEP} {\bfseries 12}
  (2014) 024}, \href{http://arxiv.org/abs/1408.3405}{{\ttfamily arXiv:1408.3405
  [hep-ph]}}. [Erratum: JHEP 11, 147 (2015)].

\bibitem{Das:2015mwa}
D.~Das and I.~Saha, ``{Search for a stable alignment limit in two-Higgs-doublet
  models},'' \href{http://dx.doi.org/10.1103/PhysRevD.91.095024}{{\em Phys.
  Rev. D} {\bfseries 91} no.~9, (2015) 095024},
  \href{http://arxiv.org/abs/1503.02135}{{\ttfamily arXiv:1503.02135
  [hep-ph]}}.

\bibitem{Bernon:2015qea}
J.~Bernon, J.~F. Gunion, H.~E. Haber, Y.~Jiang, and S.~Kraml, ``{Scrutinizing
  the alignment limit in two-Higgs-doublet models: m$_h$=125 GeV},''
  \href{http://dx.doi.org/10.1103/PhysRevD.92.075004}{{\em Phys. Rev. D}
  {\bfseries 92} no.~7, (2015) 075004},
  \href{http://arxiv.org/abs/1507.00933}{{\ttfamily arXiv:1507.00933
  [hep-ph]}}.

\bibitem{Chowdhury:2017aav}
D.~Chowdhury and O.~Eberhardt, ``{Update of Global Two-Higgs-Doublet Model
  Fits},'' \href{http://dx.doi.org/10.1007/JHEP05(2018)161}{{\em JHEP}
  {\bfseries 05} (2018) 161}, \href{http://arxiv.org/abs/1711.02095}{{\ttfamily
  arXiv:1711.02095 [hep-ph]}}.

\bibitem{Babu:2001ex}
K.~S. Babu and C.~N. Leung, ``{Classification of effective neutrino mass
  operators},'' \href{http://dx.doi.org/10.1016/S0550-3213(01)00504-1}{{\em
  Nucl. Phys. B} {\bfseries 619} (2001) 667--689},
  \href{http://arxiv.org/abs/hep-ph/0106054}{{\ttfamily arXiv:hep-ph/0106054}}.

\bibitem{deGouvea:2007qla}
A.~de~Gouvea and J.~Jenkins, ``{A Survey of Lepton Number Violation Via
  Effective Operators},''
  \href{http://dx.doi.org/10.1103/PhysRevD.77.013008}{{\em Phys. Rev. D}
  {\bfseries 77} (2008) 013008},
  \href{http://arxiv.org/abs/0708.1344}{{\ttfamily arXiv:0708.1344 [hep-ph]}}.

\bibitem{Cepedello:2017lyo}
R.~Cepedello, M.~Hirsch, and J.~C. Helo, ``{Lepton number violating
  phenomenology of d = 7 neutrino mass models},''
  \href{http://dx.doi.org/10.1007/JHEP01(2018)009}{{\em JHEP} {\bfseries 01}
  (2018) 009}, \href{http://arxiv.org/abs/1709.03397}{{\ttfamily
  arXiv:1709.03397 [hep-ph]}}.

\bibitem{Gargalionis:2020xvt}
J.~Gargalionis and R.~R. Volkas, ``{Exploding operators for Majorana neutrino
  masses and beyond},'' \href{http://dx.doi.org/10.1007/JHEP01(2021)074}{{\em
  JHEP} {\bfseries 01} (2021) 074},
  \href{http://arxiv.org/abs/2009.13537}{{\ttfamily arXiv:2009.13537
  [hep-ph]}}.

\bibitem{Herrero-Garcia:2017xdu}
J.~Herrero-Garc\'\i{}a, T.~Ohlsson, S.~Riad, and J.~Wir\'en, ``{Full parameter
  scan of the Zee model: exploring Higgs lepton flavor violation},''
  \href{http://dx.doi.org/10.1007/JHEP04(2017)130}{{\em JHEP} {\bfseries 04}
  (2017) 130}, \href{http://arxiv.org/abs/1701.05345}{{\ttfamily
  arXiv:1701.05345 [hep-ph]}}.

\bibitem{Herrero-Garcia:2014hfa}
J.~Herrero-Garcia, M.~Nebot, N.~Rius, and A.~Santamaria, ``{The
  Zee\textendash{}Babu model revisited in the light of new data},''
  \href{http://dx.doi.org/10.1016/j.nuclphysb.2014.06.001}{{\em Nucl. Phys. B}
  {\bfseries 885} (2014) 542--570},
  \href{http://arxiv.org/abs/1402.4491}{{\ttfamily arXiv:1402.4491 [hep-ph]}}.

\bibitem{Ghosal:2001ep}
A.~Ghosal, Y.~Koide, and H.~Fusaoka, ``{Lepton flavor violating Z decays in the
  Zee model},'' \href{http://dx.doi.org/10.1103/PhysRevD.64.053012}{{\em Phys.
  Rev. D} {\bfseries 64} (2001) 053012},
  \href{http://arxiv.org/abs/hep-ph/0104104}{{\ttfamily arXiv:hep-ph/0104104}}.

\bibitem{Koide:2001xy}
Y.~Koide, ``{Can the Zee model explain the observed neutrino data?},''
  \href{http://dx.doi.org/10.1103/PhysRevD.64.077301}{{\em Phys. Rev. D}
  {\bfseries 64} (2001) 077301},
  \href{http://arxiv.org/abs/hep-ph/0104226}{{\ttfamily arXiv:hep-ph/0104226}}.

\bibitem{He:2003ih}
X.-G. He, ``{Is the Zee model neutrino mass matrix ruled out?},''
  \href{http://dx.doi.org/10.1140/epjc/s2004-01669-8}{{\em Eur. Phys. J. C}
  {\bfseries 34} (2004) 371--376},
  \href{http://arxiv.org/abs/hep-ph/0307172}{{\ttfamily arXiv:hep-ph/0307172}}.

\bibitem{Babu:2013pma}
K.~S. Babu and J.~Julio, ``{Predictive Model of Radiative Neutrino Masses},''
  \href{http://dx.doi.org/10.1103/PhysRevD.89.053004}{{\em Phys. Rev. D}
  {\bfseries 89} no.~5, (2014) 053004},
  \href{http://arxiv.org/abs/1310.0303}{{\ttfamily arXiv:1310.0303 [hep-ph]}}.

\bibitem{Leveille:1977rc}
J.~P. Leveille, ``{The Second Order Weak Correction to (G-2) of the Muon in
  Arbitrary Gauge Models},''
  \href{http://dx.doi.org/10.1016/0550-3213(78)90051-2}{{\em Nucl. Phys. B}
  {\bfseries 137} (1978) 63--76}.

\bibitem{Barr:1990vd}
S.~M. Barr and A.~Zee, ``{Electric Dipole Moment of the Electron and of the
  Neutron},'' \href{http://dx.doi.org/10.1103/PhysRevLett.65.21}{{\em Phys.
  Rev. Lett.} {\bfseries 65} (1990) 21--24}. [Erratum: Phys.Rev.Lett. 65, 2920
  (1990)].

\bibitem{Bjorken:1977vt}
J.~D. Bjorken and S.~Weinberg, ``{A Mechanism for Nonconservation of Muon
  Number},'' \href{http://dx.doi.org/10.1103/PhysRevLett.38.622}{{\em Phys.
  Rev. Lett.} {\bfseries 38} (1977) 622}.

\bibitem{Ilisie:2015tra}
V.~Ilisie, ``{New Barr-Zee contributions to $\mathbf{(g-2)_\mu}$ in
  two-Higgs-doublet models},''
  \href{http://dx.doi.org/10.1007/JHEP04(2015)077}{{\em JHEP} {\bfseries 04}
  (2015) 077}, \href{http://arxiv.org/abs/1502.04199}{{\ttfamily
  arXiv:1502.04199 [hep-ph]}}.

\bibitem{Frank:2020smf}
M.~Frank and I.~Saha, ``{Muon anomalous magnetic moment in two-Higgs-doublet
  models with vectorlike leptons},''
  \href{http://dx.doi.org/10.1103/PhysRevD.102.115034}{{\em Phys. Rev. D}
  {\bfseries 102} no.~11, (2020) 115034},
  \href{http://arxiv.org/abs/2008.11909}{{\ttfamily arXiv:2008.11909
  [hep-ph]}}.

\bibitem{Cherchiglia:2016eui}
A.~Cherchiglia, P.~Kneschke, D.~St\"ockinger, and H.~St\"ockinger-Kim, ``{The
  muon magnetic moment in the 2HDM: complete two-loop result},''
  \href{http://dx.doi.org/10.1007/JHEP10(2021)242}{{\em JHEP} {\bfseries 01}
  (2017) 007}, \href{http://arxiv.org/abs/1607.06292}{{\ttfamily
  arXiv:1607.06292 [hep-ph]}}. [Erratum: JHEP 10, 242 (2021)].

\bibitem{Cherchiglia:2017uwv}
A.~Cherchiglia, D.~St\"ockinger, and H.~St\"ockinger-Kim, ``{Muon g-2 in the
  2HDM: maximum results and detailed phenomenology},''
  \href{http://dx.doi.org/10.1103/PhysRevD.98.035001}{{\em Phys. Rev. D}
  {\bfseries 98} (2018) 035001},
  \href{http://arxiv.org/abs/1711.11567}{{\ttfamily arXiv:1711.11567
  [hep-ph]}}.

\bibitem{Chang:2000ii}
D.~Chang, W.-F. Chang, C.-H. Chou, and W.-Y. Keung, ``{Large two loop
  contributions to g-2 from a generic pseudoscalar boson},''
  \href{http://dx.doi.org/10.1103/PhysRevD.63.091301}{{\em Phys. Rev. D}
  {\bfseries 63} (2001) 091301},
  \href{http://arxiv.org/abs/hep-ph/0009292}{{\ttfamily arXiv:hep-ph/0009292}}.

\bibitem{Ecker:1983dj}
G.~Ecker, W.~Grimus, and H.~Neufeld, ``{The Neutron Electric Dipole Moment in
  Left-right Symmetric Gauge Models},''
  \href{http://dx.doi.org/10.1016/0550-3213(83)90341-3}{{\em Nucl. Phys. B}
  {\bfseries 229} (1983) 421--444}.

\bibitem{BaBar:2016sci}
{\bfseries BaBar} Collaboration, J.~P. Lees {\em et~al.}, ``{Search for a
  muonic dark force at BABAR},''
  \href{http://dx.doi.org/10.1103/PhysRevD.94.011102}{{\em Phys. Rev. D}
  {\bfseries 94} no.~1, (2016) 011102},
  \href{http://arxiv.org/abs/1606.03501}{{\ttfamily arXiv:1606.03501
  [hep-ex]}}.

\bibitem{CMS:2018yxg}
{\bfseries CMS} Collaboration, A.~M. Sirunyan {\em et~al.}, ``{Search for an
  $L_{\mu}-L_{\tau}$ gauge boson using Z$\to4\mu$ events in proton-proton
  collisions at $\sqrt{s} =$ 13 TeV},''
  \href{http://dx.doi.org/10.1016/j.physletb.2019.01.072}{{\em Phys. Lett. B}
  {\bfseries 792} (2019) 345--368},
  \href{http://arxiv.org/abs/1808.03684}{{\ttfamily arXiv:1808.03684
  [hep-ex]}}.

\bibitem{Lavoura:2003xp}
L.~Lavoura, ``{General formulae for f(1) ---\ensuremath{>} f(2) gamma},''
  \href{http://dx.doi.org/10.1140/epjc/s2003-01212-7}{{\em Eur. Phys. J. C}
  {\bfseries 29} (2003) 191--195},
  \href{http://arxiv.org/abs/hep-ph/0302221}{{\ttfamily arXiv:hep-ph/0302221}}.

\bibitem{MEG:2016leq}
{\bfseries MEG} Collaboration, A.~M. Baldini {\em et~al.}, ``{Search for the
  lepton flavour violating decay $\mu ^+ \rightarrow \mathrm {e}^+ \gamma $
  with the full dataset of the MEG experiment},''
  \href{http://dx.doi.org/10.1140/epjc/s10052-016-4271-x}{{\em Eur. Phys. J. C}
  {\bfseries 76} no.~8, (2016) 434},
  \href{http://arxiv.org/abs/1605.05081}{{\ttfamily arXiv:1605.05081
  [hep-ex]}}.

\bibitem{BaBar:2009hkt}
{\bfseries BaBar} Collaboration, B.~Aubert {\em et~al.}, ``{Searches for Lepton
  Flavor Violation in the Decays tau+- ---\ensuremath{>} e+- gamma and tau+-
  ---\ensuremath{>} mu+- gamma},''
  \href{http://dx.doi.org/10.1103/PhysRevLett.104.021802}{{\em Phys. Rev.
  Lett.} {\bfseries 104} (2010) 021802},
  \href{http://arxiv.org/abs/0908.2381}{{\ttfamily arXiv:0908.2381 [hep-ex]}}.

\bibitem{HFLAV:2016hnz}
{\bfseries HFLAV} Collaboration, Y.~Amhis {\em et~al.}, ``{Averages of
  $b$-hadron, $c$-hadron, and $\tau$-lepton properties as of summer 2016},''
  \href{http://dx.doi.org/10.1140/epjc/s10052-017-5058-4}{{\em Eur. Phys. J. C}
  {\bfseries 77} no.~12, (2017) 895},
  \href{http://arxiv.org/abs/1612.07233}{{\ttfamily arXiv:1612.07233
  [hep-ex]}}.

\bibitem{SINDRUM:1987nra}
{\bfseries SINDRUM} Collaboration, U.~Bellgardt {\em et~al.}, ``{Search for the
  Decay mu+ ---\ensuremath{>} e+ e+ e-},''
  \href{http://dx.doi.org/10.1016/0550-3213(88)90462-2}{{\em Nucl. Phys. B}
  {\bfseries 299} (1988) 1--6}.

\bibitem{Hayasaka:2010np}
K.~Hayasaka {\em et~al.}, ``{Search for Lepton Flavor Violating Tau Decays into
  Three Leptons with 719 Million Produced Tau+Tau- Pairs},''
  \href{http://dx.doi.org/10.1016/j.physletb.2010.03.037}{{\em Phys. Lett. B}
  {\bfseries 687} (2010) 139--143},
  \href{http://arxiv.org/abs/1001.3221}{{\ttfamily arXiv:1001.3221 [hep-ex]}}.

\bibitem{BaBar:2010axs}
{\bfseries BaBar} Collaboration, J.~P. Lees {\em et~al.}, ``{Limits on tau
  Lepton-Flavor Violating Decays in three charged leptons},''
  \href{http://dx.doi.org/10.1103/PhysRevD.81.111101}{{\em Phys. Rev. D}
  {\bfseries 81} (2010) 111101},
  \href{http://arxiv.org/abs/1002.4550}{{\ttfamily arXiv:1002.4550 [hep-ex]}}.

\bibitem{ATLAS:2016jts}
{\bfseries ATLAS} Collaboration, G.~Aad {\em et~al.}, ``{Probing lepton flavour
  violation via neutrinoless $\tau\longrightarrow 3\mu$ decays with the ATLAS
  detector},'' \href{http://dx.doi.org/10.1140/epjc/s10052-016-4041-9}{{\em
  Eur. Phys. J. C} {\bfseries 76} no.~5, (2016) 232},
  \href{http://arxiv.org/abs/1601.03567}{{\ttfamily arXiv:1601.03567
  [hep-ex]}}.

\bibitem{LHCb:2014kws}
{\bfseries LHCb} Collaboration, R.~Aaij {\em et~al.}, ``{Search for the lepton
  flavour violating decay $\tau^{-} \to \mu^{-} \mu^{+} \mu^-$},''
  \href{http://dx.doi.org/10.1007/JHEP02(2015)121}{{\em JHEP} {\bfseries 02}
  (2015) 121}, \href{http://arxiv.org/abs/1409.8548}{{\ttfamily arXiv:1409.8548
  [hep-ex]}}.

\bibitem{Peskin:1990zt}
M.~E. Peskin and T.~Takeuchi, ``{A New constraint on a strongly interacting
  Higgs sector},'' \href{http://dx.doi.org/10.1103/PhysRevLett.65.964}{{\em
  Phys. Rev. Lett.} {\bfseries 65} (1990) 964--967}.

\bibitem{Peskin:1991sw}
M.~E. Peskin and T.~Takeuchi, ``{Estimation of oblique electroweak
  corrections},'' \href{http://dx.doi.org/10.1103/PhysRevD.46.381}{{\em Phys.
  Rev. D} {\bfseries 46} (1992) 381--409}.

\bibitem{Funk:2011ad}
G.~Funk, D.~O'Neil, and R.~M. Winters, ``{What the Oblique Parameters S, T, and
  U and Their Extensions Reveal About the 2HDM: A Numerical Analysis},''
  \href{http://dx.doi.org/10.1142/S0217751X12500212}{{\em Int. J. Mod. Phys. A}
  {\bfseries 27} (2012) 1250021},
  \href{http://arxiv.org/abs/1110.3812}{{\ttfamily arXiv:1110.3812 [hep-ph]}}.

\bibitem{Grimus:2008nb}
W.~Grimus, L.~Lavoura, O.~M. Ogreid, and P.~Osland, ``{The Oblique parameters
  in multi-Higgs-doublet models},''
  \href{http://dx.doi.org/10.1016/j.nuclphysb.2008.04.019}{{\em Nucl. Phys. B}
  {\bfseries 801} (2008) 81--96},
  \href{http://arxiv.org/abs/0802.4353}{{\ttfamily arXiv:0802.4353 [hep-ph]}}.

\bibitem{Pontecorvo:1957cp}
B.~Pontecorvo, ``{Mesonium and anti-mesonium},'' {\em Sov. Phys. JETP}
  {\bfseries 6} (1957) 429.

\bibitem{Willmann:1998gd}
L.~Willmann {\em et~al.}, ``{New bounds from searching for muonium to
  anti-muonium conversion},''
  \href{http://dx.doi.org/10.1103/PhysRevLett.82.49}{{\em Phys. Rev. Lett.}
  {\bfseries 82} (1999) 49--52},
  \href{http://arxiv.org/abs/hep-ex/9807011}{{\ttfamily arXiv:hep-ex/9807011}}.

\bibitem{Jentschura:1997tv}
U.~D. Jentschura, G.~Soff, V.~G. Ivanov, and S.~G. Karshenboim, ``{The Bound
  mu+ mu- system},'' \href{http://dx.doi.org/10.1103/PhysRevA.56.4483}{{\em
  Phys. Rev. A} {\bfseries 56} (1997) 4483},
  \href{http://arxiv.org/abs/physics/9706026}{{\ttfamily
  arXiv:physics/9706026}}.

\bibitem{Jentschura:1998vkm}
U.~D. Jentschura, V.~G. Ivanov, G.~Soff, and S.~G. Karshenboim,
  ``{Next-to-leading and higher order corrections to the decay rate of
  dimuonium},'' \href{http://dx.doi.org/10.1016/S0370-2693(98)00206-8}{{\em
  Phys. Lett. B} {\bfseries 424} (1998) 397--404},
  \href{http://arxiv.org/abs/hep-ph/9706401}{{\ttfamily arXiv:hep-ph/9706401}}.

\bibitem{Ginzburg:1998df}
I.~F. Ginzburg, U.~D. Jentschura, S.~G. Karshenboim, F.~Krauss, V.~G. Serbo,
  and G.~Soff, ``{Production of bound mu+ mu- systems in relativistic heavy ion
  collisions},'' \href{http://dx.doi.org/10.1103/PhysRevC.58.3565}{{\em Phys.
  Rev. C} {\bfseries 58} (1998) 3565--3573},
  \href{http://arxiv.org/abs/hep-ph/9805375}{{\ttfamily arXiv:hep-ph/9805375}}.

\bibitem{Clark:2003tv}
T.~E. Clark and S.~T. Love, ``{Muonium - anti-muonium oscillations and massive
  Majorana neutrinos},''
  \href{http://dx.doi.org/10.1142/S0217732304013143}{{\em Mod. Phys. Lett. A}
  {\bfseries 19} (2004) 297--306},
  \href{http://arxiv.org/abs/hep-ph/0307264}{{\ttfamily arXiv:hep-ph/0307264}}.

\bibitem{Harnik:2012pb}
R.~Harnik, J.~Kopp, and J.~Zupan, ``{Flavor Violating Higgs Decays},''
  \href{http://dx.doi.org/10.1007/JHEP03(2013)026}{{\em JHEP} {\bfseries 03}
  (2013) 026}, \href{http://arxiv.org/abs/1209.1397}{{\ttfamily arXiv:1209.1397
  [hep-ph]}}.

\bibitem{Dev:2017ftk}
P.~S.~B. Dev, R.~N. Mohapatra, and Y.~Zhang, ``{Lepton Flavor Violation Induced
  by a Neutral Scalar at Future Lepton Colliders},''
  \href{http://dx.doi.org/10.1103/PhysRevLett.120.221804}{{\em Phys. Rev.
  Lett.} {\bfseries 120} no.~22, (2018) 221804},
  \href{http://arxiv.org/abs/1711.08430}{{\ttfamily arXiv:1711.08430
  [hep-ph]}}.

\bibitem{mace}
J.~Tang.
  \url{https://www.snowmass21.org/docs/files/summaries/RF/SNOWMASS21-RF5_RF0_Jian_Tang-126.pdf},
  2020.

\bibitem{Cvetic:2005gx}
G.~Cvetic, C.~O. Dib, C.~S. Kim, and J.~D. Kim, ``{Muonium-antimuonium
  conversion in models with heavy neutrinos},''
  \href{http://dx.doi.org/10.1103/PhysRevD.71.113013}{{\em Phys. Rev. D}
  {\bfseries 71} (2005) 113013},
  \href{http://arxiv.org/abs/hep-ph/0504126}{{\ttfamily arXiv:hep-ph/0504126}}.

\bibitem{Han:2021nod}
C.~Han, D.~Huang, J.~Tang, and Y.~Zhang, ``{Probing the doubly charged Higgs
  boson with a muonium to antimuonium conversion experiment},''
  \href{http://dx.doi.org/10.1103/PhysRevD.103.055023}{{\em Phys. Rev. D}
  {\bfseries 103} no.~5, (2021) 055023},
  \href{http://arxiv.org/abs/2102.00758}{{\ttfamily arXiv:2102.00758
  [hep-ph]}}.

\bibitem{Conlin:2020veq}
R.~Conlin and A.~A. Petrov, ``{Muonium-antimuonium oscillations in effective
  field theory},'' \href{http://dx.doi.org/10.1103/PhysRevD.102.095001}{{\em
  Phys. Rev. D} {\bfseries 102} no.~9, (2020) 095001},
  \href{http://arxiv.org/abs/2005.10276}{{\ttfamily arXiv:2005.10276
  [hep-ph]}}.

\bibitem{Fukuyama:2021iyw}
T.~Fukuyama, Y.~Mimura, and Y.~Uesaka, ``{Models of Muonium to Antimuonium
  Transition},'' \href{http://arxiv.org/abs/2108.10736}{{\ttfamily
  arXiv:2108.10736 [hep-ph]}}.

\bibitem{Anastasi:2015qla}
A.~Anastasi {\em et~al.}, ``{Limit on the production of a low-mass vector boson
  in $\mathrm{e}^{+}\mathrm{e}^{-} \to \mathrm{U}\gamma$, $\mathrm{U} \to
  \mathrm{e}^{+}\mathrm{e}^{-}$ with the KLOE experiment},''
  \href{http://dx.doi.org/10.1016/j.physletb.2015.10.003}{{\em Phys. Lett. B}
  {\bfseries 750} (2015) 633--637},
  \href{http://arxiv.org/abs/1509.00740}{{\ttfamily arXiv:1509.00740
  [hep-ex]}}.

\bibitem{Alves:2017avw}
D.~S.~M. Alves and N.~Weiner, ``{A viable QCD axion in the MeV mass range},''
  \href{http://dx.doi.org/10.1007/JHEP07(2018)092}{{\em JHEP} {\bfseries 07}
  (2018) 092}, \href{http://arxiv.org/abs/1710.03764}{{\ttfamily
  arXiv:1710.03764 [hep-ph]}}.

\bibitem{BaBar:2014zli}
{\bfseries BaBar} Collaboration, J.~P. Lees {\em et~al.}, ``{Search for a Dark
  Photon in $e^+e^-$ Collisions at BaBar},''
  \href{http://dx.doi.org/10.1103/PhysRevLett.113.201801}{{\em Phys. Rev.
  Lett.} {\bfseries 113} no.~20, (2014) 201801},
  \href{http://arxiv.org/abs/1406.2980}{{\ttfamily arXiv:1406.2980 [hep-ex]}}.

\bibitem{Knapen:2017xzo}
S.~Knapen, T.~Lin, and K.~M. Zurek, ``{Light Dark Matter: Models and
  Constraints},'' \href{http://dx.doi.org/10.1103/PhysRevD.96.115021}{{\em
  Phys. Rev. D} {\bfseries 96} no.~11, (2017) 115021},
  \href{http://arxiv.org/abs/1709.07882}{{\ttfamily arXiv:1709.07882
  [hep-ph]}}.

\bibitem{Batell:2016ove}
B.~Batell, N.~Lange, D.~McKeen, M.~Pospelov, and A.~Ritz, ``{Muon anomalous
  magnetic moment through the leptonic Higgs portal},''
  \href{http://dx.doi.org/10.1103/PhysRevD.95.075003}{{\em Phys. Rev. D}
  {\bfseries 95} no.~7, (2017) 075003},
  \href{http://arxiv.org/abs/1606.04943}{{\ttfamily arXiv:1606.04943
  [hep-ph]}}.

\bibitem{Batell:2017kty}
B.~Batell, A.~Freitas, A.~Ismail, and D.~Mckeen, ``{Flavor-specific scalar
  mediators},'' \href{http://dx.doi.org/10.1103/PhysRevD.98.055026}{{\em Phys.
  Rev. D} {\bfseries 98} no.~5, (2018) 055026},
  \href{http://arxiv.org/abs/1712.10022}{{\ttfamily arXiv:1712.10022
  [hep-ph]}}.

\bibitem{Electroweak:2003ram}
{\bfseries LEP, ALEPH, DELPHI, L3, OPAL, LEP Electroweak Working Group, SLD
  Electroweak Group, SLD Heavy Flavor Group} Collaboration, t.~S. Electroweak,
  ``{A Combination of preliminary electroweak measurements and constraints on
  the standard model},'' \href{http://arxiv.org/abs/hep-ex/0312023}{{\ttfamily
  arXiv:hep-ex/0312023}}.

\bibitem{Christensen:2008py}
N.~D. Christensen and C.~Duhr, ``{FeynRules - Feynman rules made easy},''
  \href{http://dx.doi.org/10.1016/j.cpc.2009.02.018}{{\em Comput. Phys.
  Commun.} {\bfseries 180} (2009) 1614--1641},
  \href{http://arxiv.org/abs/0806.4194}{{\ttfamily arXiv:0806.4194 [hep-ph]}}.

\bibitem{Alwall:2014hca}
J.~Alwall, R.~Frederix, S.~Frixione, V.~Hirschi, F.~Maltoni, O.~Mattelaer,
  H.~S. Shao, T.~Stelzer, P.~Torrielli, and M.~Zaro, ``{The automated
  computation of tree-level and next-to-leading order differential cross
  sections, and their matching to parton shower simulations},''
  \href{http://dx.doi.org/10.1007/JHEP07(2014)079}{{\em JHEP} {\bfseries 07}
  (2014) 079}, \href{http://arxiv.org/abs/1405.0301}{{\ttfamily arXiv:1405.0301
  [hep-ph]}}.

\bibitem{OPAL:2003kcu}
{\bfseries OPAL} Collaboration, G.~Abbiendi {\em et~al.}, ``{Tests of the
  standard model and constraints on new physics from measurements of fermion
  pair production at 189-GeV to 209-GeV at LEP},''
  \href{http://dx.doi.org/10.1140/epjc/s2004-01595-9}{{\em Eur. Phys. J. C}
  {\bfseries 33} (2004) 173--212},
  \href{http://arxiv.org/abs/hep-ex/0309053}{{\ttfamily arXiv:hep-ex/0309053}}.

\bibitem{ALEPH:2013htx}
{\bfseries ALEPH, DELPHI, L3, OPAL, LEP} Collaboration, G.~Abbiendi {\em
  et~al.}, ``{Search for Charged Higgs bosons: Combined Results Using LEP
  Data},'' \href{http://dx.doi.org/10.1140/epjc/s10052-013-2463-1}{{\em Eur.
  Phys. J. C} {\bfseries 73} (2013) 2463},
  \href{http://arxiv.org/abs/1301.6065}{{\ttfamily arXiv:1301.6065 [hep-ex]}}.

\bibitem{Sjostrand:2006za}
T.~Sjostrand, S.~Mrenna, and P.~Z. Skands, ``{PYTHIA 6.4 Physics and Manual},''
  \href{http://dx.doi.org/10.1088/1126-6708/2006/05/026}{{\em JHEP} {\bfseries
  05} (2006) 026},
\href{http://arxiv.org/abs/hep-ph/0603175}{{\ttfamily arXiv:hep-ph/0603175
  [hep-ph]}}.

\bibitem{Sjostrand:2014zea}
T.~Sjöstrand, S.~Ask, J.~R. Christiansen, R.~Corke, N.~Desai, P.~Ilten,
  S.~Mrenna, S.~Prestel, C.~O. Rasmussen, and P.~Z. Skands, ``{An Introduction
  to PYTHIA 8.2},'' \href{http://dx.doi.org/10.1016/j.cpc.2015.01.024}{{\em
  Comput. Phys. Commun.} {\bfseries 191} (2015) 159--177},
\href{http://arxiv.org/abs/1410.3012}{{\ttfamily arXiv:1410.3012 [hep-ph]}}.

\bibitem{deFavereau:2013fsa}
{\bfseries DELPHES 3} Collaboration, J.~de~Favereau, C.~Delaere, P.~Demin,
  A.~Giammanco, V.~Lemaître, A.~Mertens, and M.~Selvaggi, ``{DELPHES 3, A
  modular framework for fast simulation of a generic collider experiment},''
  \href{http://dx.doi.org/10.1007/JHEP02(2014)057}{{\em JHEP} {\bfseries 02}
  (2014) 057},
\href{http://arxiv.org/abs/1307.6346}{{\ttfamily arXiv:1307.6346 [hep-ex]}}.

\bibitem{ILCgen}
\url{https://github.com/iLCSoft/ILCDelphes/blob/master/Readme.md}.

\bibitem{MuonCollider}
\url{https://github.com/delphes/delphes/blob/master/cards/delphes_card_MuonColliderDet.tcl}.

\bibitem{Belle-II:2010dht}
{\bfseries Belle-II} Collaboration, T.~Abe {\em et~al.}, ``{Belle II Technical
  Design Report},'' \href{http://arxiv.org/abs/1011.0352}{{\ttfamily
  arXiv:1011.0352 [physics.ins-det]}}.

\bibitem{Belle-II:2018jsg}
{\bfseries Belle-II} Collaboration, W.~Altmannshofer {\em et~al.}, ``{The Belle
  II Physics Book},'' \href{http://dx.doi.org/10.1093/ptep/ptz106}{{\em PTEP}
  {\bfseries 2019} no.~12, (2019) 123C01},
  \href{http://arxiv.org/abs/1808.10567}{{\ttfamily arXiv:1808.10567
  [hep-ex]}}. [Erratum: PTEP 2020, 029201 (2020)].

\bibitem{TEXONO:2010tnr}
{\bfseries TEXONO} Collaboration, M.~Deniz {\em et~al.}, ``{Constraints on
  Non-Standard Neutrino Interactions and Unparticle Physics with
  Neutrino-Electron Scattering at the Kuo-Sheng Nuclear Power Reactor},''
  \href{http://dx.doi.org/10.1103/PhysRevD.82.033004}{{\em Phys. Rev. D}
  {\bfseries 82} (2010) 033004},
  \href{http://arxiv.org/abs/1006.1947}{{\ttfamily arXiv:1006.1947 [hep-ph]}}.

\bibitem{Esteban:2020cvm}
I.~Esteban, M.~C. Gonzalez-Garcia, M.~Maltoni, T.~Schwetz, and A.~Zhou, ``{The
  fate of hints: updated global analysis of three-flavor neutrino
  oscillations},'' \href{http://dx.doi.org/10.1007/JHEP09(2020)178}{{\em JHEP}
  {\bfseries 09} (2020) 178}, \href{http://arxiv.org/abs/2007.14792}{{\ttfamily
  arXiv:2007.14792 [hep-ph]}}.

\bibitem{Lindner:2016bgg}
M.~Lindner, M.~Platscher, and F.~S. Queiroz, ``{A Call for New Physics : The
  Muon Anomalous Magnetic Moment and Lepton Flavor Violation},''
  \href{http://dx.doi.org/10.1016/j.physrep.2017.12.001}{{\em Phys. Rept.}
  {\bfseries 731} (2018) 1--82},
  \href{http://arxiv.org/abs/1610.06587}{{\ttfamily arXiv:1610.06587
  [hep-ph]}}.

\end{thebibliography}\endgroup


\providecommand{\href}[2]{#2}\begingroup\raggedright\endgroup

\end{document}